\documentclass[aps,prx,twocolumn,longbibliography,superscriptaddress,noeprint,reprint]{revtex4-2}

\usepackage{graphicx}
\usepackage{mathtools}
\usepackage{tikz}
\usepackage{siunitx}
\usepackage{xcolor}
\usepackage[normalem]{ulem}
\usepackage{hyperref}
\usepackage{wrapfig}
\usepackage{xr}
\usepackage[english]{babel}

\usepackage{float}
\usepackage{multirow}
\usepackage{sidecap}

\usepackage{xcolor}
\hypersetup{
    colorlinks,
    linkcolor={red!50!black},
    citecolor={blue!50!black},
    urlcolor={blue!80!black}
}

\usepackage{color}
\definecolor{blue}{rgb}{0.00,0.00,0.95}
\begin{document}

\title{Data-driven theory reveals protrusion and polarity interactions governing collision behavior of distinct motile cells}

\author{Tom Brandst\"atter}
\affiliation{Department of Physics and Astronomy, Vrije Universiteit Amsterdam, De Boelelaan 1105, 1081 HV Amsterdam, The Netherlands}
\affiliation{Arnold-Sommerfeld-Center for Theoretical Physics, Ludwig-Maximilians-Universität München, Theresienstr. 37, 80333 Munich, Germany}
\author{Emily Brieger}
\affiliation{Faculty of Physics and Center for NanoScience, Ludwig-Maximilians-Universität München, Geschwister-Scholl-Platz 1, 80539 Munich, Germany}
\author{David B. Brückner}
\affiliation{Institute of Science and Technology Austria, Am Campus 1, 3400 Klosterneuburg, Austria}
\author{Georg Ladurner}
\affiliation{Faculty of Physics and Center for NanoScience, Ludwig-Maximilians-Universität München, Geschwister-Scholl-Platz 1, 80539 Munich, Germany}
\affiliation{Center for Medical Physics and Biomedical Engineering, Währinger Gürtel 18-20, 1090 Wien, Austria}
\author{Joachim O. R\"adler}
\affiliation{Faculty of Physics and Center for NanoScience, Ludwig-Maximilians-Universität München, Geschwister-Scholl-Platz 1, 80539 Munich, Germany}
\author{Chase P. Broedersz}
 \thanks{Corresponding author: c.p.broedersz@vu.nl, Phone: +31 20 59 82953}
\affiliation{Department of Physics and Astronomy, Vrije Universiteit Amsterdam, De Boelelaan 1105, 1081 HV Amsterdam, The Netherlands}
\affiliation{Arnold-Sommerfeld-Center for Theoretical Physics, Ludwig-Maximilians-Universität München, Theresienstr. 37, 80333 Munich, Germany}

\begin{abstract}
The migration behavior of colliding cells is critically determined by transient contact-interactions. During these interactions, the motility machinery, including the front-rear polarization of the cell, dynamically responds to surface protein-mediated transmission of forces and biochemical signals between cells. While biomolecular details of such contact-interactions are increasingly well understood, it remains unclear what biophysical interaction mechanisms govern the cell-level dynamics of colliding cells and how these mechanisms vary across cell types. Here, we develop a phenomenological theory based on 14 candidate contact-interaction mechanisms coupling cell position, protrusion, and polarity. Using high-throughput micropattern experiments, we detect which of these phenomenological contact-interactions captures the interaction behaviors of cells. We find that various cell types - ranging from mesenchymal to epithelial cells - are accurately captured by a single model with only two interaction mechanisms: polarity-protrusion coupling and polarity-polarity coupling. The qualitatively different interaction behaviors of distinct cells, as well as cells subject to molecular perturbations of surface protein-mediated signaling, can all be quantitatively captured by varying the strength and sign of the polarity-polarity coupling mechanism. Altogether, our data-driven phenomenological theory of cell-cell interactions reveals polarity-polarity coupling as a versatile and general contact-interaction mechanism, which may underlie diverse collective migration behavior of motile cells.
\end{abstract}

\maketitle

\section{Introduction}
\begin{samepage}
	Contact-interactions between cells control the coordinated migration of tissues during fundamental physiological processes \cite{Haeger2015} in development, health, and disease \cite{Scarpa2016, Poujade2007, Friedl2009}. Already at the level of two motile cells, contact-interactions can lead to dramatic changes in the trajectories of cells after they collide \cite{Vedel2013, Abercrombie1979, Scarpa2016b, Bruckner2021}. Such interaction behavior is not merely the result of physical forces arising, for example, from membrane adhesion or deformations of the colliding cells. Instead, contact-interactions trigger a dynamical response in the intracellular biomolecular cell motility machinery \cite{Roycroft2016, Collins2015, Ladoux2017}. This machinery involves establishing a front-rear polarization \cite{Ridley2011, Rappel2017}, as well as active cytoskeleton contraction and protrusion formation, enabling cells to self-propel \cite{Danuser2013}. Contact-interactions can modify cell polarity via biochemical signaling between cells \cite{Roycroft2016, Kadir2011}, reflecting the active, responsive, and adaptive nature of cell motility and cell-cell interactions. Revealing the dominant interaction mechanisms that steer cell polarity and cell migration in response to cell-cell contacts is critical for a general understanding of the dynamics of interacting cells.
\end{samepage}

\noindent The interaction behavior of migrating cells in various processes is highly diverse. During wound healing \cite{Poujade2007}, for instance, motile epithelial cells follow each other after making contact \cite{Theveneau2013, Jain2020}. By contrast, several developmental processes rely on cells retreating from each other after forming cell-cell contacts \cite{Carmona-Fontaine2008,Theveneau2010,Davis2012,Villar-Cervino2013a}, termed Contact Inhibition of Locomotion (CIL) \cite{Abercrombie1979,Mayor2010}. Many details of the biomolecular machinery underlying these and other interaction behaviors \cite{Dalessandro2017,Hayakawa2019a} are now becoming increasingly clear. For instance, force transmission between epithelial cells is enabled by E-Cadherin mediated junctions \cite{Mendonsa2018,Maitre2013}, while receptor-ligand Eph-ephrin interactions \cite{Villar-Cervino2013a,Stramer2017,Astin2010,Kadir2011,Noren2004} enable cell-cell recognition during CIL. Both of these surface proteins enable cells to regulate their polarity machinery in response to cell-cell contacts \cite{Ladoux2017,Noren2004,Desai2009}. Such molecular pathways vary between different cell types and can change, for instance, during the epithelial-to-mesenchymal transition (EMT) \cite{Scarpa2015,wong_collective_2014}. However, it remains a major challenge to understand how contact-interactions are controlled by the molecular machinery in various cells types.\\
\\The complexity of the motility and cell-cell interaction machinery makes it difficult to gain mechanistic understanding of contact-interactions. Nevertheless, biophysical models can give mechanistic insight into single cell motility \cite{Bruckner2022,Flommersfeld2024,Danuser2013,Edelstein-Keshet2013,Sens2020} and how force transmission and cell-cell recognition give rise to the dynamics of interacting cells \cite{Camley2014,Smeets2016,Sepulveda2013,Ron2023a,Dalessandro2017,Copenhagen2018,Vercruysse2024,Kulawiak2016,Jain2020,Desai2013,Zisis2022,Alert2020,Lober2015,Vedel2013,Bertrand2024, vagne2024generictheoryinteractingspinning}. However, these bottom-up models are commonly tailored to understand specific cell types in concrete settings, making it difficult to achieve a unifying conceptual understanding of contact-interactions. 
\begin{figure}[t!]
	\includegraphics[width=0.48\textwidth]{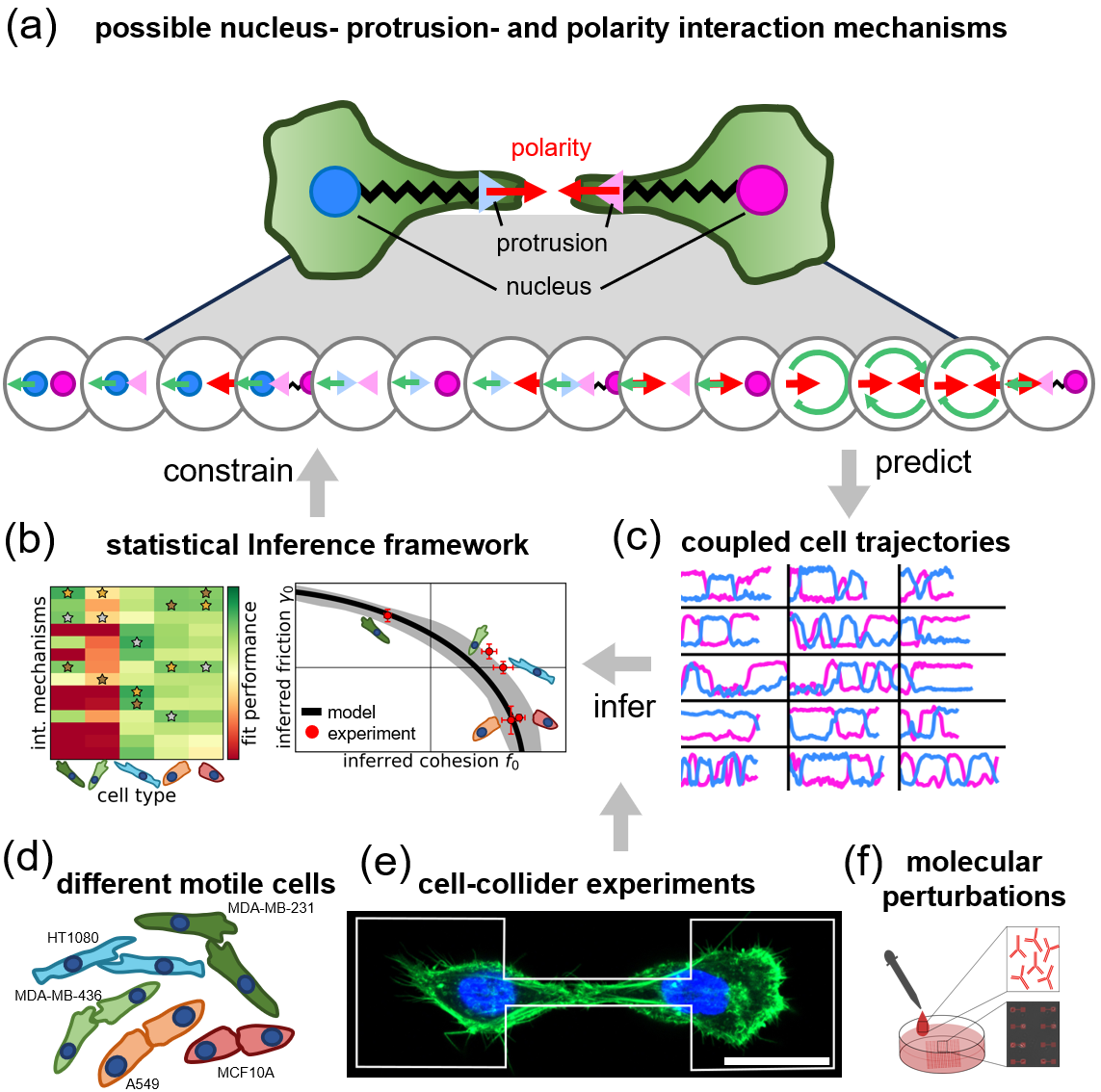}
	\centering
	\caption{\textbf{Schematic of our data-driven theoretical approach.} (a) Phenomenological interaction mechanisms coupling the nuclei, protrusions, and polarities of two interacting cells. We show a detailed analysis of these mechanisms in Fig. \ref{Fig1}. (b) Statistical inference of our overdamped model using the experimental data of each cell type. The heatmap shows the best fit performance of each interaction mechanism for each cell type. We quantify the experimental dynamics by inferring effective interaction terms that describe the underdamped nucleus dynamics of our experimental data (red points) and the phenomenological model (curves) \cite{Bruckner2021}. These terms constrain our phenomenological model. (c) We experimentally observe and predict the trajectories of the nuclei of the two cells, which quantifies cell-level interaction behavior. (d)-(f) Our experimental setup consists of cell-collider experiments, which we perform on dumbbell-shaped micropatterns. Scale bar: $25\ \si{\micro\meter}$, nuclei are stained in blue, actin is stained in green (e). In addition to considering different motile cells ranging from mesenchymal cancer cells to epithelial cells (d), we also consider various molecular perturbations of the surface protein-driven interaction machinery (f).}
	\label{Fig0}
\end{figure}
Furthermore, biophysical models for contact-interactions are rarely systematically constrained on quantitative experimental data \cite{Bruckner_2024}.\\ 
\\High-throughput experimental micropattern assays \cite{Bruckner2021,Scarpa2016b,Ron2023a,Bertrand2024} enabled the application of data-driven inference approaches to learn  dynamical descriptions of migrating and interacting cells directly from large datasets of experimental cell trajectories \cite{Bruckner2021,Bruckner2020,Bruckner_2024,Agliari2020}. In previous work, we inferred the effectively underdamped, cell level dynamics of interacting motile cells on micropatterns. The interaction behaviors of these cells could be captured by cohesion- and friction-like interactions between the cell nucleus positions \cite{Bruckner2021}. Importantly however, this approach neither considered the protrusion nor the polarity dynamics which are biophysically relevant for interacting cells. Thus, it remains unclear what interaction mechanisms involving cell shape and the polarity-driven motility and interaction machinery govern the cell-level dynamics of interacting cells and how such mechanisms vary across cell types.\\ 
\\Here, we develop a biophysical theory for contact-interaction mechanisms systematically constrained on a large data set of interacting cells for a broad range of cell types [Fig. \ref{Fig0}]. We employ a phenomenological approach using symmetry and simplicity arguments to derive 14 possible interaction mechanisms coupling the nucleus, protrusion, cell deformation, and polarity of pairs of motile cells. We combine Underdamped Langevin Inference~\cite{Bruckner2021, Bruckner2020} with a statistical inference method, which we develop here to capture the overdamped protrusion and polarity dynamics. Using this quantitative overdamped inference framework, we detect from data which contact-interaction mechanisms dominate cell-level interaction behavior across cell types and how these mechanisms change in response to molecular perturbations of the interaction machinery. To achieve this goal, we had to go beyond our previous work \cite{Bruckner2021} and increase the experimental data set by an order of magnitude, now including roughly 50,000 hours of microscopy videos of interacting cells. This data set now contains 5 distinct cell types, ranging from more mesenchymal to more epithelial, and various molecular perturbations of the cellular interaction machinery. Our data-driven theory reveals that all considered cell types can be well described by only two interaction mechanisms: Polarity-polarity coupling promoting alignment or anti-alignment of cell polarity and polarity-protrusion coupling.

\section{Results}

%------------------------------------------------------------
\subsection{
Theory of contact-interaction mechanisms}

\begin{figure*}[t!]
	\includegraphics[width=\textwidth]{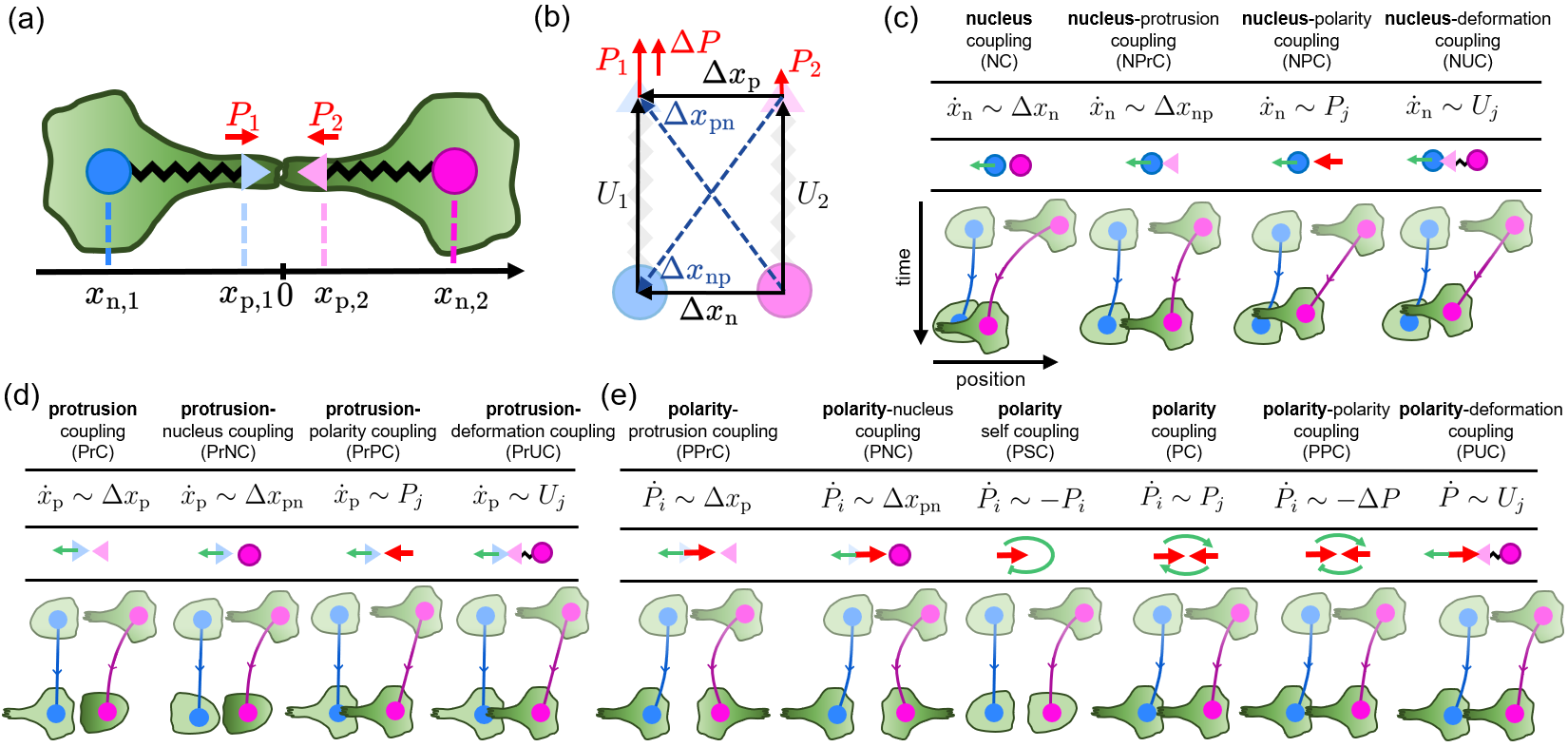}
	\centering
	\caption{
		\textbf{Phenomenological contact-interaction mechanisms.} 
		(a) To describe the interaction behavior of a whole cell, we include three dynamical degrees of freedom in our model: the position of the nucleus $x_\mathrm{n}$, the position of the protrusion $x_\mathrm{p}$, and the polarity of a cell $P$. The polarity determines a self-propulsion force, giving rise to cell migration [Appendix \ref{app:model_development}]. (b) Schematic representation of all linear vectorial quantities in our model for both cell 1 and cell 2. $\Delta P = P_i - P_j$ is the difference between polarities and $U_i = x_{\mathrm{p},i} - x_{\mathrm{n},i}$ is the deformation vector of the cells. (c)-(e) Possible candidate cell-cell interaction mechanisms acting on the nucleus (c), on the protrusion (d) and on the polarity of cells (e). In all panels (c)-(e), the upper row of the table indicates the linear coupling of these mechanisms, and the middle row shows a schematic representation of that coupling at the associated interaction strength $\epsilon$ being positive. In these schematics, the green arrows always indicate the dynamics arising from the mechanisms for the "blue" cell. The lower row of the tables shows then a schematic of the expected long-time scale interaction behaviors of both cells for one specific collision scenario. Here the different shapes of the cells indicate wether cells are elongated (nuclei and protrusions are sufficiently separated) and the green color gradient indicates the polarity of the cells. All mechanisms are described in detail in Supplementary Section 4.1}
	\label{Fig1}
\end{figure*}

To develop a dynamical theory of contact-interactions, we employ a minimal large-scale description where the cell’s nucleus position $\mathbf{x}_\mathrm{n} (t)$ and the leading protrusion position $\mathbf{x}_\mathrm{p} (t)$ are the relevant positional degrees of freedom of a migrating cell [Fig. \ref{Fig1}(a)]. This simple choice captures the cell’s position, as well as the elongated and dynamic shape of migrating cells observed widely in single-cell \cite{Bruckner2019a,Bruckner2022} and cell-cell interaction \cite{Vedel2013, Abercrombie1979, Scarpa2016b, Bruckner2021} experiments. To complete our description, we include an internal degree of freedom $\mathbf{P}(t)$ to capture cell polarity. The polarity describes the cell’s anisotropic organization of the cytoskeletal motility machinery, distinguishing front and rear \cite{Ridley2011}. This minimal level of description has proven adequate to describe single cell migration \cite{Bruckner2022,Flommersfeld2024}, thereby providing a foundation to develop a general theory for how contact-interactions determine cell protrusions, polarity, and the dynamics of migrating cells.\\
\\We define the dynamics of interacting cells by overdamped equations of motion for the positional degrees of freedom together with a stochastic description of cell polarity. For simplicity, we consider interaction behavior of cells in 1D, and write for cell $i$:

\begin{align}
	\zeta_\mathrm{n}\dot{x}_{\mathrm{n},i} &= F_{\mathrm{n}}(x_{\mathrm{n},i},x_{\mathrm{p},i}) + G_{\mathrm{n}}(\Delta x, U, P, \Delta P)\label{eq:1}\\
	\zeta_\mathrm{p}\dot{x}_{\mathrm{p},i} &= F_{\mathrm{p}}(x_{\mathrm{n},i},x_{\mathrm{p},i}) + P_i(t) + G_{\mathrm{p}}(\Delta x, U, P, \Delta P)\label{eq:2}\\
	\dot{P}_i &= F_{\mathrm{P}}(x_{\mathrm{p},i}, P_i) + G_{\mathrm{P}}(\Delta x, U, P, \Delta P) + \sigma \eta_i(t).\label{eq:3}
\end{align}
Here, $\zeta_\mathrm{n}$ and $\zeta_\mathrm{p}$ are the friction coefficients of the nucleus and the protrusion. Furthermore, the polarity directly drives the protrusion \cite{Pollard2003} and the functions $F_{\mathrm{n}}$,  $F_{\mathrm{p}}$, and $F_{\mathrm{P}}$ are one-body terms describing the single-cell behavior including a mechanical coupling between nucleus and protrusion, as well as a response of the cells to their micro-environment \cite{Bruckner2022} [Appendix \ref{app:model_development}]. These single-cell terms can be derived from a microscopic theory of cell migration \cite{Flommersfeld2024,Sens2020,Rappel2017} and are constrained by experimental data of single migrating cells \cite{Bruckner2022}. In this model, stochasticity arises in the polarity dynamics, which is modeled by Gaussian white noise $\eta_i (t)$ with amplitude $\sigma$. Taken together, by including both protrusion, cell deformation, and polarity in our overdamped description of a migrating cell, we go beyond previous models \cite{Bruckner2021, Sepulveda2013} of the underdamped dynamics of the cell as a whole, allowing us to investigate how contact-interaction mechanisms couple to the protrusion and polarity.\\
\\The main goal of our theoretical approach is to learn the two-body terms $G_{\mathrm{n}}$, $G_{\mathrm{p}}$, and $G_{\mathrm{P}}$ in [Eqs. (\ref{eq:1})-(\ref{eq:3})]. These terms encode nucleus- protrusion- and polarity interaction mechanisms and thus may depend on the distances $\Delta x$ between nucleus and protrusion, cell deformation $U=x_{\mathrm{p}} - x_{\mathrm{n}}$, polarity $P$, and polarity differences $\Delta P = P_i - P_j$ [Fig. \ref{Fig1}(b)]. Deriving these terms rigorously from detailed biophysical and biochemical signaling processes and mechanical couplings between cells is currently unfeasible due to the large number of involved molecular mechanisms. Therefore, we employ a phenomenological approach: 
\begin{figure*}[t!]
	\includegraphics[width=\textwidth]{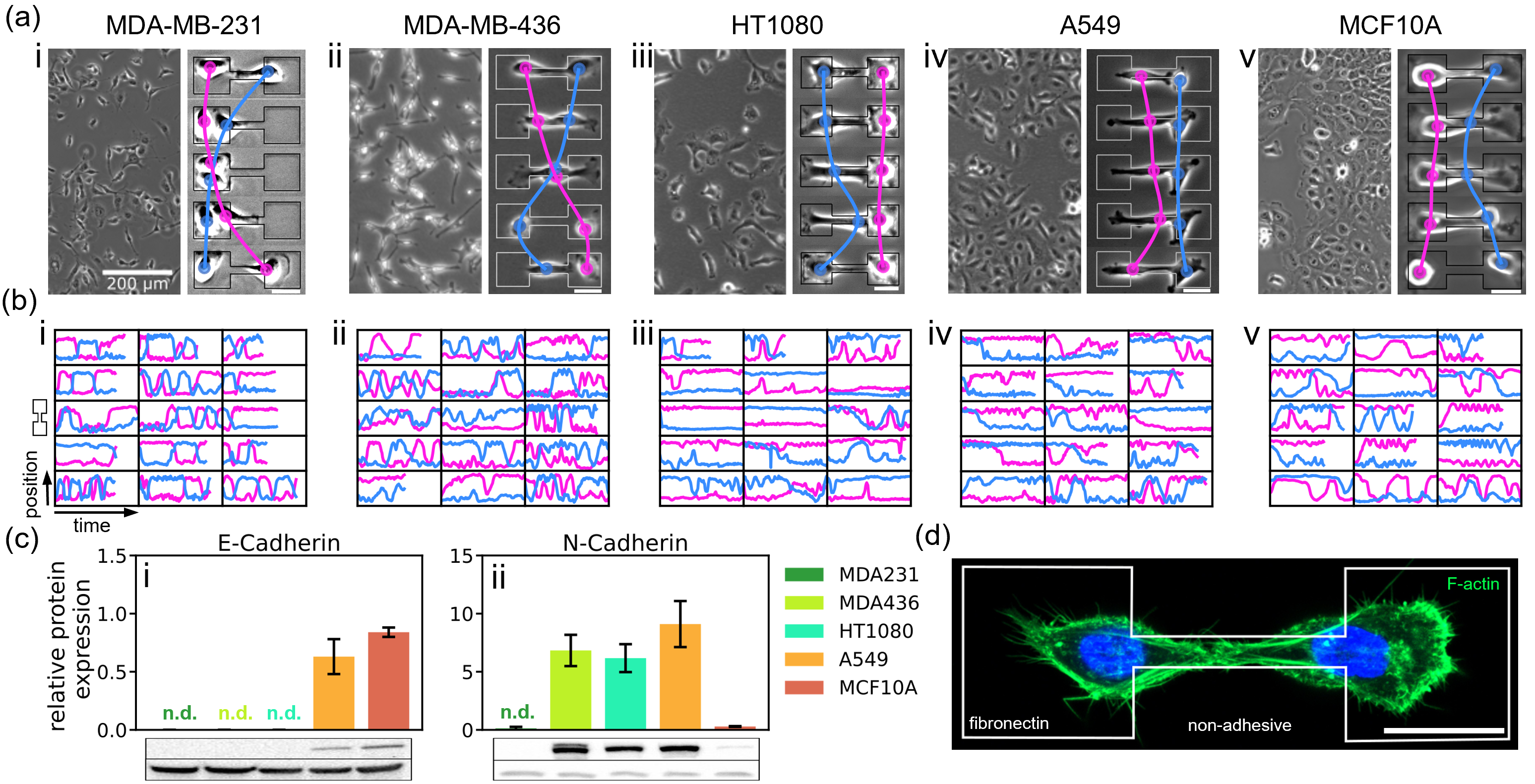}
	\centering
	\caption{
		\textbf{Studying transient cell collisions on a minimal cell collider.} (a) Brightfield microscopy images of (i) MDA-MB-231 cells, (ii) MDA-MB-436 cells, (iii) HT1080 cells, (iv) A549 cells and (v) MCF10A cells. Left panels show multicellular tissues, scale bar: $200 \ \si{\micro\meter}$ throughout. Right panels show time series of brightfield images of two cells colliding repeatedly while hopping between the islands. Scale bar: $25\ \si{\micro\meter}$. (b) Small selection of nucleus trajectories for the five different cell lines. (c) Quantitative western blot analysis of protein expression levels of E-Cadherin and N-Cadherin in the different cell lines. We show one blot for each protein with its corresponding loading control ($\beta$-actin) [Appendix \ref{app:western_blot}]. Error bars indicate the error of the mean (s.e.m) of triplicate measurements. (d) Two migrating MCF10A cells (F-actin stained in green, nucleus in blue) are confined on a dumbbell-shaped micropattern (white outline), which consists of two islands connected by a bridge. The micropattern is coated with fibronectin and is surrounded by a non-adhesive PLL-PEG layer. Scale bar: $25\ \si{\micro\meter}$. 
	}
	\label{Fig2}
\end{figure*}
In an unbiased way, we propose a large set of possible candidate cell-cell interaction mechanisms arising from the lowest-order linear couplings of the degrees of freedom that behave like vectors in arbitrary dimensions obeying rotational and translational symmetry [Fig. \ref{Fig1}(b), Appendix \ref{app:mechanisms}]. As there are 9 different linear vectorial couplings [Fig. \ref{Fig1}(b)] possibly influencing the 3 degrees of freedom in our model through $G_{\mathrm{n}}$, $G_{\mathrm{p}}$, and $G_{\mathrm{P}}$, we consider in total 27 phenomenological cell-cell interaction mechanisms. Further, we impose that interaction mechanisms cannot introduce couplings of the degrees of freedom of the same cell that are not already introduced by the single cell terms [Appendix \ref{app:mechanisms}]. This reduces the set of interactions to 14 candidates, ranging from couplings of the nuclei and protrusions via repulsion or attraction to various interactions affecting cell polarity [Fig. \ref{Fig1}(c)-(e)].\\
\\In anticipation of results below, we highlight three mechanisms: polarity-protrusion coupling (PPrC), describing how cells orient their polarity away or towards the protrusion of the other cell, polarity-polarity coupling (PPC) allowing for alignment or anti-alignment of cell polarity, and polarity-deformation coupling (PUC) allowing alignment or anti-alignment of cell polarity to the deformation vector of the other cell [Fig. \ref{Fig1}(e)]. For all interaction mechanisms, we impose a typical interaction range $r$ and interaction strength $\epsilon$. For instance, for polarity-polarity coupling, we write $G_{\mathrm{P}} = -\epsilon_{\mathrm{PPC}}e^{-|\Delta x_\mathrm{p}|/r_{\mathrm{PPC}}}\Delta P$, where $\Delta P=P_i - P_j$ and $\Delta x_\mathrm{p} = x_{\mathrm{p},i}-x_{\mathrm{p},j}$ [Supplementary Section 4.1]. Altogether, these interaction mechanisms give rise to various prototypical cell-cell interaction behaviors [Fig. \ref{Fig1}(e)]. In principle, all 14 candidate cell-cell interaction mechanisms may contribute to the behavior of interacting cells. Additionally, distinct cell types may exhibit different cell-cell interactions, further exacerbating the challenge to reveal the cell-cell interaction mechanisms underlying behavior. Therefore, we require strong quantitative model constraints from experiments to discover what interaction mechanisms are relevant to describe interaction behavior across a broad range of cell types. 

\subsection{Cell-cell collision experiments reveal diversity of interaction behaviors}

\begin{figure*}[t!]
	\includegraphics[width=\textwidth]{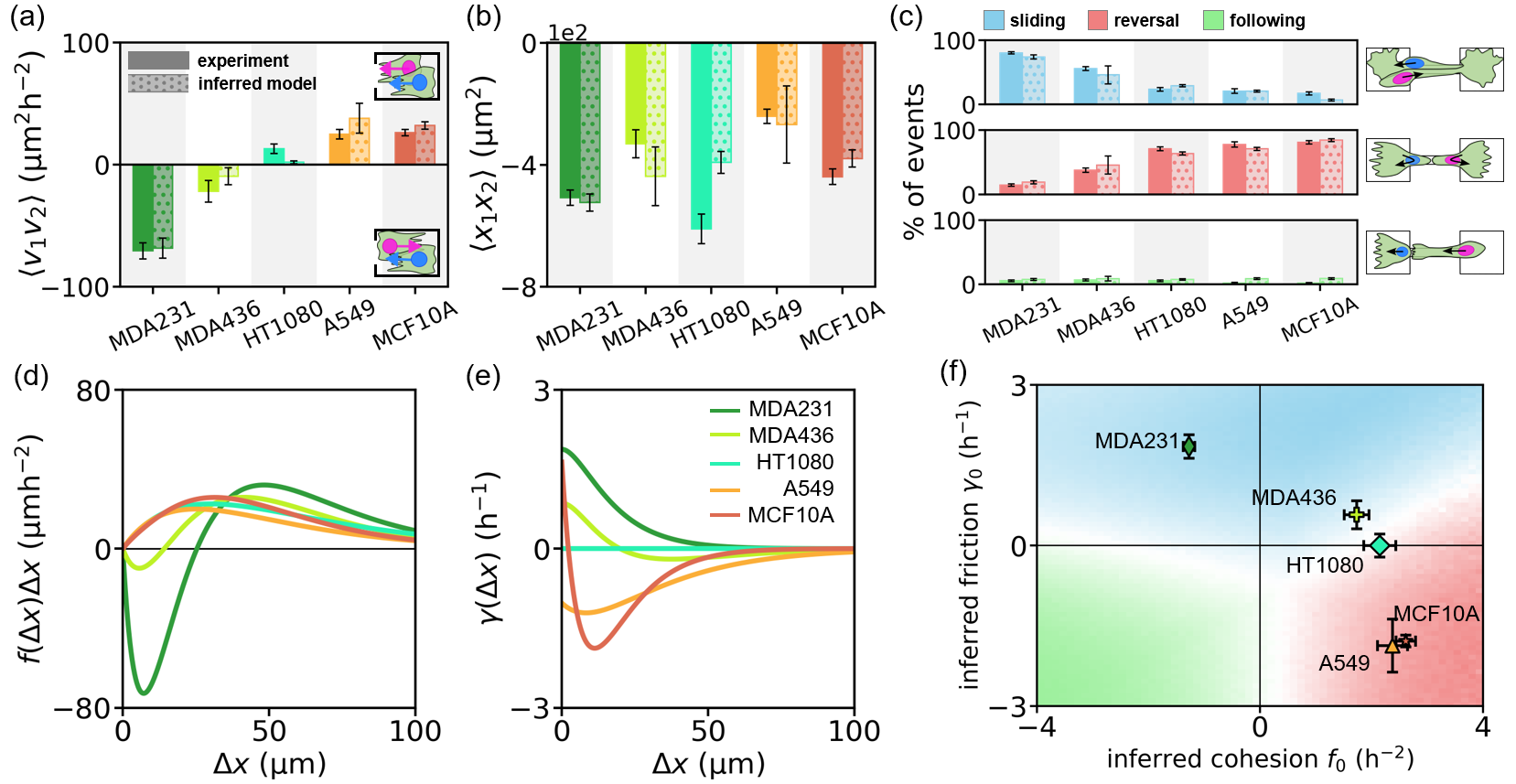}
	\centering
	\caption{
		\textbf{Quantifying interaction behavior using data-driven inference.} (a) Instantaneous velocity cross-correlation $C_V (|t-t'|=0)$ between the two cells when they occupy the same island. Solid bars show experimental results, dotted bars show the prediction of the inferred underdamped description. For panels (a)-(c), error bars show the error of the mean (s.e.m) obtained from bootstrapping the experimental data. (b) Instantaneous position cross-correlation $C_X(|t-t'|=0)$. (c) Percentages of the different collision events for the different cell lines. (d) Inferred cohesion interactions $f(\Delta x)\Delta x$ for the five different cell lines. (e) Inferred friction interactions $\gamma(\Delta x)\Delta v$ of the various cell lines. (f) Interaction behavior space spanned by the amplitudes of the inferred cohesion and friction interactions, $f_0$ and $\gamma_0$. Colors show the dominant collision event predicted in that parameter region. Throughout this figure, results for MCF10A and MDA-MB-231 cells have been obtained from data adapted from  \cite{Bruckner2021}.  
	}
	\label{Fig3}
\end{figure*}

To detect contact-interactions in experiments, we use a high-throughput assay to study the dynamics of homotypic pairs of interacting motile cells. Specifically, we employ a dumbbell-shaped micropattern as a minimal cell collider \cite{Bruckner2021} [Fig. \ref{Fig2}(d)]. This geometry effectively confines cells in 1D, isolates cell pairs, and generates many cell-cell collision events. To capture a variety of cell-cell interactions across cell types in parallel, we consider a range of distinct motile cells from human tissue. We study the two metastatic breast cancer cell lines MDA-MB-231 and MDA-MB-436, the fibrosarcoma cell line HT-1080, the non-metastatic lung cancer cell line A549, and the epithelial non-cancerous breast cell line MCF10A [Fig. \ref{Fig2}(a)]. These cell lines express different levels of adhesion proteins and exhibit distinct collective behaviors in vitro: MDA-MB-231 cells neither express the cell adhesion molecule E-cadherin nor N-Cadherin [Fig. \ref{Fig2}(c)], and do not form monolayers [Fig. \ref{Fig2}(a) i]. The breast cancer cells MDA-MB-436 and fibrosarcoma cells express N-Cadherin [Fig. \ref{Fig2}(c) ii] and also do not form monolayers [Fig. \ref{Fig2}(a) ii,iii]. These features are characteristic of a mesenchymal phenotype \cite{Blick2008}. In contrast, MCF10A and A549 cells express E-cadherin [Fig. \ref{Fig2}(c) i] and form monolayers [Fig. \ref{Fig2}(a) iv,v], characteristic of an epithelial phenotype \cite{Blick2008}. All tested cell lines are motile on our micropattern and repeatedly collide with each other [Fig. \ref{Fig2}(a), Supplementary movie 1]. However, the collision behavior of cells is variable with marked differences across the five cell lines. To quantify the behavior of pairs of colliding cells in our experiments, we use a low-dimensional representation of these interaction dynamics by tracking the 1D motion of the nucleus of both cells over time [Fig. \ref{Fig2}(b)]. This approach yields a large ensemble of experimental trajectory data reflecting the cell-level dynamics of interacting cells.\\
\\To characterize the interaction dynamics of distinct cell types, we use various interaction behavior statistics \cite{Bruckner2021}. First, we quantify how cells coordinate their behaviors in close proximity using a velocity cross-correlation function of two cells occupying the same island: $C_V(|t-t'|)=\langle v_1(t) v_2(t')\rangle_{\mathrm{same}}$ [Appendix \ref{app:behavior_statistics}]. The breast cancer and fibrosarcoma cell lines exhibit negative or very small instantaneous velocity correlations, while the epithelial cells exhibit positive instantaneous velocity correlations [Fig. \ref{Fig3}(a)]. Furthermore, we determine a position correlation function $C_X (|t-t'|)=\langle x_1 (t) x_2 (t')\rangle$. All cell types exhibit negative instantaneous position correlations [Fig. \ref{Fig3}(b)], indicating mutual exclusion behavior. To gain insight into how cells navigate each other on longer time scales, we analyze cell-cell collisions \cite{Bruckner2021}. We detect sliding events (two cells interchange positions), reversal events (at least one cell retracts) and following events (both cells transition simultaneously) [Appendix \ref{app:behavior_statistics}]. The two breast cancer cell lines mostly exhibit sliding behavior, while the epithelial and fibrosarcoma cells mostly exhibit reversal behavior [solid bars Fig. \ref{Fig3}(c)]. Altogether, these results show different dominant interaction behaviors across our cell types with marked differences between epithelial and cancer cell lines.

%%%%%%%%%%%%%%%%%%%%%%%%%%%%%%%%%%%%%
%%FIGURE 

%%%%%%%%%%%%%%%%%%%%%%%%%%%%%%%%%%%%%
%%FIGURE 
\begin{figure*}[t!]
	\includegraphics[width=\textwidth]{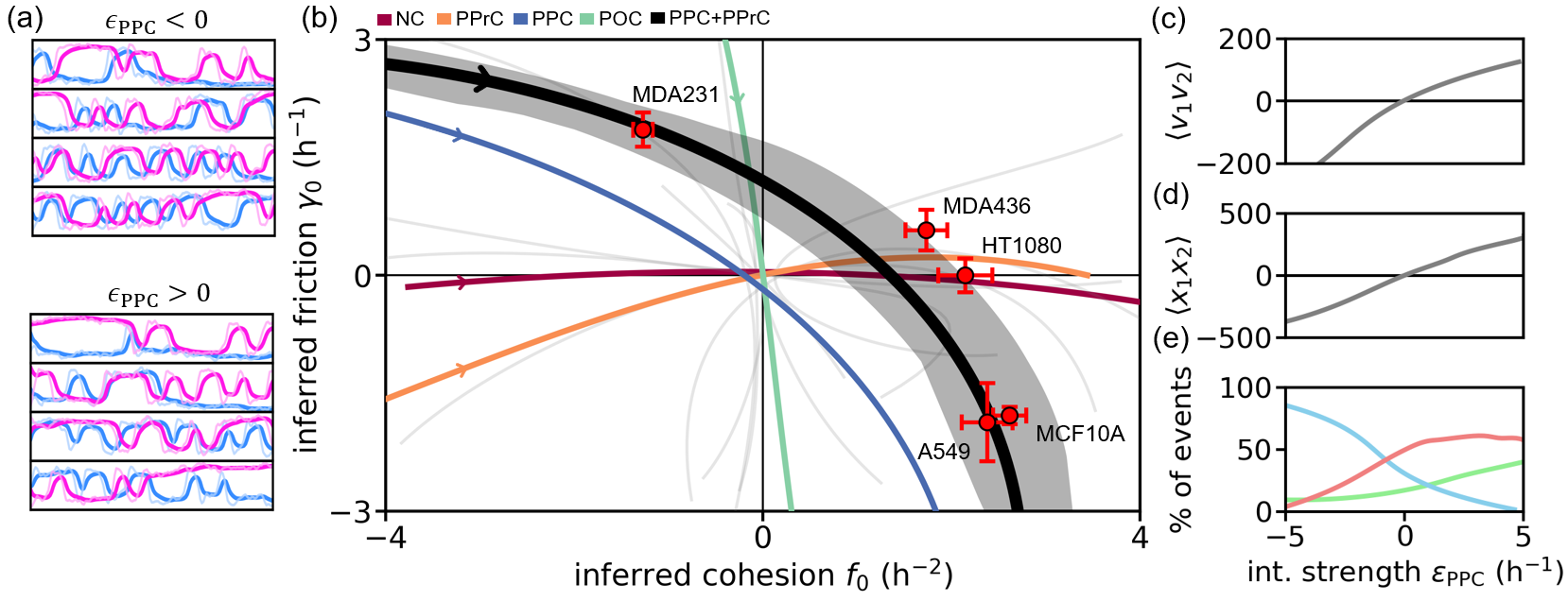}
	\centering
	\caption{
		\textbf{Underdamped dynamics of cell-cell interaction mechanisms.} (a) Nucleus- and protrusion trajectories obtained from simulating the overdamped model [Eqs. (\ref{eq:1})-(\ref{eq:3})] with polarity-polarity coupling interactions for two values of the interaction strength $\epsilon_\mathrm{PPC}$. (b) Colored and grey lines show a mapping ($f_0 (\epsilon)$, $\gamma_0 (\epsilon)$) predicted by the 14 candidate cell-cell interactions while varying the associated interaction strength $\epsilon$. We highlight 4 different candidate models by colored lines and refer to Supplementary Fig. 9 for a complete overview of the mappings of all candidate interactions. Little arrowheads are located at $\epsilon<0$ and show the direction in which we increase $\epsilon$. For all curves, $\epsilon\approx 0$ is located at the coordinate center of the IBS. Black curve shows the model prediction of PPC+PPrC that best fits the experimental data of the five different cell lines. Shaded black region around that curve indicates the spread of the model result for 20 best parameter combinations [Appendix \ref{app:statistical_analysis}]. Red symbols indicate experimental result. Error bars show the bootstrapped error of the mean of the inferred cohesion and friction coefficients. (d)-(f) Behavior statistics predicted by our mechanistic model with the candidate cell-cell interaction PPC. Panel (d) shows the instantaneous velocity correlations, panel (e) shows instantaneous position cross-correlation between the two cells, and panel (f) shows the collision statistics. Curves in panels (b)-(e) are slightly smoothed by fitting splines.  
	}
	\label{Fig4}
\end{figure*}

%------------------------------------------------------------
\subsection{Data-driven inference partially constrains cell-cell interactions}

To further quantitatively constrain the development of our theory for contact-interactions, we investigate the effective short-time dynamics of the nucleus. As a first step, we employ a data-driven inference approach to learn interacting stochastic equations of motion only from the experimentally measured cell position $x$ and cell velocity $v$ \cite{Bruckner2020}. In contrast to the general overdamped model for the coupled dynamics of the nucleus, protrusion, and polarity in Eqs. (\ref{eq:1})-(\ref{eq:3}), the equations of motion of only the nucleus dynamics are underdamped. The two levels of description can be consistent, although the overdamped description is clearly more informative about the dynamics  of the internal degrees of freedom ~\cite{Bruckner_2024}. In fact, the effective inertia in the underdamped description implicitly takes into account the protrusion and polarity dynamics. Thus, we start by using the underdamped description of the experimentally easily accessible nucleus position to partially constrain the dynamics of protrusion and polarity. For pairs of cells \cite{Bruckner2021}, we assume coupled Langevin equations of the form: 

\begin{equation}\label{eq_ULI}
	\frac{d v_i}{dt} = F(x_i, v_i) + f(\Delta x)\Delta x + \gamma(\Delta x)\Delta v + \sigma \eta_i(t).
\end{equation}

\noindent Here, $F(x_i, v_i)$ describes single cell behavior and interaction with the confining geometry, and $\sigma \eta_i(t)$ represents a Gaussian white noise with amplitude $\sigma$. Of particular importance here are the interaction terms $f(\Delta x)\Delta x$ and $\gamma(\Delta x)\Delta v$, which capture how the cell nuclei deterministically accelerate depending on the cells’ relative separation $\Delta x$ and velocity $\Delta v$. The functions $f(\Delta x)$ and $\gamma(\Delta x)$ encode the sign and spatial structure of the interactions and are inferred from experimental data. Here, $f(\Delta x)\Delta x>0$ implies effective repulsion, while negative values indicate effective attraction. In contrast, $\gamma(\Delta x)<0$ indicates effective cell-cell friction, while positive values indicate effective anti-friction. Our goal is to use these inferred dynamical terms as an additional quantitative constraint to determine which of the candidate interaction mechanisms in [Eqs. (\ref{eq:1})-(\ref{eq:3})] contribute to the nucleus dynamics of interacting cells on the dumbbell-shaped micropattern.\\
\\To infer the functions $F(x_i, v_i)$, $f(\Delta x)$, and $\gamma(\Delta x)$ from trajectory data, we use Underdamped Langevin Inference \cite{Bruckner2020} [Appendix \ref{app:inference}]. We find that single-cell terms are qualitatively similar across our cell types [Supplementary Fig. 2]. By contrast, inferred interactions vary strongly: Breast cancer cells exhibit short-range attraction and pronounced anti-friction \cite{Bruckner2021} [Fig. \ref{Fig3}(d),(e)]. The fibrosarcoma cells show repulsive interactions, but exhibit no detectable friction interactions. For the epithelial cell lines, we find repulsion and friction interactions. For all cell lines, the learned equations of motion accurately predict long-time scale behavior statistics [dotted bars in Fig. \ref{Fig3}(a)-(c), Supplementary Fig. 5].\\ 
\\To visualize the inferred dynamics of all cell types, we plot the dominant inferred amplitudes of the cohesion and friction interactions of the various cells [Fig. \ref{Fig3}(f), Appendix \ref{app:inference}]. The five different cell lines occupy different regions in this interaction behavior space (IBS) with an apparent correlation between inferred effective cohesion and friction interactions for the various cell types. Together, this inference procedure reveals a variety of different dynamics ranging from attraction and anti-friction of breast cancer cells to repulsion and friction of epithelial cells. 

%%%%%%%%%%%%%%%%%%%%%%%%%%%%%%%%%%%%%
%%FIGURE 
\begin{figure*}[t!]
	\includegraphics[width=\textwidth]{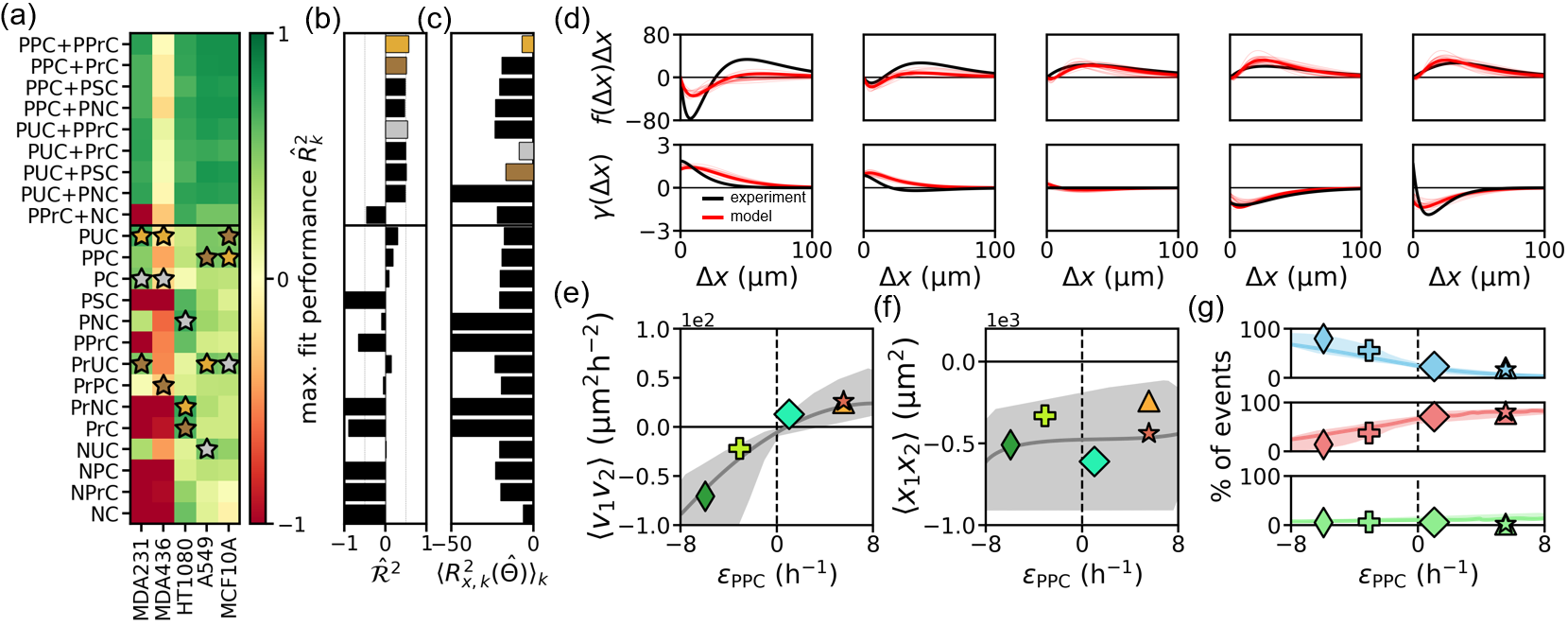}
	\centering
	\caption{
		\textbf{Overdamped statistical inference of interaction mechanisms from experimental data.} (a) Maximum fit performance as quantified by the maximum coefficient of determination (COD) $\hat{R}_k^2$ for each interaction mechanism and each cell type (index $k$) [Appendix \ref{app:fitting}]. Here, $R_k^2$ is equal to $1$ when the model perfectly fits the data and can be negative indicating bad model performance [Eq. \ref{eq:COD}]. The shown $R_k^2$ is an average over the $R^2$ values of the differences between model and experiment calculated for the inferred underdamped interactions, the velocity correlation function and the collision statistics, and is then maximized in the fit. After this fit, we predict the position correlation function in panel (c) to validate the fit. Gold, silver and bronze stars indicate the three best fitting interaction mechanisms for each cell type. We show multiple combinations of PPC and PUC with mechanisms that well fit the fibrosarcoma HT1080 cells. (b) Maximum fit performance $\hat{\mathcal{R}^2}$ of the global fit [Appendix \ref{app:fitting}]. The three models giving the highest $\hat{\mathcal{R}^2}$ for all cell types are highlighted in gold, silver and bronze. (c) Quantification of the goodness of the prediction of the position correlation function of all models. We take the best fitting parameter combination $\hat{\Theta}$ for each candidate interaction and average $R^2$ of the position correlation over all cell types to obtain $\langle R^2_{x,k}(\hat{\Theta})\rangle_k$. Again the best performing models are marked with gold, silver and bronze. (d) Inferred underdamped cohesion and friction interactions for both experiment (black) and best fitting model (red) for all the various cell lines. Model result is obtained from the best fitting candidate interaction PPC+PPrC. Thin red lines indicate model results for the 20 best fitting parameter combinations. (e)-(g) Comparison of behavior statistics between experiment (symbols) and model (curves). Experimental results are plotted at the best fitting coupling strength $\epsilon_\mathrm{PPC}$ for each cell type to facilitate comparison between model and experiment. Panels (e) and (f) show the instantaneous velocity and position correlations respectively. Panel (g) shows percentages of sliding (blue), reversal (red) and following (green) events. Curves in panels (e) and (g) are fits, the grey curve in panel (f) is a prediction as quantified by the gold bar in panel (c).   
	}
	\label{Fig5}
\end{figure*}

%------------------------------------------------------------
\subsection{Connecting cell-cell interaction mechanisms to whole cell-level dynamics}

Next, we ask if the diverse dynamics of pairs of motile cells can be captured by candidate mechanisms in our overdamped model [Eqs. (\ref{eq:1})-(\ref{eq:3})]. To analyze the interaction dynamics emerging from our candidate interaction mechanisms, we numerically simulate cell trajectories [Fig. \ref{Fig4}(a)]. First, we infer and analyze the underdamped effective interactions from simulations of each individual candidate interaction. Specifically, varying the amplitude and sign $\epsilon$ of each interaction mechanism and employing our inference procedure yields a mapping $f_0(\epsilon)$ and $\gamma_0(\epsilon)$, which we depict in the IBS [Fig. \ref{Fig4}(b)]. Intuitively, varying the nucleus coupling strength $\epsilon_{\mathrm{NC}}$ and thus tuning between nucleus attraction and repulsion corresponds mainly to varying the underdamped cohesion parameter $f_0$ [red curve in Fig. \ref{Fig4}(b)]. By contrast, interactions between cell protrusions and polarities give rise to more complicated underdamped dynamics, including friction or anti-friction [grey curves in Fig. \ref{Fig4}(b)]. Interestingly, polarity-polarity coupling can qualitatively predict the experimentally observed correlations between inferred cohesion and friction interactions when tuning between anti-alignment ($\epsilon_{\mathrm{PPC}}<0$) and alignment ($\epsilon_{\mathrm{PPC}}>0$) [blue curve in Fig. \ref{Fig4}(b)]. Furthermore, with PPC, we can capture a switch from anti-correlated sliding to correlated reversal behavior when tuning the interaction strength [Fig. \ref{Fig4}(c),(e), Supplementary Fig. 8(c),(e), Supplementary movie 5], as experimentally observed [Fig. \ref{Fig3}(a),(c)]. These results show that inferring the underdamped nucleus dynamics can provide strong quantitative constraints on our phenomenological nucleus- protrusion- and polarity interaction mechanisms. However, the underdamped dynamics shown in the IBS provides a simplified low-dimensional description of cell interaction behavior. In general, this simple description does not uniquely map onto the overdamped cell-interaction mechanisms. Thus, to fully constrain the interaction mechanisms, we need to take both the detailed underdamped nucleus dynamics [Fig. \ref{Fig3}(d),(e)] and the full behavior statistics [Fig. \ref{Fig3}(a)-(c)] into account.  

%------------------------------------------------------------
\subsection{Polarity-polarity coupling tunes between the behavior of various cell types}

To quantitatively detect and determine cell-cell interaction mechanisms, we next apply statistical inference of our overdamped phenomenological model [Eqs. (\ref{eq:1})-(\ref{eq:3})] to the experimental data of each cell type. To this end, we perform a fit, in which we vary the interaction strength $\epsilon$ and range $r$ of the individual interaction mechanisms. Then, we quantify and minimize the difference between the inferred underdamped interactions [Fig. \ref{Fig3}(d),(e)], as well as between the velocity correlations and collision statistics [Fig. \ref{Fig3}(a),(c)] of model and experiment using an $R^2$ value [Eq. \ref{eq:COD}]. Afterwards, we validate this fit by predicting the position correlation function [Appendix \ref{app:fitting}]. Initially, we take a minimal approach and allow only a single possible interaction mechanism. Furthermore, we first consider an individual fit, allowing the interaction mechanism and all parameters to vary between cell types. This inference procedure shows that we can confidently rule out many interaction mechanisms for all cell types [Fig. \ref{Fig5}(a)]. Further, we find that epithelial and breast cancer cells are best captured by mechanisms like PPC, PUC, PC, or PrOC that allow either anti-alignment or alignment between the polarities or deformation vectors of the cells [Fig. \ref{Fig1}(d),(e)]. Fibrosarcoma (HT1080) cells are best described by mechanisms that couple either the polarity or the protrusion to the position of the other cell, reminiscent of CIL [Fig. \ref{Fig5}(a)].\\ 
\\In addition to this individual fit, we consider a global fit, where we use the same interaction mechanism for all cell types, allowing only the interaction strength $\epsilon$ to vary. Interestingly, the global fit reveals that both PPC and PUC best capture the dynamics of all cell types, while only varying their coupling strength $\epsilon$ between cells [Fig. \ref{Fig5}(b), Supplementary Fig. 14,15]. However, both PPC and PUC alone do not quantitatively reproduce the mutual exclusion behavior, the percentage of following events, and the effective repulsion interactions of epithelial and breast cancer cells [Fig. \ref{Fig4}b,d,e, Supplementary Fig. 14,15].\\
\\To fully capture the dynamics of epithelial and breast cancer cells, additional interaction mechanisms are required. Indeed, combining PPC and PUC with mechanisms that couple the polarity or protrusion to the position of the other cell provides an improved individual and global fit to these cells. From these combinations, PPC and PUC both together with PPrC best capture the experimental interaction behavior across cells [Fig. \ref{Fig5}(a),(b)]. However, PPC together with PPrC better captures both the underdamped interactions and the experimental behavior statistics of all cell types [Fig. \ref{Fig5}(d)-(g), Supplementary Fig. 18,19, Supplementary movie 6]. In particular, PPC+PPrC better predicts the position correlation function, which we have not used in the fit [Fig. \ref{Fig5}(c), Supplementary Fig. 12]. Here, PPrC with $\epsilon_{\mathrm{PPrC}}>0$ adds an additional repulsive component to the dynamics, induces mutual exclusion behavior, and suppresses following behavior [Supplementary Fig. 17]. Remarkably, the combined fit of PPC and PPrC is successful even if we allow only the strength of the polarity coupling $\epsilon_{\mathrm{PPC}}$ to vary between different cell types.\\ 
\\Taken together, we find that the two breast cancer cell lines exhibit anti-alignment with $\epsilon_{\mathrm{PPC}}<0$. By contrast, the fibrosarcoma cell line exhibits very weak alignment interactions and its dynamics are dominated by PPrC. Finally, the epithelial cells exhibit alignment interactions with $\epsilon_{\mathrm{PPC}}>0$, yielding correlated velocities and dominant reversal behavior [Supplementary Fig. 16]. These results do not depend sensitively on the details of the overdamped statistical inference procedure [Supplementary Fig. 13], on the single cell behavior [Supplementary Fig. 21], or the chosen dumbbell-shaped geometry [Supplementary Fig. 22]. Furthermore, considering examples of different pairs of interactions or more than two does not provide a better description of the dynamics of these cell lines [Fig. \ref{Fig5}(a), Supplementary Fig. 20]. Taken together, PPC and PPrC enable cells to adapt their polarity relative to both polarity and position of the other cell, providing a large range of possible behavioral responses to cell-cell collisions. Thus, with these interaction mechanisms we can quantitatively capture a wide variety of interaction behaviors across a broad range of distinct cell types.    

%%%%%%%%%%%%%%%%%%%%%%%%%%%%%%%%%%%%%
%%FIGURE 
\begin{figure*}[t!]
	\includegraphics[width=\textwidth]{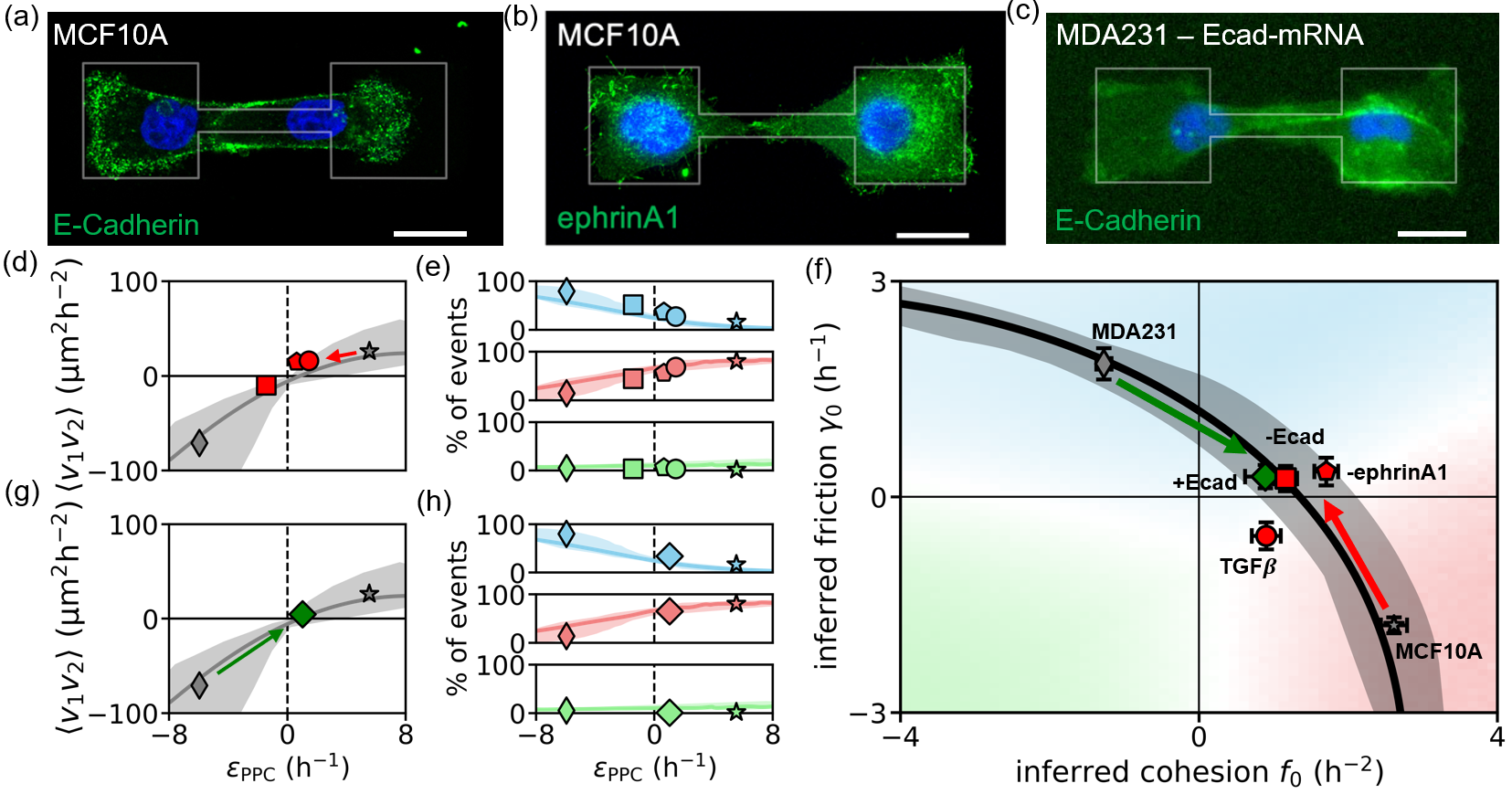}
	\centering
	\caption{
		\textbf{Capturing molecular perturbations with interaction mechanisms.} (a),(b) Fluorescence images of E-Cadherin expression (a) and ephrinA1 expression (b) in MCF10A cells. (c) E-Cadherin expression in transfected MDA-MB-231 cells while confined on the dumbbell-shaped micropattern. Nuclei stained in blue and scale bar for all panels (a)-(c): $20\ \si{\micro\meter}$. (d) and (g) Instantaneous velocity cross-correlation between two cells, when they occupy the same island. Throughout, the solid lines indicates our model result obtained from simulating the phenomenological model with PPC and PPrC varying only $\epsilon_\mathrm{PPC}$ but using the parameters obtained from the global fit to the various cell lines as presented in Fig. \ref{Fig5}. Symbols show experimental results of the various perturbed cell lines plotted at the best fitting interaction strength $\epsilon_\mathrm{PPC}$. (e) and (h) Percentages of the three different collision events for the five different cell lines. Again, symbols show experimental results and solid line shows model result. (f) Interaction behavior space shows the change of the dynamics of the MCF10A and MDA-MB-231 cell lines due to various molecular perturbations. Solid black curve indicates model result from the best global fit of the experimental dynamics of the various cell lines as presented in Fig. \ref{Fig4}(c).    
	}
	\label{Fig6}
\end{figure*}

%------------------------------------------------------------
\subsection{Modeling molecular perturbations of cell-cell interaction pathways}

We hypothesize that the varying polarity-polarity coupling strength across our cell types originates in different molecular expression profiles of several relevant proteins [Fig. \ref{Fig2}(b)]. To test this, we disrupt specific cell-cell interaction pathways through molecular perturbations, which we would then expect to tune the polarity-polarity coupling strength. Specifically, we use antibodies to block E-Cadherin bonds that are established upon contact between two epithelial MCF10A cells [Fig. \ref{Fig6}(a)]. These bonds are known to be adhesive and are crucial for collective cell migration of epithelial tissues \cite{Jain2020}. In addition, we consider perturbations of Eph-ephrin interactions, which directly regulate cell polarity through up- or downregulation of polarity cues such as Cdc42 or RhoA \cite{Noren2004}. Epithelial MCF10A cells express the ligand ephrinA1 [Fig. \ref{Fig6}(b)], which we block using antibodies. Furthermore, we investigate the behavioral shift in MCF10A cells undergoing an epithelial-mesenchymal transition (EMT). This transition changes the coordinated interaction behavior of epithelial cells to that of more individually migrating cells \cite{Milano2016,Theveneau2013,Scarpa2015}. Finally, we induce E-Cadherin in the breast cancer cell line MDA-MB-231 via mRNA transfection [Fig. \ref{Fig6}(c), Supplementary Section 2.4,3.3]. Importantly, all perturbations do not greatly affect single cell behavior [Supplementary Fig. 3], and the resulting interaction dynamics are insensitive to the used transfection protocol [Supplementary Fig. 4].\\
\\In all these perturbation experiments [Supplementary movie 3,4], we can capture the changing interaction dynamics with our phenomenological model and the combination of polarity-polarity coupling and polarity-protrusion repulsion. Specifically, we observe that blocking of E-Cadherin and Eph-ephrin pathways, as well as EMT in epithelial MCF10A cells reduces the polarity-polarity coupling strength $\epsilon_{\mathrm{PPC}}$, thereby inhibiting polarity alignment interactions between MCF10A cells. Consequently, these perturbations reduce positive velocity correlations [Fig. \ref{Fig6}(d)], the number of reversal events observed in MCF10A cells [Fig. \ref{Fig6}(e)], and the amplitudes of the inferred underdamped cohesion and friction interactions [Fig. \ref{Fig6}(f)]. Similarly, our model captures the changing dynamics upon transfecting MDA-MB-231 breast cancer cells with E-Cadherin by a reduction of polarity-polarity coupling [Fig. \ref{Fig6}(f)], indicating that E-Cadherin can inhibit anti-alignment interactions in these cells. The E-Cadherin transfected cells exhibit no significant velocity correlations anymore [Fig. \ref{Fig6}(g)] and fewer sliding events [Fig. \ref{Fig6}(h)]. Taken together, our model captures the changing dynamics of cell-cell interactions upon molecular perturbations of the underlying interaction pathways by tuning the strength of the polarity-polarity coupling mechanism.

%------------------------------------------------------------
\section{Discussion}

To unravel protrusion- and polarity driven interaction mechanisms governing the cell-level dynamics of interact-
ing cells, we proposed a phenomenological theory for contact-interactions. Previous work focused on only few cell types, included ad hoc bottom-up models \cite{Zadeh2022,Vedel2013,Villar-Cervino2013a,Desai2013,Ron2023a,Bertrand2024}, was limited to qualitative analyses of behavior \cite{Camley2014,Scarpa2016b, vagne2024generictheoryinteractingspinning}, or yielded only effective whole-cell dynamical descriptions of interacting cells \cite{Bruckner2021, Sepulveda2013}. Here, we proposed 14 unbiased interaction mechanisms coupling the nuclei, protrusions and polarities of interacting cells. We quantitatively constrain these mechanisms using trajectory data from cell collision experiments with distinct cell types. We find that a combination of only two mechanisms - polarity-polarity coupling (PPC) and polarity-protrusion coupling (PPrC) – is sufficient to describe these cells: Tuning PPC from anti-alignment to alignment of cell polarity quantitatively captures the qualitatively distinct behaviors observed across our cell types. We identify anti-alignment interactions between cell polarity as a novel interaction mode, which promotes sliding behavior of breast cancer cells. This interaction is reminiscent of contact enhancement of locomotion \cite{Dalessandro2017} and may cause doublet-rotations observed for MDCK cells \cite{lu_polarity-driven_2024,Camley2014}. In contrast, polarity alignment promotes correlated reversal behavior of epithelial cells and may be related to mechanisms of epithelial tissue migration \cite{Jain2020,Serra-Picamal2012,Tambe2011,Ron2023a,LoVecchio2024}. PPrC, which causes polarities to grow away from the other cell similar to contact inhibition of locomotion \cite{Roycroft2016,Stramer2017}, captures the mutual exclusion behavior of all our cells. Despite the complexity of the cellular interaction machinery, our results reveal that the emergent large-scale behaviors of distinct interacting cells are captured by two simple polarity interaction mechanisms.\\
\\To challenge our theory, we performed various molecular perturbations of surface proteins that likely signal to the polarity machinery of cells. By blocking the function of E-Cadherins \cite{Desai2009,Nakagawa2001,Milano2016} and cell-cell recognition via Eph-ephrin interactions \cite{Astin2010,Kadir2011}, we demonstrated that these surface proteins set the strength of polarity-polarity coupling. Furthermore, activating expression of E-Cadherin inhibits anti-alignment in cancerous MDA-MB-231 cells, which may be interpreted as a partial reversion of their mesenchymal phenotype \cite{Chao2010}. Given our finding that tuning the sign and strength of polarity-polarity coupling captures various cell types, multiple molecular perturbations, and complex processes like the epithelial-mesenchymal transition, polarity-polarity coupling may be a robust and conserved mechanism underlying contact-interactions between cells.\\
\\Polarity-polarity coupling has been invoked in toy models for active matter \cite{Vicsek1995} and captures the tissue-scale flocking behavior of epithelial MDCK cells on 1D tracks \cite{Ron2023a,Bertrand2024,LoVecchio2024}. Here, we showed in an unbiased way that such polarity interactions accurately capture the collision dynamics of epithelial and breast cancer cells on a two-cell level. As our model readily generalizes to 2D and 3D, polarity-polarity coupling may introduce polar order on the level of two cells within small migrating epithelial cell clusters \cite{Copenhagen2018,Ron2023a,Bertrand2024,Vercruysse2024}, 2D epithelial tissues \cite{Poujade2007,Streichan2018,Serra-Picamal2012,Peyret2019}, or collective rotations of 3D organoids \cite{Brandstatter2021}. In contrast, the loss of polar order is associated with cancer progression and cancer cells exhibit nonaligned disordered motion in 2D sheets \cite{wong_collective_2014,Lee2021}. This loss of order may be related to anti-alignment interactions, which could lead to disordered cell spreading \cite{Dalessandro2017} or fluid like disordered collective behavior \cite{Malinverno2017,Kang2021,Palamidessi2019}. Thus, our phenomenological theory could give insight into how contact-interactions at the cellular level control collective cell migration of epithelial and mesenchymal cells.

%------------------------------------------------------------
\subsection*{Acknowledgements} 
We thank Johannes Flommersfeld, Bram Hoogland, and Ricard Alert for helpful discussions. We thank Gerlinde Schwake for producing the E-Cadherin mRNA. Funded by the Deutsche Forschungsgemeinschaft (DFG, German Research Foundation) - Project-ID 201269156 - SFB 1032 (Project B01 and B12).\\
\\T.B. and E.B. contributed equally to this work. E.B. and J.R. designed experiments; E.B. and G.L. performed experiments; T.B., D.B.B., and E.B. analyzed data; T.B., D.B. developed model, performed simulations and statistical inference; T.B., E.B., D.B.B., J.R. and C.P.B. interpreted the results and wrote the manuscript. J.R. and C.P.B. conceived this study and led the research. All authors edited and approved the manuscript.  \\\

\appendix
\section{Experimental Methods}
\subsection{Micropatterning and sample preparation }
\label{sec:Micropatterning}
For the cell collision experiments, we employ a micropattern structure with a dumbbell shaped design with two $35\times35\ \si{\micro\meter}$ squares connected by a $40\times7\ \si{\micro\meter}$ bridge \cite{Bruckner2021}. For micropatterning, we employ a photopatterning technique using the PRIMO module (Alvéole). For background passivation of the µ-dish (ibidi), a drop of $0.01\%$ (w/v) PLL (Sigma-Aldrich) is added and incubated for 30 min. Afterwards, the sample is rinsed with HEPES buffer (Sigma-Aldrich) and 100mg/ml PEG-SVA (LaysanBio) is incubated for 1 h at room temperature. The passivated dish is photopatterned by employing the PRIMO module mounted on an automated inverted microscope (Nikon Ti Eclipse). The photoinitiator PLPP (Alvéole) is added to the dish. The dumbbell shaped pattern was placed on top of the dish via the Leonardo software (Alvéole) and illuminated with UV-light with a dose of $15\ \si{\milli\joule}/\si{\milli\meter}^2$. The dish is washed with milliQ water and rehydrated with PBS for 5 min followed by an incubation with 20 µg/ml of labelled Fibronectin-Alexa647 (YO-protein, ThermoFisher) for 15 minutes at room temperature. For antibody blocking experiment fibronectin micropatterns are made by microplasma-initiated protein patterning as described in \cite{Bruckner2019a}. 

\subsection{Cell culture}
In this study, we analyzed the behavior of the cell lines MDA-MB-231, MDA-MB436, HT-1080, A549 and MCF10A. The individual culture condition are shown in Supplementary table 1. Cells are grown at $37^\circ\text{C}$ in an atmosphere containing $5\%$ CO2. For passaging, cells are being washed and treated with accutase for $5$ min. The cell solution is centrifuged at $800$ r.c.f. for $3$ min and the cells are resuspended in medium. Approximately $10 000$ cells are added into the micropatterned µ-dish and left to adhere for up to $4\ \mathrm{h}$ in the incubator. After this incubation period, the medium is exchanged for phenol red free medium and $25\ \si{\nano\mol}$ Hoechst 33342 (Invitrogen) is added to stain the nuclei when needed.  

\subsection{Inhibitors and antibody blocking} 
In order to block the function of E-Cadherin or ephrinA1, blocking antibody CD324 (functional grade, Invitrogen) or  anti-ephrinA1 antibody (Invitrogen, ThermoFisher) was added to the dish after cells adhered to the pattern at a concentration of $5 \ \si{\micro\gram}/\si{\milli\liter}$ and $1 \ \si{\micro\gram}/\si{\milli\liter}$ respectively. After an incubation period of 1h, time-lapse measurements were started. EMT was induced by $10\ \si{\nano\gram}/\si{\milli\liter}$ treatment with TGF$\beta$ (ThermoFisher) for up to 7 days. 

\subsection{Transfection}

Prior to the transfection, cells were seeded in an µ-dish (ibidi) without patterning. At $90\%$ confluency the cells were transfected with LipofectamineTM 2000 (Invitrogen, Germany) complexes containing E-Cadherin mRNA (Supplementary Section 2.4). First, $2\ \si{\micro\liter}$ LipofectamineTM 2000 reagent are mixed with $398\ \si{\micro\liter}$ OptiMEM (Invitrogen, Germany) and incubated for 5 min at room temperature. Simultaneously, $2\ \si{\micro\liter}$ mRNA ($1735\si{\nano\gram}/\si{\micro\liter}$) is diluted in $198\ \si{\micro\liter}$ OptiMEM. The mRNA mix is added onto $200\ \si{\micro\liter}$ of the LipofectamineTM 2000 dilution and incubated for 20min at room temperature. Cells are washed once with OptiMEM and then the lipoplexes are added to the dish. After 1h, cells are washed again and re-incubated with normal medium. For control experiments, lipoplexes were created with either GFP-mRNA or substituted with milliQ water at the same ratio. For more details see Supplementary Section 2.4. Note that we investigate the effect of the transfection protocol on the single cell and interaction dynamics in Supplementary Section 3.2. Specifically, we consider the control experiments with only GFP transfection and show that our results do not depend sensitively on the transfection protocol [Supplementary Fig. 4].

\subsection{Microscopy and cell tracking}
All measurements are performed in time-lapse mode for 48 h on a Nikon Eclipse Ti microscope or on a Nikon Eclipse Ti2 microscope using a 10× objective. The samples are kept in a heated chamber (Okolab) at $37^\circ\text{C}$ at $5\%$ CO2 throughout the measurements. Every 10 min a bright-field image and a fluorescence image of the stained nuclei are acquired. Cell tracking of  the Nuclei is performed by using TrackPy. Track length varied between 8 h and 48h. For more details see Supplementary Section 2.2 and 2.3.

\subsection{Western blots}
\label{app:western_blot}
Cells were harvested and lysed in RIPA lysis buffer supplemented with 1M PMSF and a protease inhibitor cocktail (ThermoFisher). Lysates were centrifuged at 14 000g for 20min at $4^\circ\text{C}$. Supernatant was transferred and stored at $-80^\circ\text{C}$. Protein concentration were determined by Bradford assay and an equal amount of protein was loaded onto the precast SDS-gel (BioRad). Proteins were separated by gel electrophoresis. The transfer was performed on Immuno-Blot polyvinylidene difluoride (PVDF) membranes (BioRad) with the Transblot turbo transfer system (BioRad) during 7 min. After blocking for 1h with $5\%$ non-fat dried milk (ThermoFisher) in PBS $0.1\%$ Tween 20 (Roth) the membranes were probed with primary antibodies mouse anti-ECAD (ThermoFisher) (1:1000) and rabbit anti-NCAD (ThermoFisher) (1:2000) overnight at $4^\circ\text{C}$. Secondary antibodies HRP-anti-mouse IgG (1:10000) and HRP-anti-rabbit IgG (1:10000) were incubated on the membrane for 1h at room temperature. Development was performed using Pierce western ECL substrate (ThermoFisher) and the ChemiDoc Imaging System (BioRad). The intensity of the band was quantified via densitometry using ImageJ and  protein amount was normalized to a $\beta$-actin loading control on the same membrane. 

\subsection{Immunohistochemistry}
The cells were fixed after the experiment with $4\%$ PFA for 10min. Cells were washed three times with PBS. Afterwards the cells were permeabilized for 5 min in $0.1\%$ Triton-X-100 solu- tion (Sigma) and blocked for 1h in cold $4\%$ BSA (ThermoFisher). The cells were rinsed once with cold $1\%$ BSA. The excess liquid was removed and the cells were subjected to primary antibodies mouse anti-ECAD (ThermoFisher) (1:100) and rabbit anti-ephrinA1 (ThermoFisher) (1:100) diluted in $1\%$ BSA overnight at $4^\circ\text{C}$. Three washes with $1\%$ BSA were carried out before the adding of the secondary antibody AlexaFluor 488 goat anti-mouse (ThermoFisher) (1:1000) or Alexa 647 goat anti-rabbit (ThermoFisher) (1:1000) for 1h in the dark. The solution was then removed and washed 3 times with PBS. F-Actin staining was done with rhodamine-phalloidin (ThermoFisher, 1:1000). Cells were imaged on a Confocal LSM980 microscope using the airy scan mode with a 40x water immersion objective. 

\section{Theoretical Methods}
\label{app:theoretical_methods}

\subsection{Model development}
\label{app:model_development}
Based on previous work \cite{Alert2020,Smeets2016,Sepulveda2013,Szabo2016a,Belmonte2008}, we employ a generalized active particle model to describe pairs of interacting cells. This model aims to describe the large scale cell-level dynamics in terms of three minimal key degrees of freedom. First, we describe the nucleus, marking the COM of the cells. Secondly, we include the position of the leading protrusion, allowing us to capture the dynamic and elongated shape of cells. Thirdly, we capture cell polarity as an internal stochastic degree of freedom, providing the active pushing force onto the leading protrusion. To this end, we generalize a previously employed mechanistic model for single migrating cells in confinement \cite{Bruckner2022}: 

\begin{align}
	\zeta_\mathrm{n}\dot{x}_{\mathrm{n},i} &= -\frac{k}{\gamma(x_{\mathrm{n},i})}\left(x_{\mathrm{n},i} - x_{\mathrm{p},i}\right) + G_{\mathrm{n}}(\Delta x, U, P, \Delta P)\\
	\zeta_\mathrm{p}\dot{x}_{\mathrm{p},i} &= k\left(x_{\mathrm{n},i} - 
	x_{\mathrm{p},i}\right) + P_i(t) + G_{\mathrm{p}}(\Delta x, U, P, \Delta P) \\ 
	&+ F_{\mathrm{boundary}}(x_{\mathrm{p},i}) \nonumber\\
	\dot{P}_i &= -\alpha(x_{\mathrm{p},i})P_i - \beta P_i^3 + G_{\mathrm{P}}(\Delta x, U, P, \Delta P) + \sigma \eta_i(t).
\end{align}
Here, the nucleus and the protrusion of the same cell are coupled by a linear spring and we define $k_\mathrm{n}=k/\zeta_\mathrm{n}$ and $k_\mathrm{p}=k/\zeta_\mathrm{p}$ as the spring constant $k$ reduced by the friction coefficients of the cell nucleus $\zeta_\mathrm{n}$ and the cell protrusion $\zeta_\mathrm{p}$, respectively. Here, $\gamma(x_{\mathrm{n},i})$ is a dimensionless rescaling factor of the friction coefficient $\zeta_\mathrm{n}$ that depends on the position of the cell nucleus with a minimum when the cell nucleus is on the bridge [Fig. \ref{Fig1_app}(a)]. This models the reduced adhesive area accessible to the cell on the bridge, which is a key feature of confined cell migration on our dumbbell-shaped micropattern \cite{Bruckner2022}. In agreement with previous work \cite{Bruckner2022}, we choose:
\begin{equation}\label{equ:gamma}
    \gamma(x_{\mathrm{n},i}) = \frac{1-\gamma_{\mathrm{min}}}{2}\left(1 - \cos\left(\frac{x_{\mathrm{n},i}\pi}{L_{\mathrm{system}}}\right)\right) + \gamma_{\mathrm{min}}
\end{equation}
where $L_{\mathrm{system}}$ is half of the size of the dumbbell-shaped micropattern and $\gamma_{\mathrm{min}}$ is the minimum rescaling factor reached while the nucleus is on the bridge [Fig. \ref{Fig1_app}(a)]. The cell’s protrusion is confined on the micropattern by a boundary force $F_{\mathrm{boundary}}(x_{\mathrm{p},i})$, which represents a soft repulsive force at the boundaries of the micropattern. In this model, the polarity of the cell is guided by the geometry of the micropattern \cite{Bruckner2022,Flommersfeld2024}. Specifically, the polarity switches from a negative feedback loop on the island to a positive feedback loop on the bridge of the micropattern, causing the polarity to grow into the bridge as cells transitions between the islands \cite{Bruckner2022}. This is implemented by the function $\alpha(x_{\mathrm{p},i})$, which switches sign dependent on $x_{\mathrm{p},i}$:

\begin{equation}\label{equ:alpha}
    \alpha(x_{\mathrm{p},i}) = -\frac{\alpha_0 - \alpha_{\mathrm{min}}}{2}\cos\left(\frac{x_{\mathrm{p},i} \pi}{L_{\mathrm{sys}}}\right) + \frac{\alpha_{\mathrm{min}} + \alpha_0}{2}
\end{equation}

where $\alpha(x_{\mathrm{p},i}) = \alpha_0$ on the island and $\alpha(x_{\mathrm{p},i}) = \alpha_{\mathrm{min}} < 0$ on the bridge [Fig. \ref{Fig1_app}(b)]. The higher order term $- \beta P_i^3$ prevents unbound growth of the polarity and gives rise to a preferred polarity when $\alpha(x_{\mathrm{p},i})$ is negative. The interaction terms $G_\mathrm{n}$, $G_\mathrm{p}$, and $G_\mathrm{P}$ encode phenomenological interaction mechanisms, which we derive in an unbiased way. The term $\sigma \eta_i(t)$ encodes uncorrelated Gaussian white noise, modeling the inherent stochasticity of cell polarity and cell migration. For details of the numerical implementation refer to Supplementary Section 4.2. 

\subsection{Single cell parameters}
Throughout, we make use of a set of parameters ($\sigma = 100\ \mu\mathrm{m}\mathrm{h}^{-\frac{1}{2}}$, $k_\mathrm{n} = 0.6\ \mathrm{h}^{-1}$, $k_\mathrm{p} = 1.2\ \mathrm{h}^{-1}$, $\beta = 0.0001\ \mu\mathrm{m}^{-2}\mathrm{h}$, $\alpha_0 = 10\ \mathrm{h}^{-1}$, $\alpha_{\mathrm{min}} = -6\ \mathrm{h}^{-1}$, $\gamma_{\mathrm{min}} \approx 0.25$, $L_{\mathrm{system}} = 52.5\ \mu\mathrm{m}$) which has been constrained for MDA-MB-231 cells using a data-driven inference approach and best describes the detailed dynamics of the hopping behavior of single MDA-MB-231 cells on the dumbbell-shaped micropattern used here \cite{Bruckner2022}. In Supplementary Section 5.5, we describe how these parameters vary across cell types and investigate the role of single cell aspects on the interaction dynamics of cells.

\subsection{Phenomenological derivation of interaction mechanisms}
\label{app:mechanisms}

To construct phenomenological interaction mechanisms, we assume (i) that our model obeys rotational and translational symmetry. Thus, cell-cell interactions in our model can not depend on the absolute position of cells and can only depend on quantities in our model that behave as vectors in arbitrary dimensions. These quantities are the intercellular distances $\Delta x_\mathrm{n} = x_{\mathrm{n},i} - x_{\mathrm{n},j}$, $\Delta x_\mathrm{np} = x_{\mathrm{n},i} - x_{\mathrm{p},j}$, $\Delta x_\mathrm{p} = x_{\mathrm{p},i} - x_{\mathrm{p},j}$, $\Delta x_\mathrm{pn} = x_{\mathrm{p},i} - x_{\mathrm{n},j}$, the two deformation vectors of the cells $U_i = x_{\mathrm{p},i} - x_{\mathrm{p},i}$, $U_j = x_{\mathrm{p},j} - x_{\mathrm{p},j}$, and the polarities $P_i$, $P_j$ [Fig. \ref{Fig1}(b)]. We further consider differences between these vectorial quantities only if the vectorial quantities themselves are not already differences of the degrees of freedom. This adds the potentially biophysically relevant polarity difference $\Delta P = P_i - P_j$. 
\begin{figure}[t!]
	\includegraphics[width=0.48\textwidth]{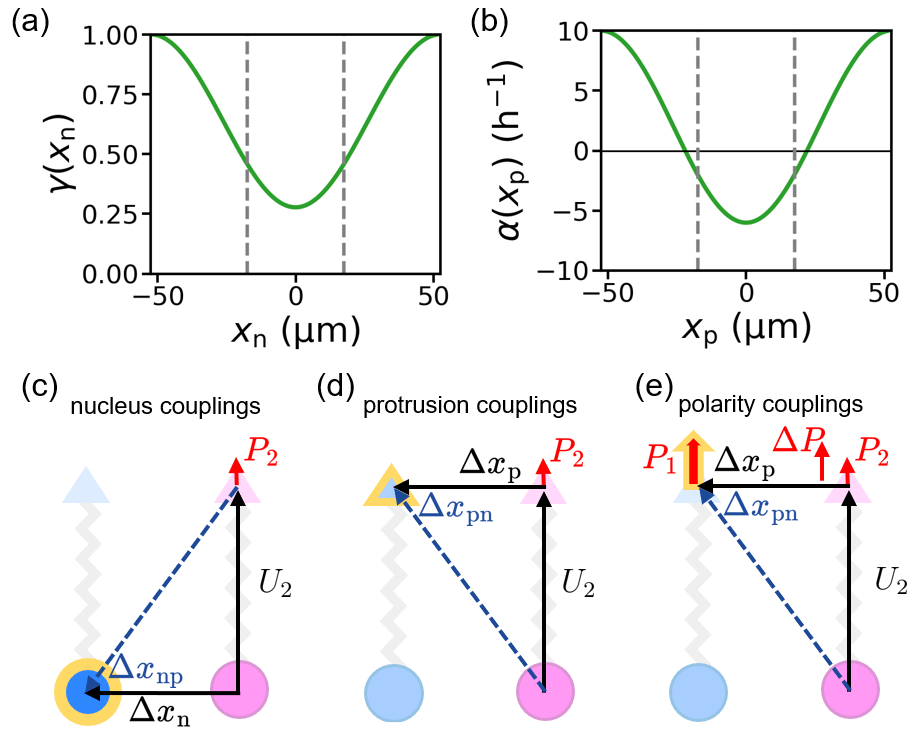}
	\centering
	\caption{\textbf{Model details} (a) Rescaling factor $\gamma(x_{\mathrm{n}})$ used in the model. Vertical dashed lines indicate the boundary of the two islands and the bridge of the dumbbell-shaped micropattern. (b) Geometry-dependent growth rate of the polarity $\alpha(x_{\mathrm{p}})$. (c)-(e) Schematic of all couplings between the nucleus of cell $i$ and cell $j$ (c), between the protrusion of cell $i$ and cell $j$ (d), and between the polarity of cell $i$ and DOFs of cell $j$ (e). With gold color we highlight the degree of freedom of cell 1 that gets coupled to the other cell.}
	\label{Fig1_app}
\end{figure}
Further, we employ a minimal approach and propose (ii) that interaction terms are linear in the vectorial quantities and impose a characteristic interaction range by assuming that (iii) interactions exponentially decay on a length scale $r$. Thus, we can use the following functional dependencies to construct interaction mechanisms:
\begin{align}
    &\epsilon e^{-|\Delta x|/r}\Delta x\\
    &\epsilon e^{-|\Delta x|/r}U\\
    &\epsilon e^{-|\Delta x|/r}P\\
    &\epsilon e^{-|\Delta x|/r}\Delta P
\end{align}
Here, $\epsilon$ is the interaction strength in our phenomenological model and $\Delta x$ is either $\Delta x_\mathrm{n}$, $\Delta x_\mathrm{np}$, $\Delta x_\mathrm{p}$ or $\Delta x_\mathrm{pn}$. Also $P$ and $U$ are either equal to $P_i$ or $P_j$ and $U_i$ or $U_j$ respectively. To construct interaction mechanisms and encode them in $G_\mathrm{n}$, $G_\mathrm{p}$, and $G_\mathrm{P}$, we need to specify the allowed couplings and thus the allowed dependencies of the interaction terms. We assume (iv) that interactions acting on the positional degrees of freedom ($G_\mathrm{n}$, $G_\mathrm{p}$) do not introduce additional couplings between the degrees of freedom of the same cells that are not already introduced by the single cell terms. In practice, this means that $G_\mathrm{n}$ can only depend on $\Delta x_\mathrm{n}$, $\Delta x_\mathrm{np}$, $U_j$, or $P_j$ (and not on the vectorial quantities that involve the protrusion position or polarity of cell $i$). Thus we couple the nucleus of cell $i$ to the nucleus, protrusion, deformation and polarity of cell $j$ [Fig. \ref{Fig1_app}(c)]. Likewise, $G_\mathrm{p}$ only depends on $\Delta x_\mathrm{p}$, $\Delta x_\mathrm{pn}$, $U_j$, or $P_j$ coupling the protrusion of cell $i$ to the nucleus, protrusion, deformation and polarity of cell $j$ [Fig. \ref{Fig1_app}(d)]. For the polarity couplings encoded in $G_\mathrm{P}$ we do not make assumption (iv). This is because the polarity is an internal degree of freedom. Thus, in order to fulfill the assumption that interactions decay in space, we have to allow $G_\mathrm{P}$ to depend on the relative position of cells even though this makes $G_\mathrm{P}$ dependent on the DOFs of cell $i$. This means that $G_\mathrm{P}$ may depend on $\Delta x_\mathrm{p}$ or $\Delta x_\mathrm{pn}$, $U_j$, $P_i$, $P_j$, $\Delta P$, coupling the polarity of cell $i$ to the nucleus, protrusion, deformation, and polarity of cell $j$ as well as to its own polarity [Fig. \ref{Fig1_app}(e)]. These assumptions give rise to 14 different interaction mechanisms, which we show in [Fig \ref{Fig1}] and describe in detail in [Supplementary Section 4.1].

\subsection{Behavior statistics}
\label{app:behavior_statistics}
To quantify interaction behavior of pairs of motile cells, we compute several behavior statistics introduced in the literature \cite{Bruckner2021}. First, we find the correlation functions $C_V(|t-t'|)=\langle v_1(t) v_2(t')\rangle_{\mathrm{same}}$ and $C_X (|t-t'|)=\langle x_1 (t) x_2 (t')\rangle$. For both these correlation functions, we define $t'=t+\Delta$ and then average the products of the velocities or positions of the two cells over time $t$ and all pairs of cells. For $C_V$, we condition our averaging on both cells being located on the same island. For easy visualization, we show in the main text figures the instantaneous correlations $(|t-t'|=0)$ and show the full result in Supplementary Fig. 5. Furthermore, we detect collision events in our cell trajectories by defining a threshold $\Delta x<\Delta x_c$, where $\Delta x$ is the intercellular distance. Then we analyze the trajectories within a time frame of $dT$ following the first time when $\Delta x<\Delta x_c$. If cells simultaneously transition from one island to the other, we detect a following event. If cells switch positions at least once during $dT$, we detect a sliding event. Finally, if cells do not switch positions during $dT$, we detect a reversal event. The collision statistics is robust against the choice of parameters of the detection algorithm \cite{Bruckner2021}. 

\subsection{Data-driven inference}
\label{app:inference}
We perform a data-driven inference approach to learn effective dynamical interaction terms that govern the short time scale dynamics of the nuclei of interacting cells \cite{Bruckner2020,Bruckner2021,Bruckner_2024}. These dynamics implicitly capture the dynamics of protrusion and polarity \cite{Bruckner_2024}. This approach involves proposing coupled underdamped Langevin equations for the nucleus trajectories $x(t)$ of the two cells \cite{Bruckner2021} (Eq. (\ref{eq_ULI})). The key idea of this approach is to estimate from the experimentally measured nucleus trajectory $x(t)$ both the instantaneous velocity $\hat{v}(t)$ and the instantaneous acceleration $\hat{a}(t)$, while considering errors introduced by the discrete sampling and possible measurement errors. Then, assuming that the Langevin equation can capture the dynamics of these cells, we can rigorously infer the deterministic single cell term $F(x_i,v_i)$ and the interaction terms $f(\Delta x)\Delta x$ and $ \gamma(\Delta x)\Delta v$ on the right hand side of equation \ref{eq_ULI} using stochastic estimators \cite{Bruckner2020}. Briefly, this is done by fitting the experimental accelerations, measured over the phase space of our system, by the deterministic terms expanded in sets of basis functions. Throughout, we expand the interaction terms into exponentials of the form $b_n(|\Delta x|)=e^{-|\Delta x|/nr}$, where $n=1,…,N$. Thus, for instance $f(\Delta x)\approx \sum_n u_n b_n(|\Delta x|$) and the coefficients $u_n$ get estimated rigorously from the trajectory data \cite{Bruckner2020}. This procedure is robust against the choice of basis function \cite{Bruckner2021}, but the experimental data is best captured at certain values of $N$. For MCF10A, MDA231, and MDA436 cells, we use $N=3$ and $r=20\ \mathrm{\mu m}$. For A549 cells we use $N=2$ and $r=25\ \mathrm{\mu m}$ and for HT1080 cells, we use $N=1$ and $r=30\ \mathrm{\mu m}$. Note that for HT1080 cells, we found that the best Langevin equation is one without the term $\gamma(\Delta x)\Delta v$. Finally, to predict the interaction behavior space (IBS), we fix $N=1$ and $r=30\ \mathrm{\mu m}$ and vary manually $u_n$ which we defined as either $f_0$ or $\gamma_0$ for $f(\Delta x)$ and $\gamma(\Delta x)$ respectively. For each parameter combination, we simulate trajectories and find the dominant collision behavior, which we depict as colormap in [Fig. \ref{Fig3}g]. For the position of the experimental cell lines, we perform the inference procedure with $N=1$ and $r=30\ \mathrm{\mu m}$ giving $f_0$ or $\gamma_0$. 

\subsection{Overdamped statistical inference procedure}
\label{app:fitting}
To perform our statistical inference procedure of our overdamped model to the experimental data, we perform a fitting procedure. Specifically, we sweep over the interaction strength $\epsilon$ and interaction range $r$ of each candidate interaction mechanism as well as for each pair of mechanisms. For each parameter combination, we predict the velocity cross correlation $C_V (|t-t'|)$, the position cross correlation $C_X(|t-t'|)$, and find the collision distribution. Note that in the fitting procedure, we only consider $|t-t'|<\tau$ to avoid fitting the correlations that are close to zero in the experiment. For $C_V$, we choose $\tau=0.5 \mathrm{h}$ and for $C_X$ we choose $\tau=10 \mathrm{h}$. This choice does not greatly influence the final results of the inference procedure described in Fig. \ref{Fig5}a,b [Supplementary Fig. 13 d-f]. Furthermore, we perform our data-driven inference method to find the underdamped cohesion $f(\Delta x)\Delta x$ and friction interactions $ \gamma(\Delta x)\Delta v$ from the simulated data. For each of these five statistics, we compute the so called Coefficient of determination (COD) defined by:
\begin{equation}\label{eq:COD}
    R^2=1-SS_\mathrm{res}/SS_\mathrm{tot}
\end{equation}
, where $SS_\mathrm{res}$ is the sum of squared deviations between model and experiment: $SS_\mathrm{res}=\sum_i(y_i - \hat{y}_i)^2$ with $y_i$ being the experimental result and $\hat{y}_i$ being the model result for the five different statistics. The index $i$ runs over the discrete values of the numerically computed statistics. $SS_\mathrm{tot}=\sum_i(y_i - \overline{y})^2$ is proportional to the variance of the experimental data as $\overline{y}$ is the mean of the experimental data. Consequently, $R^2$ is equal to $1$ if the model perfectly captures the experimental data but equal to $0$ if the model merely captures the mean of the experimental data. $R^2$ is negative when the model performs even worse than that.\\ 
\\A simultaneous fit to all five statistics is challenging given that our models do not well predict the position correlation function $C_X(|t-t'|)$. Thus, the COD of the position correlation is strongly negative, rendering an average of the $R^2$ values over all statistics biased [Supplementary Fig. 12b]. Thus, throughout, we average $R^2$ over the underdamped interactions, the velocity correlation function and the collision statistics, excluding the position correlation. However, we later predict this position correlation using the model coming out of the fitting procedure. This way, we have a strong way to validate the fit [Supplementary Fig. 12]. Further, the weight we put on each statistics in this average does not greatly affect the final results described in the main text [Supplementary Fig. 13 a-c]. Furthermore, note that the inferred dynamics and behavior statistics act effectively as independent degrees of freedom given our overdamped model [Supplementary Fig. 11], making it possible to strongly constrain our overdamped model. For the individual fit, we maximize the resulting average $R^2(\epsilon,r)$ for each cell type individually. This gives us optimal parameters $\hat{\epsilon}$ and $\hat{r}$ for each cell type. We show the maximum COD $\hat{R}^2$ for each cell type and each candidate interaction mechanism in Fig. \ref{Fig5}a. For the global fit, we only allow the interaction strength $\epsilon$ to vary between cell types, while fixing all other parameters. Thus, we maximize $R^2$ averaged over all statistics and cell types, giving us a single optimal value $\hat{r}$, but varying interaction strengths for each cell type $\hat{\epsilon}$. We show the maximum global COD $\hat{\mathcal{R}}$ for each candidate interaction in Fig. \ref{Fig5}b. Throughout, we adjust the bounds of the parameters $\epsilon$ and $r$ such that we at least include a local maximum of $R^2(\epsilon,r)$ [Supplementary Fig. 10]. For more details refer to Supplementary Section 4.3. 

\subsection{Statistical analysis}
\label{app:statistical_analysis}
Throughout, error bars indicate the standard error of the mean (s.e.m) which we obtain for experimental data by bootstrapping. For the error bars of the statistics predicted by the inferred Langevin equation as presented by dotted bars in Fig. 4(a)-(c), we again bootstrap the experimental data, perform the inference procedure and re-simulate the inferred equation to predict these statistics. The error bars then again represent the error of the mean across these bootstrapped subsets of the data. For the phenomenological model, we indicate the variability of the predicted behavior statistics and inferred underdamped interaction terms by showing the spread of the 20 best fitting parameter combinations. For model results, we typically simulate $N=300$ trajectories. The number of experimental trajectories is given in Supplementary table 1.

\bibliography{library}
 
\end{document}

%\includepdf[pages=-]{main_text_final_revised.pdf}

%\foreach \x in {1,...,16}
%{%
%	\clearpage
%	\includepdf[pages={\x}]{../submission_arxiv/main_text_manuscript_arxiv.pdf}
%}
%
%\newpage

\centerline{\bf{\LARGE{Supplementary Information:}}}\vspace{0.1 cm}
\begin{center}
\bfseries\Large
Data-driven theory reveals protrusion and polarity interactions governing collision  behavior of distinct motile cells
\end{center}
\vspace{1cm}
\centerline{\normalsize{Tom Brandstätter*, Emily Brieger*, David B. Brückner, Georg Ladurner, Joachim Rädler, Chase P. Broedersz$^\dagger$}}
\vspace{0.4 cm}

* These authors contributed equally to this work: Tom Brandstätter, Emily Brieger\\
$\dagger$ Corresponding author: c.p.broedersz@vu.nl, Phone: +31 20 59 82953

\newpage
\tableofcontents
\newpage

\section{Supplementary movie captions}

\paragraph{Supplementary movie 1: } Bright field movies of various different migrating pairs of cells on a dumbbell-shaped micropattern. The nuclei of the cells are stained in blue. The dumbbell-shaped micropattern is stained in red. Scale bar is equal to $30\ \mu\mathrm{m}$. Time code is in units of hours ($\mathrm{h}$).

\paragraph{Supplementary movie 2: } Bright field movies of various single migrating cells on a dumbbell-shaped micropattern. The nuclei of the cells are stained in blue. The dumbbell-shaped micropattern is stained in red. Scale bar is equal to $30\ \mu\mathrm{m}$. Time code is in units of hours ($\mathrm{h}$).

\paragraph{Supplementary movie 3: } Bright field movies of epithelial MCF10A cells with various different biomolecular perturbations on a dumbbell-shaped micropattern. Upper left shows the wildtype, upper right shows the cells with ephrinA1 antibodies, lower left shows MCF10A cells with TGF$\beta$-induced epithelial-mesenchymal transition, and lower right shows MCF10A cells with E-Cadherin antibodies. The nuclei of the cells are stained in blue. The dumbbell-shaped micropattern is stained in red. Scale bar is equal to $30\ \mu\mathrm{m}$. Time code is in units of hours ($\mathrm{h}$).

\paragraph{Supplementary movie 4: } Bright field movies of breast cancer MDA-MB-231 cells with two different mRNA transfections. Left shows the wildtype, upper right shows the MDA-MB-231 cells with the E-Cadherin transfection, and lower right shows MDA-MB-231 cells with the GFP transfection for control. The nuclei of the cells are stained in blue, and E-Cadherin is stained in green. The dumbbell-shaped micropattern is stained in red. Scale bar is equal to $30\ \mu\mathrm{m}$. Time code is in units of hours ($\mathrm{h}$).

\paragraph{Supplementary movie 5: } Animation of the overdamped model with polarity-polarity coupling for two values of the coupling strength $\epsilon_\mathrm{PPC}$. Left shows polarity anti-alignment ($\epsilon_\mathrm{PPC}<0$), right shows polarity alignment ($\epsilon_\mathrm{PPC}>0$). Circles indicate the nuclei of the cells and triangles indicate the protrusions of the cells. The grey bar indicates the mechanical coupling of these two degrees of freedom.

\paragraph{Supplementary movie 6: } Animation of the overdamped model with polarity-polarity coupling (PPC) and polarity-protrusion coupling (PPrC) for the five best fitting parameter combinations for the five different cell types as shown in Fig. \ref{fig:global_fit}. Circles indicate the nuclei of the cells and triangles indicate the protrusions of the cells. The grey bar indicates the mechanical coupling of these two degrees of freedom.

\section{Supplementary experimental details}

\subsection{Cell culture conditions}
For the five cell lines utilized in the experiments, different culture condition and different types of media are used. The exact culture conditions and medium supplements needed for the five different cell lines are listed in Table \ref{cellcultur}.
\begin{table}[h!]
\centering
\begin{tabular}{ |p{5cm}|p{5cm}|p{5cm}|}
 %\multicolumn{3}{|c|}{Cell culture conditions} \\
 \hline
 Cell line & Medium &Origin\\
 \hline
 MDAMB231 H2B mCherry   & L15 (Gibco) + 10\% FBS (Gibco)   &Gift from Betz Lab\\
 \hline
 MDAMB436  &  L15 (Gibco) + 10\% FBS (Gibco) & ATCC\\
 \hline
 HT1080   &     MEM (Gibco) + 10\% FBS (Gibco) + 5\% CO2 & Authenticated by CLS\\
 \hline
 A549 & RPMI (Gibco)+ 10\% FBS (Gibco) 5\% CO2  & DSMZ\\
\hline
 MCF10A & DMEM/F-12 (Gibco) + 5\% horse serum(Gibco), 20 ng/ml human epidermal growth factor (Sigma), 100 ng/ml cholera toxin (Sigma), 10 $\mu$g/ml insulin (Sigma) and 500ng/ml hydrocortisone (Sigma) + 5\% CO2  & ATCC\\
 \hline
\end{tabular}
\caption{Cell culture conditions of the different cell lines and their origin.}
\label{cellcultur}
\end{table}

\subsection{Cell exclusion criteria}
For each cell line we track the nuclei of a large number of cell pairs confined to the dumbbell shaped micropattern. Following previous work \cite{Bruckner2021}, we apply the following criteria for cells to be eligible for tracking: 
\begin{enumerate}
    \item Two-cell trajectories are obtained by either selecting a cell that undergoes division during the duration of the experiment, or by selecting an already attached cell pair at the beginning of the experiment. Tracking is terminated when one of the two cells rounds up for division. 
    \item The tracks have to be at least 50 consecutive frames (500 min) long.
    \item The whole cell including its protrusion has to be confined inside the micropattern.
    \item The cells have to be healthy. Abnormalities such as multiple nuclei, disrupted nuclei and occurrences like cell death or detachment from the substrates leads to exclusion of the tracks. 
\end{enumerate}

\subsection{Tracking procedure and resulting cell trajectories}
During the time-lapse measurement (48h), every 10 min a brightfield image and a fluorescence image of the stained nuclei are acquired. Since the micropattern is not visible in the brightfield, one fluorescence image of the stained dumbbells for each position is taken before the start of the time-lapse measurement. The cell trajectories are determined by using TrackPy (Python Version 3.10.5). After application, the trajectories are inspected manually to correct for tracking mistakes. In the cases where the position of the two cells are mixed up, this is corrected manually. The fluorescence image of the micropattern is used to determine the boundaries of the micropattern and to determine the coordinate origin, which is set at the center of the bridge. The resulting trajectory data sets are summarized in table \ref{table:all_data_sets_two_cell} for all two-cell experiments and in table \ref{table:all_data_sets_single_cell} for all single cell experiments.

\begin{table}[h!]
\setlength{\arrayrulewidth}{0.5mm}
\renewcommand{\arraystretch}{1.5}
\centering
\begin{tabular}{| m{2.5cm} | m{2cm}| m{2.1cm} | m{0.5cm} | m{1.5cm} | m{4.5cm} |} 
 \hline
 \textbf{cell line} & \textbf{perturbation} & \textbf{micropattern}  & \textbf{$N$} & \textbf{Figure} & \textbf{comment} \\ [0.5ex] 
 \hline\hline
 MCF10A & wildtype & dumbbell & $251$ & Fig. 2-6 & we perform new experiments and include data adapted from \cite{Bruckner2021} \\ 
 \hline
 MCF10A & wildtype & rectangular & $96$ & SI Fig. 22 & \\
 \hline
 MCF10A & E-Cadherin blocking & dumbbell & $89$ & Fig. 7 & \\
 \hline
 MCF10A & EphrinA1 & dumbbell & $103$ & Fig. 7 & \\
 \hline
 MCF10A & TGF$\beta$ & dumbbell & $101$ & Fig. 7 & \\
 \hline
 A549 & wildtype & dumbbell & $100$ & Fig. 2-6 & \\
 \hline
 HT1080 & wildtype & dumbbell & $87$ & Fig. 2-6 & \\
 \hline
 MDA-MB-436 & wildtype & dumbbell & $102$ & Fig. 2-6 & \\
 \hline
 MDA-MB-231 & wildtype & dumbbell &  $185$ & Fig. 2-6 & we perform new experiments and include data adapted from \cite{Bruckner2021}\\ 
 \hline
 MDA-MB-231 & wildtype & rectangular &  $86$ & SI Fig. 22 & data adapted from \cite{Bruckner2021}\\ 
 \hline
 MDA-MB-231 & E-Cadherin transfection & dumbbell &  $63$ & Fig. 7, SI Fig. 4 & \\ 
 \hline
 MDA-MB-231 & GFP transfection & dumbbell &  $97$ & SI Fig. 4 & \\ [1ex] 
 \hline
\end{tabular}
\caption{Overview over the various two-cell experiments conducted in this study. The column $N$ indicates the number of trajectories extracted from the experiment.}
\label{table:all_data_sets_two_cell}
\end{table}

\phantom{blabla}
\newpage

\begin{table}[h!]
\setlength{\arrayrulewidth}{0.5mm}
\renewcommand{\arraystretch}{1.5}
\centering
\begin{tabular}{| m{2.5cm} | m{2cm}| m{2.1cm} | m{0.5cm} | m{1.5cm} | m{4.5cm} |}  
 \hline
 \textbf{cell line} & \textbf{perturbation} & \textbf{micropattern}  & \textbf{$N$} & \textbf{Figure} & \textbf{comment} \\ [0.5ex] 
 \hline\hline
 MCF10A & wildtype & dumbbell & $215$ & SI Fig. 2,3 & data adapted from \cite{Bruckner2019c} \\ 
 \hline
 MCF10A & wildtype & rectangular & $23$ & SI Fig. 23 & \\
 \hline
 MCF10A & E-Cadherin blocking & dumbbell & $60$ & SI Fig. 3 & \\
 \hline
 MCF10A & EphrinA1 & dumbbell & $43$ & SI Fig. 3 & \\
 \hline
 MCF10A & TGF$\beta$ & dumbbell & $63$ & SI Fig. 3 & \\
 \hline
 A549 & wildtype & dumbbell & $100$ & SI Fig. 2 & \\
 \hline
 HT1080 & wildtype & dumbbell & $98$ & SI Fig. 2 & \\
 \hline
 MDA-MB-436 & wildtype & dumbbell & $102$ & SI Fig. 2 & \\
 \hline
 MDA-MB-231 & wildtype & dumbbell &  $149$ & SI Fig. 2,3 & data adapted from \cite{Bruckner2019c}\\ 
 \hline
 MDA-MB-231 & wildtype & rectangular &  $33$ & SI Fig. 23 & data adapted from \cite{Bruckner2022}\\ 
 \hline
 MDA-MB-231 & E-Cadherin transfection & dumbbell &  $45$ & SI Fig. 3 & \\ 
 \hline
 MDA-MB-231 & GFP transfection & dumbbell &  $90$ & SI Fig. 3 & \\ [1ex] 
 \hline
\end{tabular}
\caption{Overview over the various single cell experiments conducted in this study. The column $N$ indicates the number of trajectories extracted from the experiment.}
\label{table:all_data_sets_single_cell}
\end{table}

\subsection{Experimental details of perturbations}
In this section, we expand on experimental details of various molecular perturbations.

\subsubsection{EMT in MCF10A cells}
\begin{figure}[h!]
    \centering
    \includegraphics[width=0.9\linewidth]{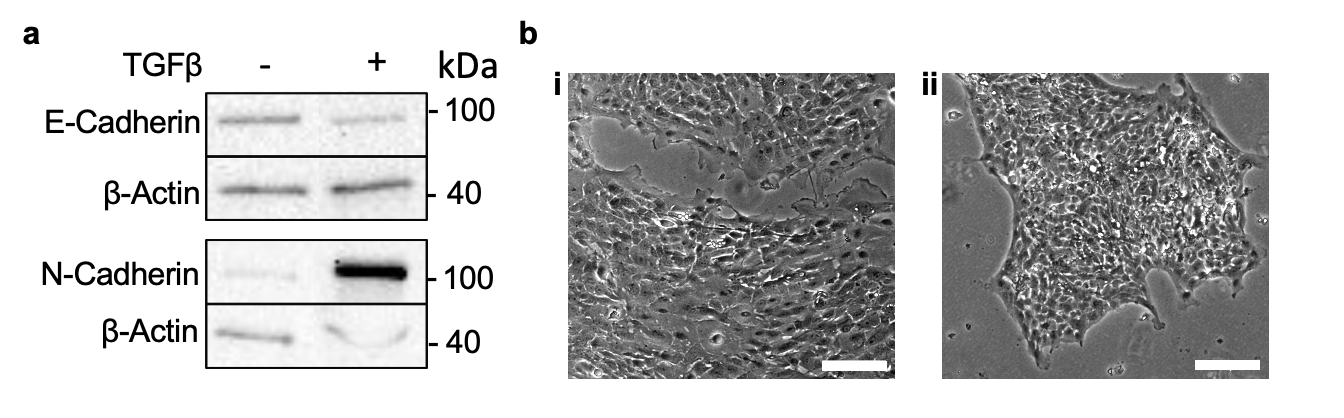}
    \caption{EMT in MCF10A. a) Western blot for E-Cadherin, N-Cadherin in untreated and TGF$\beta$-treated MCF10A cells. b) Brightfield images of TGF$\beta$-treated MCF10A cells (i) and untreated MCF10A cells (ii) after 7 days of culture.} 
    \label{tgfbeta}
\end{figure}

MCF10A cells were seeded in a T25 flask and treated with 10 ng/ml TGF$\beta$ (Thermofisher) for 7 days. Every other day the medium was exchanged including TGF$\beta$. Western blot analysis shows the decrease in E-Cadherin expression and increase in N-Cadherin expression in TGF$\beta$-treated cells (figure \ref{tgfbeta}a). In culture, TGF$\beta$-treated cells are more elongated and have a ruffled border at the edge compared to the untreated MCF10A cells (figure \ref{tgfbeta}b).

\subsubsection{mRNA construction}
In this section, we give further details for the mRNA transfection of E-Cadherin into the breast cancer cell line MDA-MB-231.
\paragraph{1. Plasmid vector:}
hE-cadherin-pcDNA3 was a gift from Barry Gumbiner (Addgene plasmid \#45769). The pSTI-A120-vector was a friendly gift from Dr. Carsten Rudolph (ethris GmbH) and has a 120-bp poly(A) tail and a 3’ untranslated region (UTR) from human $\beta$-globin enabling in vitro transcription of polyadenylated RNA.
To generate the desired vector, a Gibson DNA Assembly reaction was conducted by using the NEBuilder HiFi DNA Assembly kit (NEB, NEBuilder HiFi DNA Assembly Cloning Kit) as described. Briefly, the coding region for hE-cadherine was PCR amplified using the Q5 High-Fidelity PCR Kit (NEB). The oligonucleotide primers for the hE-cadherine-fragment were designed so that the DNA fragments to be assembled overlap each other by at least 30 bp at the ends.:\\
Primer (fwd): tacgactcactatagggcgagggagactgccaccATGGGCCCTTGGAGCCGC\\ 
Primer (rev): atctgcacgcctccttgcttgcttgaattcGTGGTCCTCGCCGCCTCC)\\ 
\\The assembly reaction contained 100 pmol of insert DNA and 35 pmol of the SpeI-linearized pSTI-A120-vector and was incubated at 50°C for 20 min. After the GDA reaction was completed, the reaction mix was transformed into chemically competent E.coli cells (NEB 5-alpha Competent E. coli (High Efficiency), NEB). 100 $\mu$L of the transformed LB (Lysogeny Broth)-E.coli-mix were plated onto LB/Kanamycin plates, followed by 37°C incubation overnight. Single colonies were picked and positive clones were further verified by DNA sequencing. To eliminate the SapI-restriction site in the hE-cadherine gene, we carried out a mutagenesis and exchange leucin to isoleucine by using Q5® Site-Directed Mutagenesis Kit (NEB) as described. Positive clones were verified by DNA sequencing. 
\paragraph{2. mRNA production:}
To generate in vitro-transcribed mRNA (IVT RNA), the plasmid was linearized downstream of the poly(A) tail by SapI digestion and purified with the NEB Monarch PCR \& DNA Cleanup Kit (NEB). The linearized vector (1 $\mu$g) was used as a template for the in vitro transcription reaction using the Biozym Kit (MessageMAX™ T 7 ARCA-Capped Message Transcription Kit), having 100\% of the Anti-Reverse Cap Analog in the produced IVT RNA in the correct orientation, increasing the translation efficiency of the IVT RNA. The complete IVT mix was incubated at 37°C for 45 min followed by a DNA digestion with DNaseI for 15 min at 37°C. RNA was precipitated with ammonium acetate and washed with 70\% EtOH. The washing step was performed twice. Finally, the RNA pellet was re-suspended in RNAse-free water. 
%gfp and E-Cadherin

\section{Supplementary analysis of the experimental data}
In this section, we describe supplementary analysis results of the experimental data that support the conclusions drawn in the main text. 

\subsection{Single cell behavior of the different cell types}

An important aspect of two cells colliding is the way single cells migrate on their own. Thus, for all of our two-cell experiments, we additionally investigate single cells migrating on our dumbbell-shaped micropattern (Supplementary movie 2). Specifically, we track individual cells and analyze the statistical properties of the resulting cell trajectories. We will use knowledge about the single-cell behavior in three places: i) we show that our cells exhibit different levels of "invasivenes" in section \ref{sec:invasiveness_single_cells}. This reveals that our selection of cells covers invasive and non-invasive cell lines. ii) We investigate in section \ref{sec:model_single_cells} the influence of single cell migration aspects on two-cell dynamics within our overdamped model. iii) In section \ref{sec:perturbations_single_cells}, we investigate how our molecular perturbations affect the single cell behavior of our cell lines.

\begin{figure}[t!]
	\centering
	\includegraphics[width=\textwidth]{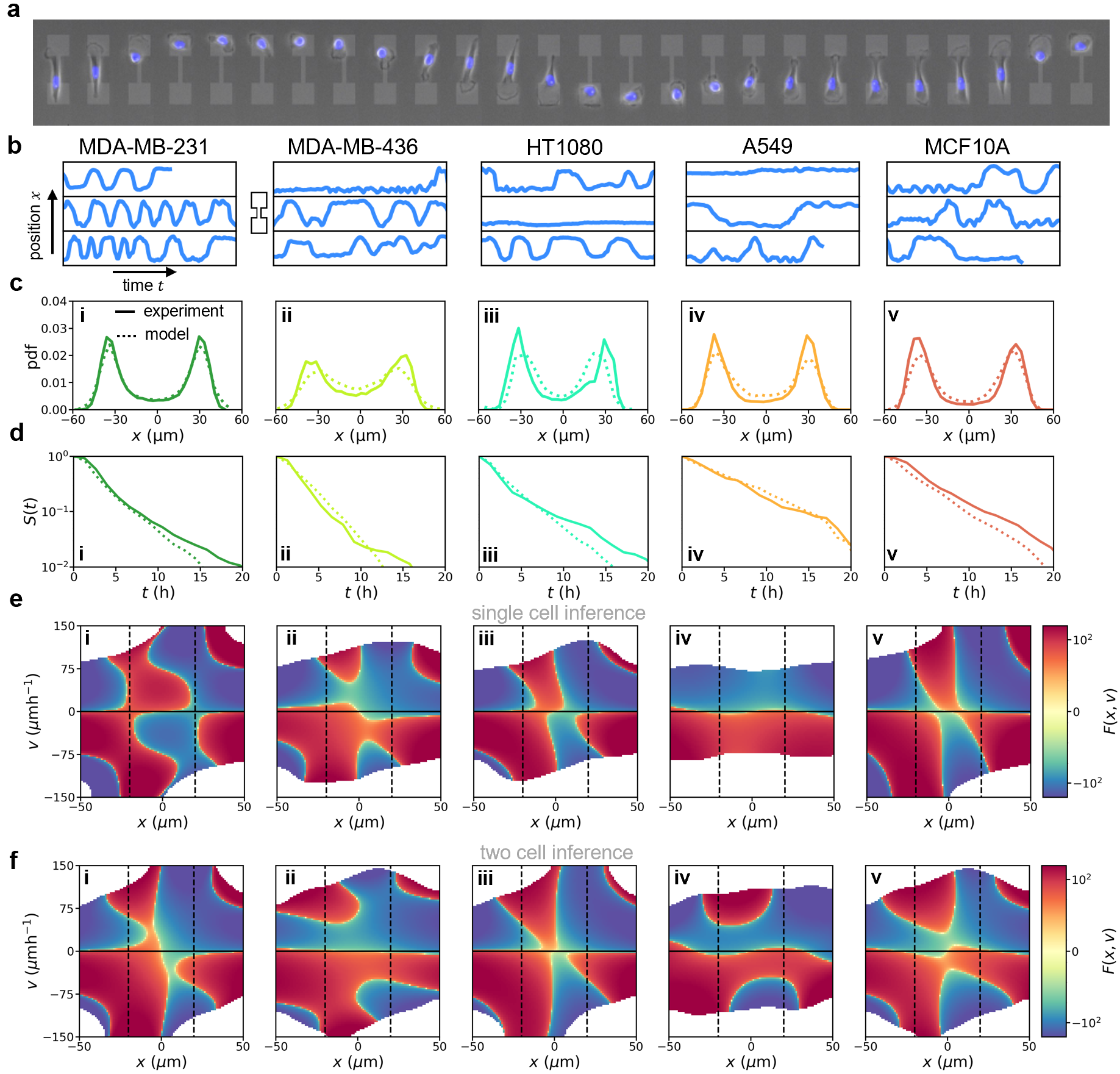}
	\caption{\textbf{Underdamped dynamics of single cell behavior.} \small \textbf{a)} Time series of brightfield images of a single MDA-MB-231 cell hopping between the two islands of our pattern. \textbf{b)} Small selection of nucleus trajectories for the five different cell lines. \textbf{c)} Steady state probability distribution of the position of the cell nucleus. \textbf{d)} Survival probability $S(t)$ for the five different cell types. Both in (c) and (d), solid lines show experimental results and dotted lines show model results obtained from simulating the inferred equation of motion \ref{eq:single_cell}. \textbf{e)} Inferred effective force $F(x,v)$ for all five cell types. $F_s(x,v)$ encodes the average acceleration of a cell nucleus given a certain position $x$ and velocity $v$. Positive values indicates positive accelerations to right on our pattern. Vertical black dashed lines indicate the two islands separated by the bridge on our pattern. \textbf{f)} Inferred effective force $F(x,v)$ for all five cell types from our two-cell inference procedure.}
	\label{fig:single_cell_invasiveness}
\end{figure}

\subsubsection{Dynamics of single cell behavior of cell types}\label{sec:invasiveness_single_cells}

To quantify the single cell behavior of our considered cell types, we analyze how single cells transition between the two islands of our dumbbell-shaped geometry (Fig. \ref{fig:single_cell_invasiveness}a). Specifically, we track the cell nucleus (Fig. \ref{fig:single_cell_invasiveness}b) and find the steady-state probability distribution of the nucleus position. In agreement with \cite{Bruckner2019c}, this pdf reveals that single cells prefer to be on one of the two islands of the dumbbell-shaped micropattern (Fig. \ref{fig:single_cell_invasiveness}c). In addition, we analyze the dynamical features of cell hopping. Specifically, we measure the survival probability of a cell $S(t)$. This is the probability that after a time $t$, the cell has not transitioned yet. We expect that this probability decays for all cell types as all cells show prominent hopping behavior (Fig. \ref{fig:single_cell_invasiveness}b) and thus it is less likely for cells to stay long on one island. Interestingly, we observe differences in the survival probability of the various different cell types. The two breast cancer cell lines (MDA-MB-231 and MDA-MB-436) as well as the fibrosarcoma cell line HT1080 exhibit lower survival probabilities than the epithelial (A549 and MCF10A) cell types. This indicates that breast cancer cell lines and the fibrosarcoma cell line transition more frequently and thus are more invasive than the epithelial cell lines. To gain more detailed insight into the dynamics of single cell behavior, we infer the dynamics of the nucleus using ULI (Methods). As no cell-cell interactions are present, we impose
\begin{equation}\label{eq:single_cell}
    \dot{v} = F_s(x,v) + \sigma \eta(t)
\end{equation}
Here $F_s(x,v)$ 
describes the deterministic driving capturing how the nucleus of a single cell is on average accelerated at a given position and velocity like used in the main text. This effective driving force encodes how the migrating cell interacts with the geometric confinement of the micropattern and gives rise to the hopping dynamics. To capture the stochasticity of the cell trajectories, we include a dynamical noise term $\sigma \eta(t)$, where $\sigma$ is the noise amplitude and $\eta(t)$ is a Gaussian white noise as defined in the Methods. We infer $F_s(x,v)$ and simulate our inferred equation of motion. The analysis of the simulated trajectories shows that the inferred effective force terms $F_s(x,v)$ capture the dynamics of hopping of all our cell types (dotted lines in Fig. \ref{fig:single_cell_invasiveness}c,d). Of note are that all (except the lung epithelial cells A549) deterministically accelerate as they move onto the bridge (red region at $v>0$ and $x \approx 0$) (Fig. \ref{fig:single_cell_invasiveness}e). Taken together, these results show that our cells exhibit  qualitatively similar underdamped hopping dynamics, but cell types have quantitatively distinct survival probabilities. 

\subsubsection{Robustness of the single cell behavior}\label{sec:self_consistency_single_cells}

Here we investigate if the inferred single cell dynamics is altered in the presence of cell-cell interactions.
To this end, we consider the single cell terms $F(x,v)$ that we infer from experimental data on two cells interacting as discussed in the main text. We show these single cell terms in Fig. \ref{fig:single_cell_invasiveness}f. Furthermore, we consider the inferred single cell terms $F_s(x,v)$ from our single cell analysis presented in section \ref{sec:invasiveness_single_cells}, where we know that there are no cell-cell interactions involved in determining the nucleus dynamics of the cells. Thus, if the single cell behavior is unaffected by the presence of a second cell, we expect that $F_s(x,v)\approx F(x,v)$. A comparison of the two terms for each cell type (Fig. \ref{fig:single_cell_invasiveness}e,f) reveals that the single cell behavior of all our cell types is not significantly affected by the presence of cell-cell interactions with other cells. 

\subsubsection{Single cell behavior after perturbations} \label{sec:perturbations_single_cells}

\begin{figure}[t!]
	\centering
	\includegraphics[width=\textwidth]{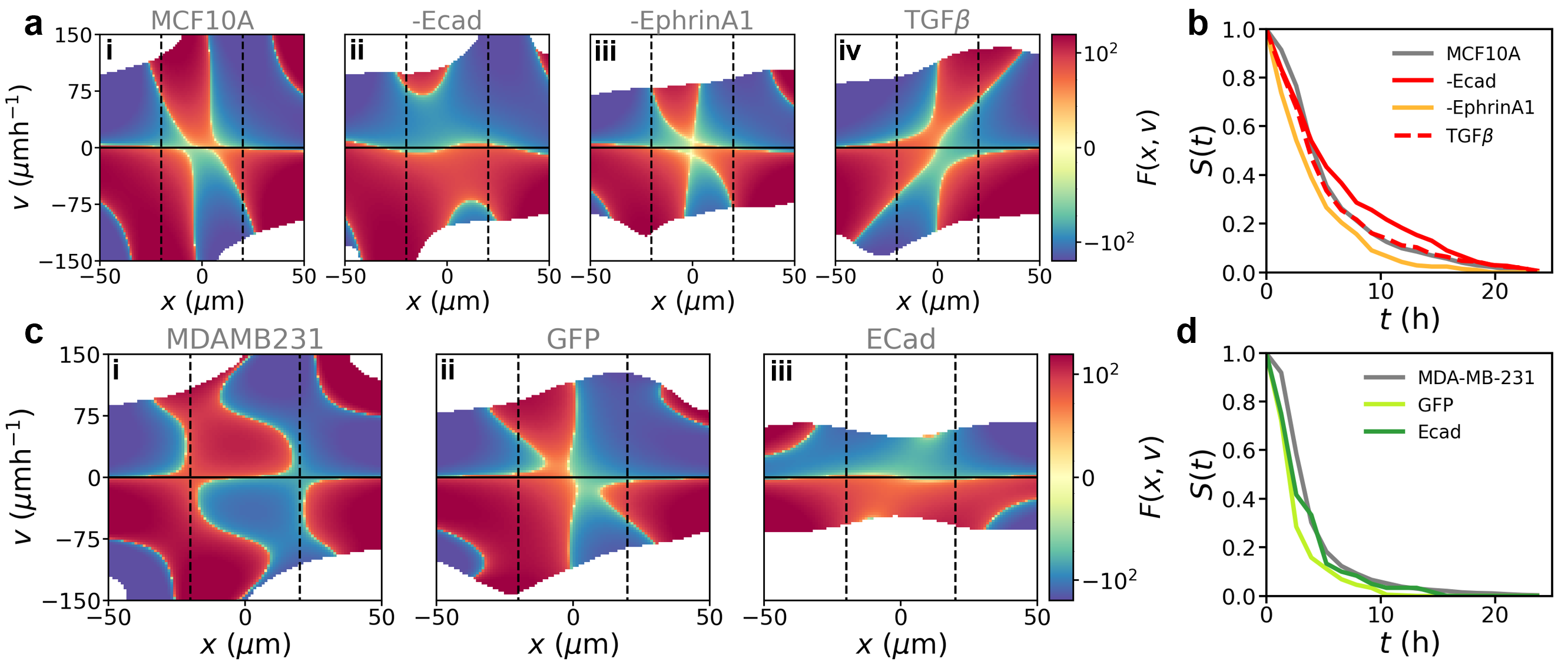}
	\caption{\textbf{Single cell behavior after molecular perturbations.} \small \textbf{a)} Inferred effective force $F(x,v)$ for wild type MCF10A cells (i) and the three considered molecular perturbations (ii-iv). $F_s(x,v)$ encodes the average acceleration of a cell nucleus given a certain position $x$ and velocity $v$. Positive values indicates positive accelerations to right on our pattern. Vertical black dashed lines indicate the two islands separated by the bridge on our pattern. \textbf{b)} Survival probability $S(t)$ of MCF10A cells and the considered molecular perturbations. \textbf{c)} Inferred effective force $F(x,v)$ for wild type MDA-MB-231 cells (i) and the two transfected MDA-MB-231 cell lines (ii,iii). \textbf{d)} Survival probability $S(t)$ of MDA-MB-231 cells and the considered transfections.}
	\label{fig:single_cells_perturbations}
\end{figure}

In this section, we analyze the impact of our molecular perturbations on the single cell behavior of our cell types. Specifically, we compare the inferred single cell dynamics $F_s(x,v)$ and the survival probability $S(t)$ of the wild type epithelial MCF10A and breast cancer MDA-MB-231 cells with their respective molecular perturbations. We find for MCF10A cells that for the anti-body blockings of E-Cadherin and EphrinA1 the inferred single cell dynamics $F_s(x,v)$ are qualitatively similar to the wild type (Fig. \ref{fig:single_cells_perturbations}a i-iii). Furthermore, also the survival probability of the MCF10A cells is roughly similar to the MCF10A cells with anti-body blockings of E-Cadherin and EphrinA1 (Fig. \ref{fig:single_cells_perturbations}b). For the TGF$\beta$ treated cells, we find that the single cell dynamics changed and are qualitatively similar to that of cancerous MDA-MB-231 cells (Fig. \ref{fig:single_cells_perturbations}a iv, c i) while the survival probabilities remain almost unchanged (Fig. \ref{fig:single_cells_perturbations}b). For MDA-MB-231 cells we analyze the single cell behavior of our transfection experiment. We find that MDA-MB-231 cells with transfected GFP does neither greatly affect the single cell dynamics nor the survival probability of these cells (Fig. \ref{fig:single_cells_perturbations}c,d). In contrast, the single cell behavior of MDA-MB-231 cells transfected with E-Cadherin does change to the single cell behavior observed in more epithelial A549 cells. Taken together, we show that antibody blockings do not greatly affect single cell behavior, but more complex perturbations like inducing EMT in MCF10A cells or transfecting MDA-MB-231 cells with E-Cadherin does change aspects of single cell behavior. 

\subsection{Controlling E-Cadherin transfection}\label{sec:GFP_control}
In this section, we compare the dynamics of wildtype MDA-MB-231 cells to MDA-MB-231 cells that have been transfected with GFP. We conducted this transfection in order to control the effects of the protocol that we used to transfect MDA-MB-231 cells with E-Cadherin. Analyzing the correlation functions of velocities and positions as well as the collision statistics as in the main text, we find that both wildtype MDA-MB-231 cells and GFP-transfected MDA-MB-231 cells exhibit very similar behavior statistics. Furthermore, we infer similar underdamped cohesion and friction interactions from the trajectories of these two cell types. Thus, we conclude that the transfection protocol does not greatly affect the two-cell dynamics of MDA-MB-231 cells.

\begin{figure}[t!]
	\centering
	\includegraphics[width=\textwidth]{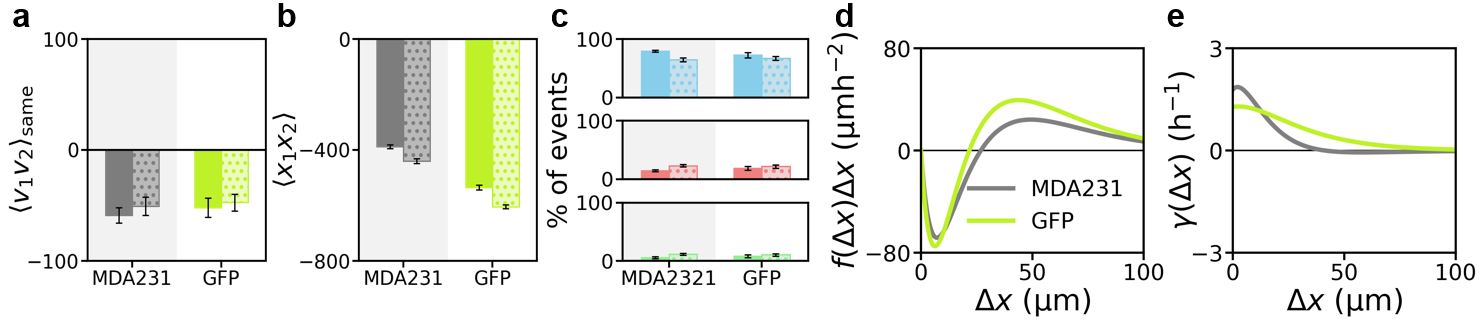}
	\caption{\textbf{Controlling the E-cadherin transfection.} \small \textbf{a)} Instantaneous velocity cross-correlation between the two cells when they occupy the same island. Solid bars show experimental results, dotted bars show the prediction of the inferred underdamped description. For panels (a)-(c), error bars show the error of the mean (s.e.m) obtained from bootstrapping. \textbf{b)} Instantaneous position cross-correlation between the two cells. \textbf{c)} Percentages of the three different collision events for the five different cell lines. \textbf{d)} Inferred effective cohesion interactions for the five different cell lines. Colors correspond to the two cell types as shown in panels (a) and (b). \textbf{f)} Inferred effective friction interactions for the five different cell lines. }
	\label{fig:GFP_control}
\end{figure}

\subsection{Complete quantification of the experimental nucleus dynamics}

\begin{figure}[t!]
	\centering
	\includegraphics[width=\textwidth]{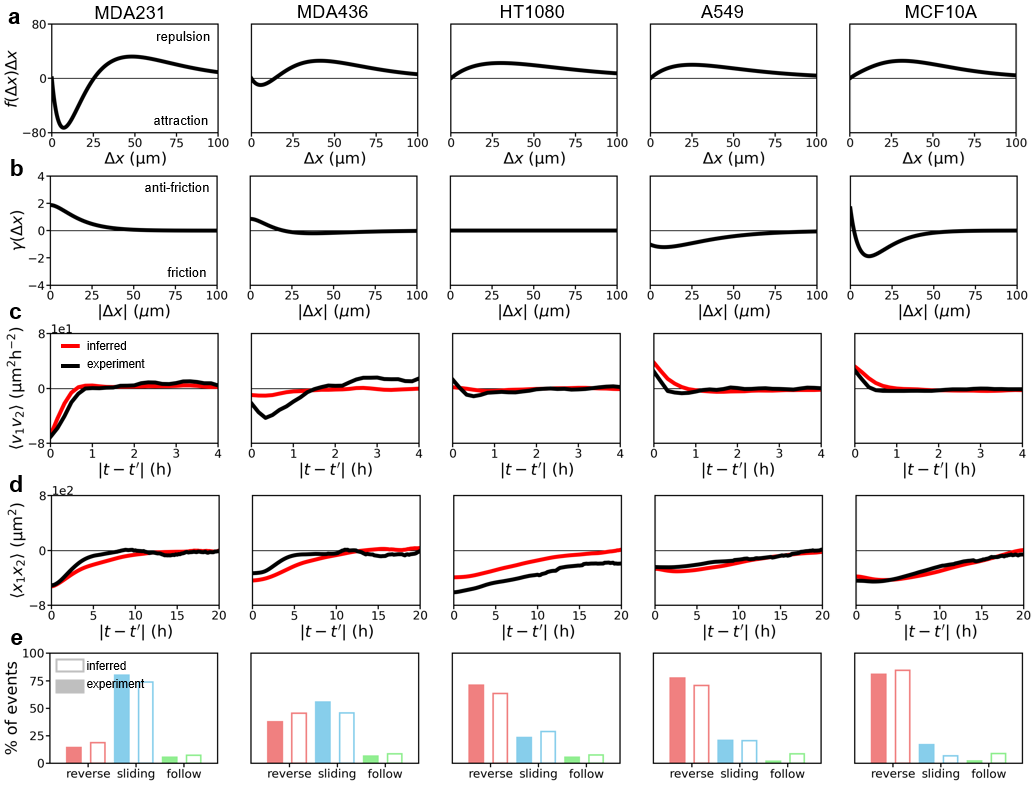}
	\caption{\textbf{Behavior statistics and inferred interactions for all cell types} \small \textbf{a)},\textbf{b)} Underdamped cohesion (a) and friction (b) interactions for all cell types \textbf{c)}-\textbf{e)} Behavior statistics for all cell types. Black curve shows experimental results, red curves show the behavior predicted by the inferred langevin equation including the underdamped interactions shown in panels (a) and (b). Here, (c) shows the velocity correlation function $\langle v_1(t)v_2(t')\rangle_{\mathrm{same}}$ when cells are on the same island. (d) show the position correlation function $\langle x_1(t)x_2(t')\rangle$. (e) shows the collision statistics for all different cell types.}
	\label{fig:full_dynamics}
\end{figure}

In the main text, we only plot the instantaneous velocity correlation $C_V(|t-t'| = 0)$ and the instantaneous position correlation $C_X(|t-t'| = 0)$. In Fig. \ref{fig:full_dynamics} we show the full result of these behavior statistics and show also long time scale correlations of the various cell types. Also these long time scale correlations are well captured by the inferred Langevin equation (red curves in Fig. \ref{fig:full_dynamics}c,d).  

\newpage
\section{Model development and fitting procedure}

In this section, we expand on details of the development and implementation of the overdamped model for cell-cell interactions used in the main text to capture the interaction behavior of various different cell types.
  
\subsection{Complete description of cell-cell interaction mechanisms}

Here, we describe in detail all cell-cell interaction mechanisms:
\begin{itemize}
    \item \textbf{NC} (nucleus coupling): $G_\mathrm{n}(\Delta x_\mathrm{n}) = \epsilon_\mathrm{NC} e^{-|\Delta x_\mathrm{n}|/r_\mathrm{NC}}\Delta x_\mathrm{n}$, modeling either repulsion ($\epsilon_{NC}>0$) or attraction ($\epsilon_{NC}<0$) interactions between the cell nuclei. 
    \item \textbf{NPrC} (nucleus-protrusion coupling): $G_\mathrm{n}(\Delta x_\mathrm{np}) = \epsilon_\mathrm{NPrC} e^{-|\Delta x_\mathrm{np}|/r_\mathrm{NPrC}}\Delta x_\mathrm{np}$, modeling either repulsion ($\epsilon_\mathrm{NPrC}>0$) or attraction ($\epsilon_\mathrm{NPrC}<0$) interactions between the cell nucleus and the protrusion of the other cell.
    \item \textbf{NPC} (nucleus-polarity coupling): $G_\mathrm{n}(\Delta x_\mathrm{np}, P_j) = \epsilon_\mathrm{NPC} e^{-|\Delta x_\mathrm{np}|/r_\mathrm{NPC}}P_j$, modeling a potential direct influence of the polarity of cell $j$ on the nucleus position of cell $i$. For $\epsilon_\mathrm{NPC}<0$, we have that the nucleus reacts opposite to the polarity of the other cell.
    \item \textbf{NUC} (nucleus-deformation coupling): $G_\mathrm{n}(\Delta x_\mathrm{np}, U_j) = \epsilon_\mathrm{NUC} e^{-|\Delta x_\mathrm{np}|/r_\mathrm{NPC}}U_j$, modeling a potential direct influence of the deformation of cell $j$ on the nucleus position of cell $i$. For $\epsilon_\mathrm{NUC}<0$, we have that the nucleus reacts opposite to the deformation of the other cell.
    \item \textbf{PrC} (protrusion coupling): $G_\mathrm{p}(\Delta x_\mathrm{p}) = \epsilon_\mathrm{PrC} e^{-|\Delta x_\mathrm{p}|/r_\mathrm{PrC}}\Delta x_\mathrm{p}$, modeling either repulsion ($\epsilon_\mathrm{PrC}>0$) or attraction ($\epsilon_\mathrm{PrC}<0$) interactions between the cell protrusions.
    \item \textbf{PrNC} (protrusion-nucleus coupling): $G_\mathrm{p}(\Delta x_\mathrm{pn}) = \epsilon_\mathrm{PrNC} e^{-|\Delta x_\mathrm{pn}|/r_\mathrm{PrNC}}\Delta x_\mathrm{pn}$, modeling either repulsion ($\epsilon_\mathrm{PrNC}>0$) or attraction ($\epsilon_\mathrm{PrNC}<0$) interactions between the cell protrusion and the nucleus of the other cell.
    \item \textbf{PrPC} (protrusion-polarity coupling): $G_\mathrm{p}(\Delta x_\mathrm{pn},P_j) = \epsilon_\mathrm{PrPC} e^{-|\Delta x_\mathrm{pn}|/r_\mathrm{PrPC}} P_j$, modeling a direct influence of the polarity of cell $j$ on the protrusion of cell $i$. For $\epsilon_\mathrm{PrPC}<0$, we have that the protrusion reacts opposite to the polarity of the other cell.
    \item \textbf{PrUC} (protrusion-oreintation coupling): $G_\mathrm{p}(\Delta x_\mathrm{pn},U_j) = \epsilon_\mathrm{PrUC} e^{-|\Delta x_\mathrm{pn}|/r_\mathrm{PrPC}} U_j$, modeling a direct influence of the deformation of cell $j$ on the protrusion of cell $i$. For $\epsilon_\mathrm{PrUC}<0$, we have that the protrusion reacts opposite to the deformation of the other cell.
    \item \textbf{PPrC} (polarity-protrusion coupling): $G_\mathrm{P}(\Delta x_\mathrm{p}) = \epsilon_\mathrm{PPrC} e^{-|\Delta x_\mathrm{p}|/r_\mathrm{PPrC}}\Delta x_\mathrm{p}$ models that the growth rate of the polarity is dependent on the separation between the cell protrusions as well as on the orientation of cells with respect to each other. For $\epsilon_\mathrm{PPrC} > 0$, the polarity grows away from the protrusion of the other cell while for $\epsilon_{PPrC} < 0$, the polarity grows towards the other protrusion. Note that this interaction has been used to model contact inhibition of locomotion \cite{Bertrand2024}.
    \item \textbf{PNC} (polarity-nucleus coupling): $G_\mathrm{P}(\Delta x_\mathrm{pn}) = \epsilon_\mathrm{PNC} e^{-|\Delta x_\mathrm{pn}|/r_\mathrm{PNC}}\Delta x_\mathrm{pn}$ models that the growth rate of the polarity is dependent on the separation between the cell's protrusion and the other cell's nucleus. For $\epsilon_\mathrm{PNC} < 0$, the polarity grows towards the other nucleus.
    \item \textbf{PSC} (polarity self coupling): $G_\mathrm{P}(\Delta x_\mathrm{p}, P_i) = -\epsilon_\mathrm{PSC} e^{-|\Delta x_\mathrm{p}|/r_\mathrm{PSC}}P_i$. This interaction models the polarity shrinking (or growing dependent on the sign of $\epsilon_{PSC}$) dependent on the relative separation between cells but irrespective of the relative orientations of the cells. Note that this interaction is non-reciprocal.
    \item \textbf{PC} (polarity  coupling): $G_\mathrm{P}(\Delta x_\mathrm{p}, P_j) = \epsilon_\mathrm{PC} e^{-|\Delta x_\mathrm{p}|/r_\mathrm{PC}}P_j$. This interaction is mathematically very similar to PSC, but now the dynamics of polarity $P_i$ depends on $P_j$. This interaction thus models how the polarity of cell $i$ grows in or against the direction of the polarity of the other cell $j$ with a growth rate given by the magnitude of $P_j$. 
    Here $\epsilon_\mathrm{PC} > 0$ indicates that the polarity grows in the same direction. Note that this interaction has been called an alignment interaction in the literature \cite{Bertrand2024}. Further, we prevent infinite growth of the polarities by the non-linear term in the single cell dynamics of the polarity (main text Eq. (3)). 
    \item \textbf{PPC} (polarity-polarity coupling) $G_\mathrm{P}(\Delta x_\mathrm{p}, \Delta P) = -\epsilon_{PPC} e^{-|\Delta x_\mathrm{p}|/r_{PPC}}\Delta P$, which models the cell's polarity aligning to the polarity of the other cell dependent on the relative separation between the cell protrusions. For $\epsilon_{PPC} < 0$, we have anti-alignment of cell polarity. Note that PPC and PC both describe how the polarity of cell $i$ grows dependent on the polarity of cell $j$. The difference between these interactions is only given by the final polarities after interacting. For PPC, the final polarities are as equal as possible. For PC, the final polarities are given by the non linear term in main text Eq. (3).
    \item \textbf{PUC} (polarity deformation coupling): $G_\mathrm{P}(\Delta x_\mathrm{p}, U_j) = \epsilon_\mathrm{PUC} e^{-|\Delta x_\mathrm{p}|/r_\mathrm{PUC}}U_j$. This interaction is conceptually very similar to PC. However, in this interaction mechanism, the polarity of cell $i$ grows in or against the direction of the deformation vector of the other cell $U_j$.
\end{itemize}

\subsection{Model implementation}

We numerically solve main text Eqs. (1)-(3) by a Euler-Maruyama forward integration scheme. To this end, we initialize random positions for the cell nucleus on the pattern and choose close but random positions of the protrusion. The polarity of cells is also randomly initialized. For each candidate interaction we then simulate $N = 3000$ timesteps with a time step of $dt = 1/60\ \mathrm{h}$. We oversample and only retain every 10th timestep to obtain a final simulation time of $T = 50\ \mathrm{h}$ with a timedelta of $\Delta t = 1/6\ \mathrm{h}$. These values are similar to the experimental values. These numerical simulations yield the model nucleus trajectories shown in main text Fig. 5a and are analyzed in a similar way like the experimental data. 

\subsection{Overdamped statistical inference of candidate interactions}\label{sec:model_fitting}

To infer which candidate interaction best captures the cell-level dynamics of interacting cell, we perform a fit of our overdamped model to the experimental data. Thus, for each candidate cell-cell interaction, we vary the set of parameters $\Theta$ in our model and numerically simulate $N = 150$ two-cell trajectories. Here $\Theta$ consists out of $\epsilon$ and $r$ for the respective candidate interaction. Note that if we combine candidate cell-cell interactions, we have four parameters in $\Theta$. We then infer the underdamped cell-cell interactions and measure the behavior statistics predicted by our overdamped model. In the following, we lay out how we fit our overdamped model to the experimental data. We then show supplementary results of the fitting procedure including a comprehensive comparison of our candidate interactions in section \ref{sec:model_comparison} and section \ref{sec:inference_results}.  
\subsubsection{The coefficient of determination}

To quantify the goodness of fit to the experimental data, we compute the so called \textit{coefficient of determination} (COD), which is defined as 
\begin{equation}
    R^2 = 1 - \frac{SS_{res}}{SS_{tot}}
\end{equation}
Here, $SS_{res}$ is defined as the sum of the squared deviations between the model and the experiment:
\begin{equation}
    SS_{res} = \sum_i (y_i-\hat{y}_i)^2 
\end{equation}
with $\hat{y}$ being either the model result for the underdamped cohesion interaction $f(\Delta x)\Delta x$, the underdamped friction interaction $\gamma(\Delta x)$ or one of the three behavior statistics we employ in the main text. Here, $y$ is the experimental result and the index $i$ indicates the discrete positions/categories where the statistics are measured. $SS_{tot}$ is then proportional to the variance of the experimental data: 
\begin{equation}
    SS_{tot} = \sum_i (y_i - \overline{y}_i)^2 
\end{equation}
$R^2$ is equal to $1$ if the model matches the experiment so that $SS_{res}=0$. Furthermore, $R^2$ is around $0$ if the model predicts at least the mean of the experimental data $\overline{y}$ and negative if the model does worse than that. Varying all parameters in the model gives the COD as a function of the model parameters for each candidate cell-cell interaction, for each cell type and for each statistics: $R^2_{j,k}(\Theta)$. Here $j$ indicates the statistics of which we want to asses the goodness of fit. Thus, $R^2_{d,k}(\Theta)$ indicates the COD of the model fit to the underdamped interactions, while $R^2_{v,k}(\Theta)$ and $R^2_{x,k}(\Theta)$ is the COD of capturing the correlation functions of the velocities and positions, respectively. $R^2_{c,k}(\Theta)$ is the COD of the collision statistics. The index $k$ indicates different cell types. 

\subsubsection{Individual fit of the dynamics}\label{sec:fitting}

To fit the dynamics, we need to maximize the COD for each cell type and across the various different statistics. Let's define

\begin{equation}\label{eq:COD_average}
    R_k^2(\Theta) = \frac{1}{2} R^2_{d,k}(\Theta) + \frac{1}{4}R^2_{v,k}(\Theta) + \frac{1}{4}R^2_{c,k}(\Theta)
\end{equation}

as the mean COD across the underdamped dynamics, the velocity correlations and the collisions statistics for one individual cell type. Note, that in Eq. \ref{eq:COD_average} we put equal weight on all the statistics ($R^2_{d,k}(\Theta)$ contains already both the underdamped cohesion and friction interactions, thus the factor $\frac{1}{2}$). In Fig. \ref{fig:robustness_fit} we investigate the robustness of our inference procedure on this choice. Note that in Eq. \ref{eq:COD_average}, we exclude $R^2_{x,k}(\Theta)$ of the position correlations. This is done because across all cell types and all candidate cell-cell interactions, we observed that the position correlation function $C_X (|t-t'|)$ is not well matched and thus $R^2_{x,k}(\Theta)$ is very negative (Fig. \ref{fig:position_correlation}b), which would skew the average defined in Eq. \ref{eq:COD_average}. Thus, we will use $R^2_{x,k}(\Theta)$ to later check if our model can consistently predict $C_X (|t-t'|)$ (Fig. \ref{fig:position_correlation}b,c).\\ 
\\We first maximize $R_k^2(\Theta)$ individually for the different cell types without any constraints on the parameters. Thus, for each candidate cell-cell interaction, we find a unique set of parameters for each cell type as: 
\begin{equation}
    \hat{\Theta}_k = \argmax_{\Theta}R^2_k(\Theta)
\end{equation}
with 
\begin{equation}\label{equation1}
    \hat{R}^2_k = \max_{\Theta}R^2_k(\Theta)
\end{equation}
To compare models, we plot $\hat{R}^2_k$ for each candidate interaction mechanism in main text Fig. 6a. We show in Fig. \ref{fig:individual_fit} the results of that procedure for each cell type for the best fitting candidate cell-cell interaction PPC+PPrC. 

\subsubsection{Global fit of the dynamics}

Because we want to further challenge our theory and test if we can find a model that captures the experimental data with less parameters than the individual fit, we attempt a global fit of all cell types. Specifically, we try to capture the dynamics of all different cell lines with a single candidate interaction, while varying only a single parameter. This single parameter then has to tune between all the different behaviors that we observe in the experiment. For the single candidate cell-cell interactions we keep the interaction range $r$ fixed and vary $\epsilon$. This is because we observed that $\epsilon$ affects the resulting dynamics significantly more than the interaction range. To this end, for each candidate interaction, we first vary $\epsilon$ and maximize the COD for each individual cell type for each value of $r$: 
\begin{equation}
    \hat{\epsilon}_k(r) = \argmax_{\epsilon}R^2_k(\epsilon,r)
\end{equation}
this leaves us with optimal $\hat{\epsilon}_k(r)$ for each cell type and each value of $r$. Then, for the global fit, we consider the COD at the optimal $\hat{\epsilon}_k$ for each cell type:
\begin{equation}\label{equation2}
    \hat{R}^2_k(r) = \max_{\epsilon}R^2_k(\epsilon,r)
\end{equation}
Note that in equation \ref{equation1}, $\hat{R}^2_k$ is a scalar, while $\hat{R}^2_k$ is a function of $r$ in equation \ref{equation2}. We then average over different cell types to obtain a global COD defined as 
\begin{equation}\label{eq:COD}
    \mathcal{R}^2(r) = \langle\hat{R}^2_k(r) \rangle_k
\end{equation}
Finally, we maximize $\mathcal{R}^2(r)$ to obtain 
\begin{equation}\label{eq:global_interaction_range}
    \hat{r} = \argmax_{r}\mathcal{R}^2(r)
\end{equation}
with maximal global COD
\begin{equation}\label{global_COD}
    \hat{\mathcal{R}}^2 = \max_{r}\mathcal{R}^2(r) = \max_{r}\langle\max_{\epsilon}R^2_k(\epsilon,r) \rangle_k
\end{equation}
This gives us for each cell type a pair of parameters $(\hat{\epsilon}_k(\hat{r}), \hat{r})$ with fixed $\hat{r}$. Thus, in this global fit, we effectively only use two parameters to capture the experimental data compared to the $10$ parameters in the individual fit. In addition, for the pairs of interactions, we always allow the polarity-polarity-coupling strength $\epsilon_{\mathrm{PPC}}$ (or the polarity-deformation coupling strength $\epsilon_{\mathrm{PUC}}$) to vary, while keeping all the other parameters (the interaction strength of the other interaction $\epsilon$ and the two interaction ranges for the two candidate interactions) fixed. This gives us a global COD $\mathcal{R}^2(\epsilon,r_{\mathrm{PPC}}, r)$, where $\epsilon$ and $r$ are the strength and range of the other interaction. We show $\mathcal{R}^2(\epsilon,r_{\mathrm{PPC}}, r)$ at a fixed value of $r_{\mathrm{PPC}}$ in Fig. \ref{fig:COD} o-v. Finally, of important note is that for all candidate interactions we sweep over the parameter ranges in which we can identify at least a local maximum in the COD (Fig. \ref{fig:COD}). This way we make sure that the overdamped model with all interaction mechanisms provides the best possible fit to the experimental data. 

\section{Supplementary model results}
In this SI section we describe supplementary model results that support the conclusions drawn in the main text. 

\subsection{Comparing the dynamics of candidate cell-cell interactions}\label{sec:model_comparison}
In this subsection, we show a detailed analysis and subsequent comparison of the underdamped nucleus dynamics that all our candidate cell-cell mechanisms predict.
\subsubsection{Underdamped dynamics of candidate cell-cell interactions}\label{sec:model_comparison_qual}
\begin{figure}[t!]
	\centering
	\includegraphics[width=0.9\textwidth]{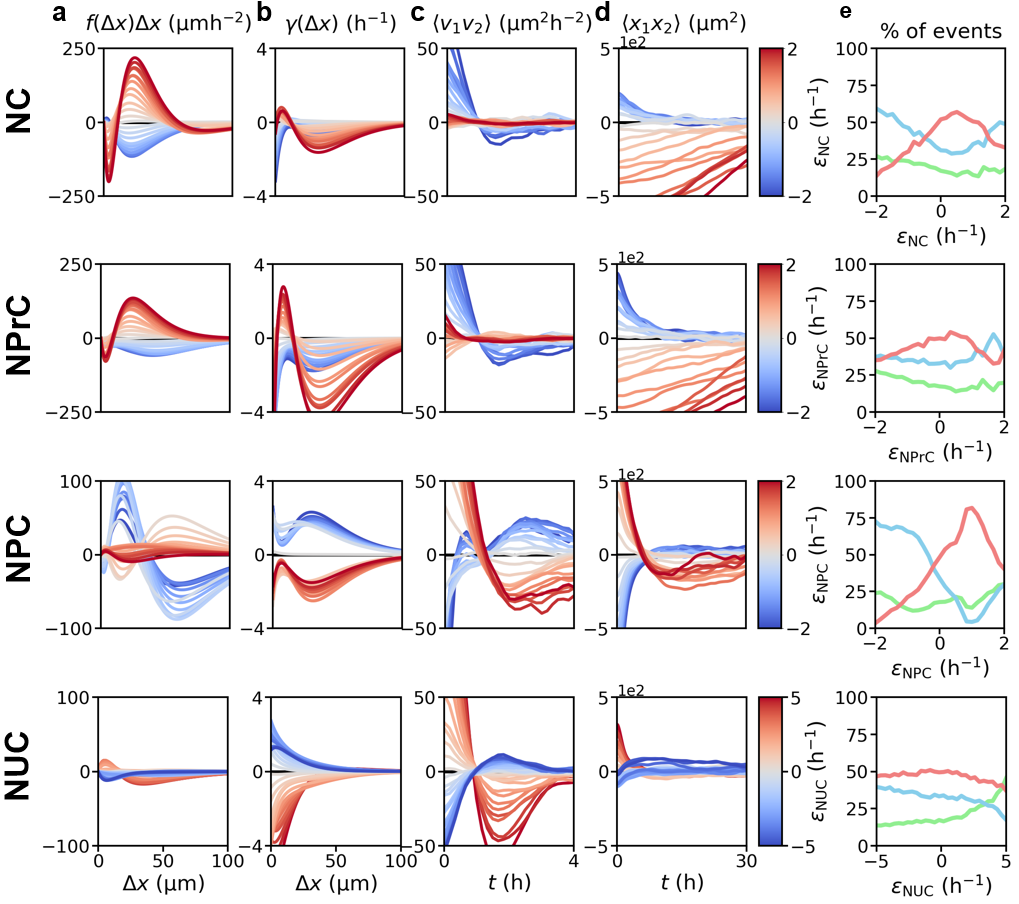}
	\caption{\textbf{Underdamped dynamics of candidate cell-cell interactions acting on the nucleus position.} \small In each row, we show the underdamped dynamics of one of the positional candidate cell-cell interaction as indicated. From left to right, we show: \textbf{a)} the inferred cohesion interactions between cells, \textbf{b)} the inferred friction interactions, \textbf{c)} the velocity cross-correlation between the two cells as they are on the same island. \textbf{d)} The position correlation function between cells, characterizing the mutual exclusion behavior of cells and \textbf{e)} the collision statistics. For (a)-(d), we vary the interaction strength $\epsilon$ as indicated on the colorbars. In (e) we vary the interaction strength $\epsilon$ on the x-axis.}
	\label{fig:model_comparison_nucleus}
\end{figure}

\begin{figure}[t!]
	\centering
	\includegraphics[width=0.9\textwidth]{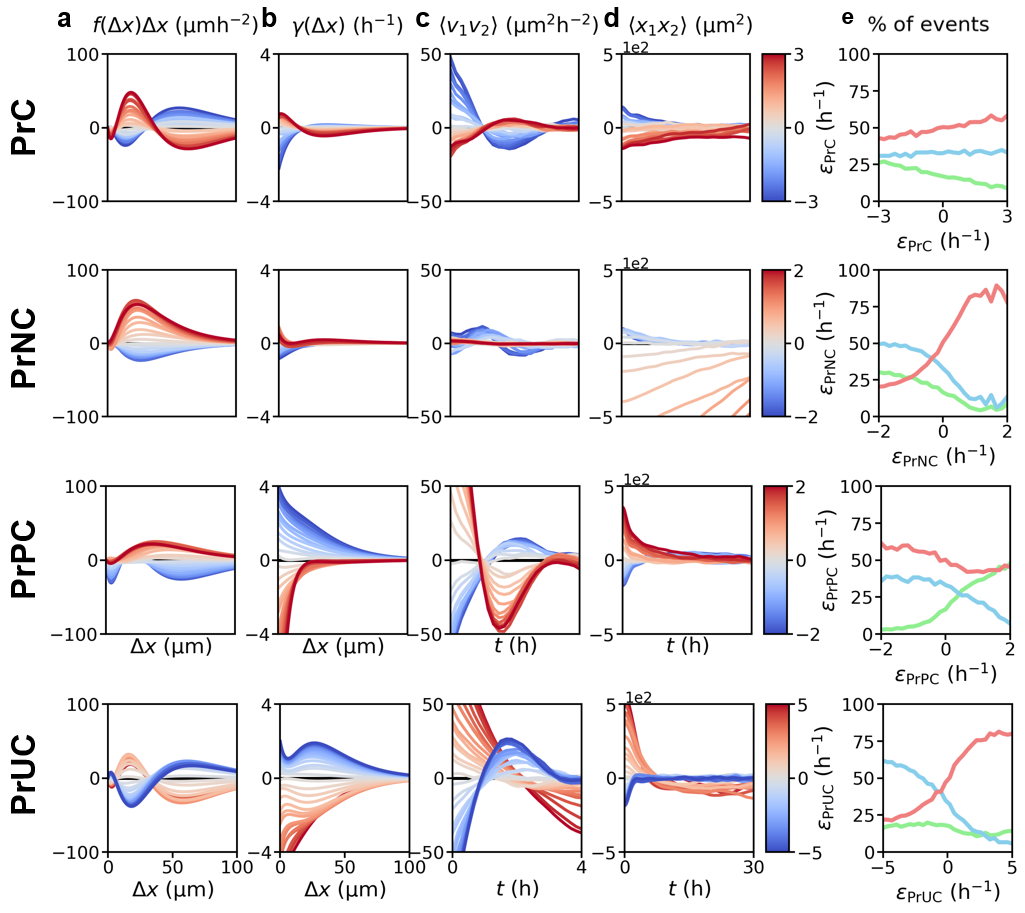}
	\caption{\textbf{Underdamped dynamics of candidate cell-cell interactions acting on the nucleus position.} \small In each row, we show the underdamped dynamics of one of the positional candidate cell-cell interaction as indicated. From left to right, we show: \textbf{a)} the inferred cohesion interactions between cells, \textbf{b)} the inferred friction interactions, \textbf{c)} the velocity cross-correlation between the two cells as they are on the same island. \textbf{d)} The position correlation function between cells, characterizing the mutual exclusion behavior of cells and \textbf{e)} the collision statistics. For (a)-(d), we vary the interaction strength $\epsilon$ as indicated on the colorbars. In (e) we vary the interaction strength $\epsilon$ on the x-axis.}
	\label{fig:model_comparison_protrusion}
\end{figure}

\begin{figure}[t!]
	\centering
	\includegraphics[width=0.9\textwidth]{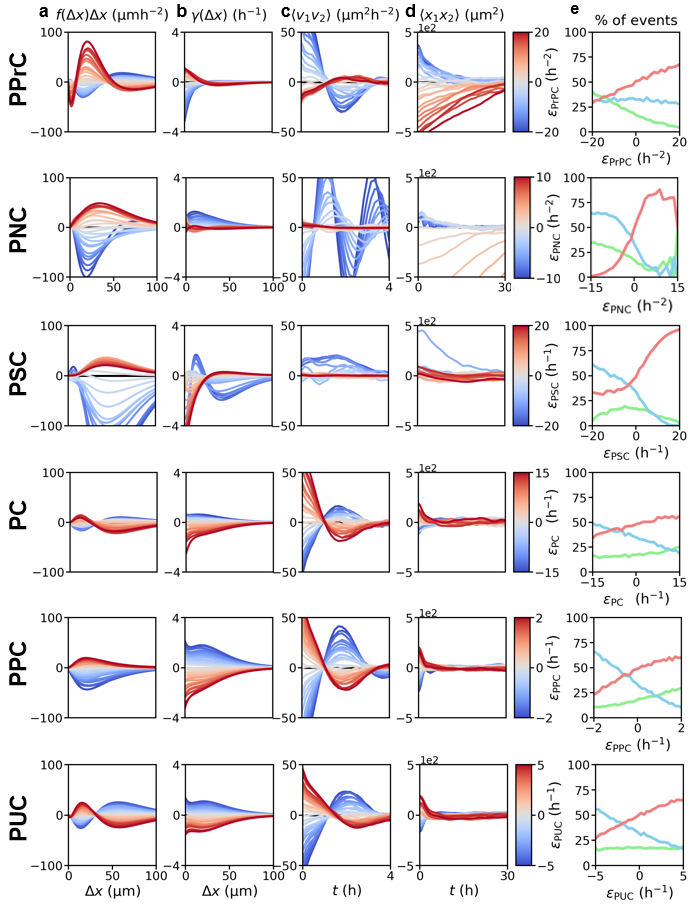}
	\caption{\textbf{Underdamped dynamics of polarity candidate cell-cell interactions.} \small In each row, we show the underdamped dynamics of one of the polarity candidate cell-cell interaction as indicated. From left to right, we show: \textbf{a)} the inferred cohesion interactions between cells, \textbf{b)} the inferred friction interactions, \textbf{c)} the velocity cross-correlation between the two cells as they are on the same island. \textbf{d)} The position correlation function between cells, characterizing the mutual exclusion behavior of cells and \textbf{e)} the collision statistics. For (a)-(d), we vary the interaction strength $\epsilon$ as indicated on the colorbars. In (e) we vary the interaction strength $\epsilon$ on the x-axis.}
	\label{fig:model_comparison_pol}
\end{figure}

To this end, we first discuss qualitative features of the various candidate cell-cell interactions. Specifically, we will analyze what underdamped cell-cell interactions and what behavior statistics the candidate cell-cell interactions can predict. For visualization and to show the best possible model predictions, we fix the interaction range $r$ to $\hat{r}$, which is the value for the interaction range that provides the best global fit to the underdamped dynamics and behavior statistics of all cell types as defined in equation \ref{eq:global_interaction_range}. For each candidate interaction we then vary the interaction strength $\epsilon$. While doing so, we infer the underdamped cell-cell interactions and infer the various behavior statistics. We note that in the experiments, we observed strong correlations between the various behavior statistics and inferred interactions. Specifically, in the experiment we observed the combinations (Fig. \ref{fig:full_dynamics}): 
\begin{itemize}
    \item \textbf{cohesion}, \textbf{friction}, \textbf{positive} velocity correlations, \textbf{negative} position correlations, and dominant \textbf{reversal} behavior (in epithelial cells)
    \item \textbf{attraction}, \textbf{anti-friction}, \textbf{negative} velocity correlations, \textbf{negative} position correlations, and dominant \textbf{sliding} behavior (in breast cancer cells)
\end{itemize}
In the following, we summarize qualitative features of the various candidate cell-cell interaction mechanisms and discuss them in relation to the experimental data. 
\paragraph{NC and NPrC}: For both nucleus coupling and nucleus protrusion coupling, we find that our overdamped model predicts both repulsive and attractive underdamped interactions, but is unable to predict significant anti-friction interactions (NC and NPrC in Fig. \ref{fig:model_comparison_nucleus}a,b). Furthermore, both candidates do not predict negative velocity correlations at short time scales (NC and NPrC in Fig. \ref{fig:model_comparison_nucleus}c) and thus are unable to capture both combinations as outlined above. However, these candidate interactions do predict mutual exclusion behavior and dominant reversal and sliding behavior (NC and NPrC in Fig. \ref{fig:model_comparison_nucleus}d,e).
\paragraph{NPC}: For nucleus polarity coupling, we find very week underdamped repulsion for ($\epsilon > 0$) but strong short-range repulsion and long-range attraction interactions for ($\epsilon < 0$) (NPC in Fig. \ref{fig:model_comparison_nucleus}a). This candidate cell-cell interaction is able to predict both friction and anti-friction (NPC in Fig. \ref{fig:model_comparison_nucleus}b). However, note that here strong short-range repulsion interactions are correlated with anti-friction, which we do not observe in the experimental data. Also this candidate interaction is able to predict both negative and positive velocity correlations and the switch from sliding behavior to reversal behavior (NPC in Fig. 
\ref{fig:model_comparison_nucleus}c,e). 
\paragraph{NUC}: For nucleus deformation coupling, we find only weak underdamped repulsion but significant attraction (NUC in Fig. \ref{fig:model_comparison_nucleus}a). For NUC, repulsion is correlated with underdamped friction and attraction is correlated with underdamped ant-friction (NUC in Fig. \ref{fig:model_comparison_nucleus}b) as observed in the experimental data. Further, NUC can predict both negative and positive velocity correlations (NUC in Fig. \ref{fig:model_comparison_nucleus}b), which are both correlated with the underdamped interactions like in the experiment. However, NUC does not predict significant position correlations and only predicts dominant reversal behaviors (NUC in Fig. \ref{fig:model_comparison_nucleus}d,e). 
\paragraph{PrC and PrNC}: For protrusion coupling and protrusion nucleus coupling, we find very similar dynamics. We find that our models predicts significant underdamped repulsion and attraction interactions, but no significant underdamped anti-friction interactions (PrC and PrNC in Fig. \ref{fig:model_comparison_protrusion}a,b). Also, neither PrC nor PrNC are able to predict negative velocity correlations at short time scales (PrC and PrNC in Fig. \ref{fig:model_comparison_protrusion}c). Furthermore, PrC does not predict dominant sliding behavior while PrNC does (PrC and PrNC in Fig. \ref{fig:model_comparison_protrusion}e). 
\paragraph{PrPC}: For protrusion polarity coupling, we find repulsion and long range attraction interactions not observed in the experimental data (PrPC in Fig. \ref{fig:model_comparison_protrusion}a). We furthermore observe strong underdamped friction and anti-friction interactions (PrPC in Fig. \ref{fig:model_comparison_protrusion}b) and both positive and negative velocity correlations (PrPC in Fig. \ref{fig:model_comparison_protrusion}c). PrPC is unable to predict mutual exclusion behavior and does not give rise to dominant sliding events (PrPC in Fig. \ref{fig:model_comparison_protrusion}d,e). 
\paragraph{PrUC}: For protrusion deformation coupling, we find short range repulsion/long range attraction as well as short range attraction/long range repulsion (PrUC in Fig. \ref{fig:model_comparison_protrusion}a). These underdamped cohesion interactions are correlated with friction and anti-friciton, respectively (PrUC in Fig. \ref{fig:model_comparison_protrusion}b). These are the correlations we also observe in the experimental data. Finally, PrUC predicts a switch between anti-correlated velocites and sliding behavior to correlated velocites and reversal behavior (PrUC in Fig. \ref{fig:model_comparison_protrusion}c,e) However, again PrUC does not predict significantly negative position correlations (PrUC in Fig. \ref{fig:model_comparison_protrusion}d). 
\paragraph{PPrC}: For polarity protrusion coupling, we observe a switch from short-range repulsion and long-range attraction to short-range attraction and long-range repulsion (PPrC in Fig. \ref{fig:model_comparison_pol}a). Here, repulsion interactions are correlated with anti-friction. In general, PPrC predcits dynamics that are very similar to PrC. Both mechanisms do not capture the correlations of the various behavior statistics and underdamped interactions as observed in the experimental data. 
\paragraph{PNC}: For polarity nucleus coupling, we find clear repulsion and attraction interactions, but no friction interactions (PNC in Fig. \ref{fig:model_comparison_pol}a,b). While this candidate cell-cell interaction is able to predict mutual exclusion behavior and both reversal and sliding behavior, it is unable to capture significant positive velocity correlations at short time scales (PNC in Fig. \ref{fig:model_comparison_pol}c-e).
\paragraph{PSC}: For polarity self coupling, we find weak underdamped repulsion interactions and very weak underdamped anti-friction interactions (PSC in Fig. \ref{fig:model_comparison_pol}a,b). Furthermore, PSC does not predict negative velocity or position correlations, but is able to capture a switch from dominant sliding behavior to dominant reversal behavior (PSC in Fig. \ref{fig:model_comparison_pol}c-e).   
\paragraph{PC}: For polarity coupling, we find that our model predicts a switch from short-range repulsion and long-range attraction to short-range attraction and long-range repulsion (PC in Fig. \ref{fig:model_comparison_pol}a). Furthermore, repulsion interactions are correlated with friction interactions and attraction interactions are correlated with anti-friction interactions (PC in Fig. \ref{fig:model_comparison_pol}a,b) as observed in the experimental data. PC is able to predict both negative and positive velocity correlations as well as both sliding and reversal behavior (PC in Fig. \ref{fig:model_comparison_pol}c-e).   
\paragraph{PPC}: For polarity-polarity coupling, we observe a switch from repulsion and friction to attraction and anti-friction (PPC in Fig. \ref{fig:model_comparison_pol}a,b). Furthermore, PPC can predict a switch from positive to negative velocity correlations and a switch from dominant reversal to dominant sliding behavior (PPC in Fig. \ref{fig:model_comparison_pol}c,e). These dynamics and behaviors are correlated like in the experimental data. However, note that PPC is unable to predict pronounced position correlations (PPC in Fig. \ref{fig:model_comparison_pol}d). Thus, PPC does not lead to mutual exclusion behavior of cells. 
\paragraph{PUC}: For polarity-deformation coupling, we observe very similar dynamics like for PPC. However, instead of pure repulsion/attraction like for PPC, for PUC we find short range repulsion/long range attraction and short range attraction/long range repulsion. 

\newpage\phantom{blabla}
\newpage

\begin{figure}[t!]
	\centering
	\includegraphics[width=\textwidth]{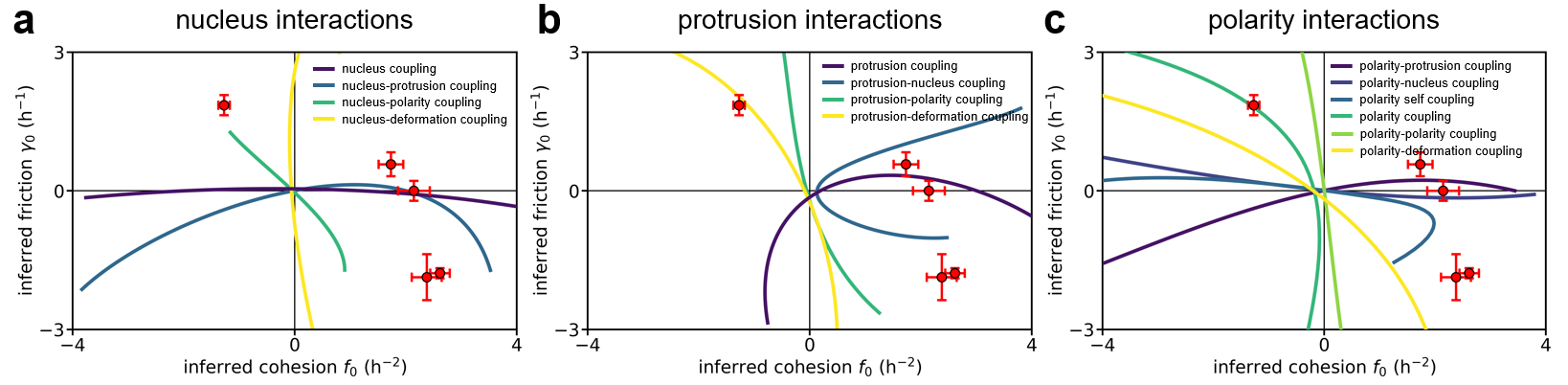}
	\caption{\textbf{Underdamped inferred interaction amplitudes of all interaction mechanisms} \small Overview over all mappings $(f_0(\epsilon),\gamma_0(\epsilon))$ obtained from simulating our overdamped model and performing underdamped Langevin inference on the obtained nucleus trajectories. \textbf{a)} shows the mapping for all interaction mechanisms acting on the nucleus, \textbf{b)} all interaction mechanisms acting on the protrusion, \textbf{c)} all interaction mechanisms acting on the polarity of cells}
	\label{fig:model_comparison_IBS}
\end{figure}

\subsubsection{Summary of qualitative comparison of candidate cell-cell interactions}
To summarize the qualitative features ouf our candidate cell-cell interactions, we count how many qualitative features of the two experimentally observed feature sets 
\begin{itemize}
    \item \textbf{epithelial set}: cohesion, friction, positive velocity correlations, negative position correlations, and dominant reversal behavior 
    \item \textbf{cancer cell set}: attraction, anti-friction, negative velocity correlations, negative position correlations, and dominant sliding behavior 
\end{itemize}
can be predicted by our candidate cell-cell interactions. The rules for counting are that we consider the dynamics of our candidate cell-cell interactions and then choose either positive or negative interaction strength $\epsilon$ to maximize the number of features our candidate cell-cell interaction can predict of a certain feature set. For example, for nucleus repulsion (NC), we see that at positive $\epsilon_{\mathrm{NC}}$, our model predicts repulsion, friction, no velocity correlations, negative position correlations, and reversal behavior. Thus, we can match 4 out of the 5 qualitative features of the epithelial feature set. For negative $\epsilon_{\mathrm{NC}}$, we find that only attraction and sliding is captured out of the cancer cell feature set. We summarize these features in table \ref{table:all_models_features}. Taken together, we find that only PC and PPC can predict  simultaneously $4$ out of $5$ qualitative features of the dynamics of both epithelial and cancer cells. These two candidate cell-cell interactions thus stand out above the other candidate cell-cell interactions as a prime candidate to provide an accurate description of the dynamics of all considered cell types. Note that PC and PPC are mathematically very similar as both implement different versions of alignment or anti-alignment interactions. In PC, we impose that $\dot{P}_i \sim P_j$, while in PPC, we impose that $\dot{P}_i \sim \Delta P$. Thus, in both cases, the polarity of cell $i$ adjusts to the polarity of cell $j$. 
\newpage

\begin{table}[h!]
\setlength{\arrayrulewidth}{0.5mm}
\renewcommand{\arraystretch}{1.5}
\centering
\begin{tabular}{| m{6cm} | m{4cm}| m{4cm} |} 
 \hline
 \textbf{candidate interaction} & \textbf{epithelial set} & \textbf{cancer cell set} \\ [0.5ex] 
 \hline\hline
 nucleus repulsion & $4$ (repulsion, friction, exclusion, reversal) & $2$ (attraction, sliding) \\ 
 \hline
 nucleus protrusion repulsion & $4$ (repulsion, friction, exclusion, reversal) & $1$ (attraction) \\ 
 \hline
 nucleus polarity repulsion & $3$ (friction, positive vel.corr., reversal) & $4$ (attraction, anti-friction, negative vel.corr., sliding) \\ 
 \hline
 protrusion repulsion & $3$ (repulsion, exclusion, reversal) & $2$ (attraction, anti-friction) \\ 
 \hline
 protrusion nucleus repulsion & $3$ (repulsion, exclusion, reversal) & $2$ (attraction, sliding) \\ 
 \hline
 protrusion polarity repulsion & $4$ (repulsion, friction, positive vel.corr., reversal) & $3$ (attraction, anti-friction, negative vel.corr.) \\ 
 \hline
 polarity repulsion & $3$ (repulsion, exclusion, reversal) & $1$ (attraction) \\ 
 \hline
 polarity nucleus repulsion & $3$ (repulsion, exclusion, reversal) & $4$ (attraction, anti-friction, negative vel.corr., sliding) \\ 
 \hline
 polarity shrinking & $3$ (repulsion, friction, reversal) & $2$ (attraction, sliding) \\ 
 \hline
 polarity cross shrinking & $4$ (repulsion, friction, positive vel.corr., reversal) & $4$ (attraction, anti-friction, negative vel.corr., sliding) \\ 
 \hline
 polarity alignment & $4$ (repulsion, friction, positive vel.corr., reversal) & $4$ (attraction, anti-friction, negative vel.corr., sliding) \\ 
 \hline
\end{tabular}
\caption{Summary of the number of qualitative features that our candidate cell-cell interactions can predict out of the experimentally observed feature sets. We always find the maximum number of features that each candidate cell-cell interaction can predict at either positive or negative interaction strengths.}
\label{table:all_models_features}
\end{table}

\newpage

\subsection{Supplementary results of the statistical inference procedure}\label{sec:inference_results}

In this section, we describe supplementary results of the statistical inference procedure of the candidate cell-cell interactions as described in section \ref{sec:model_fitting}.

\subsubsection{Fitting results}

\begin{figure}[t!]
	\centering
	\includegraphics[width=\textwidth]{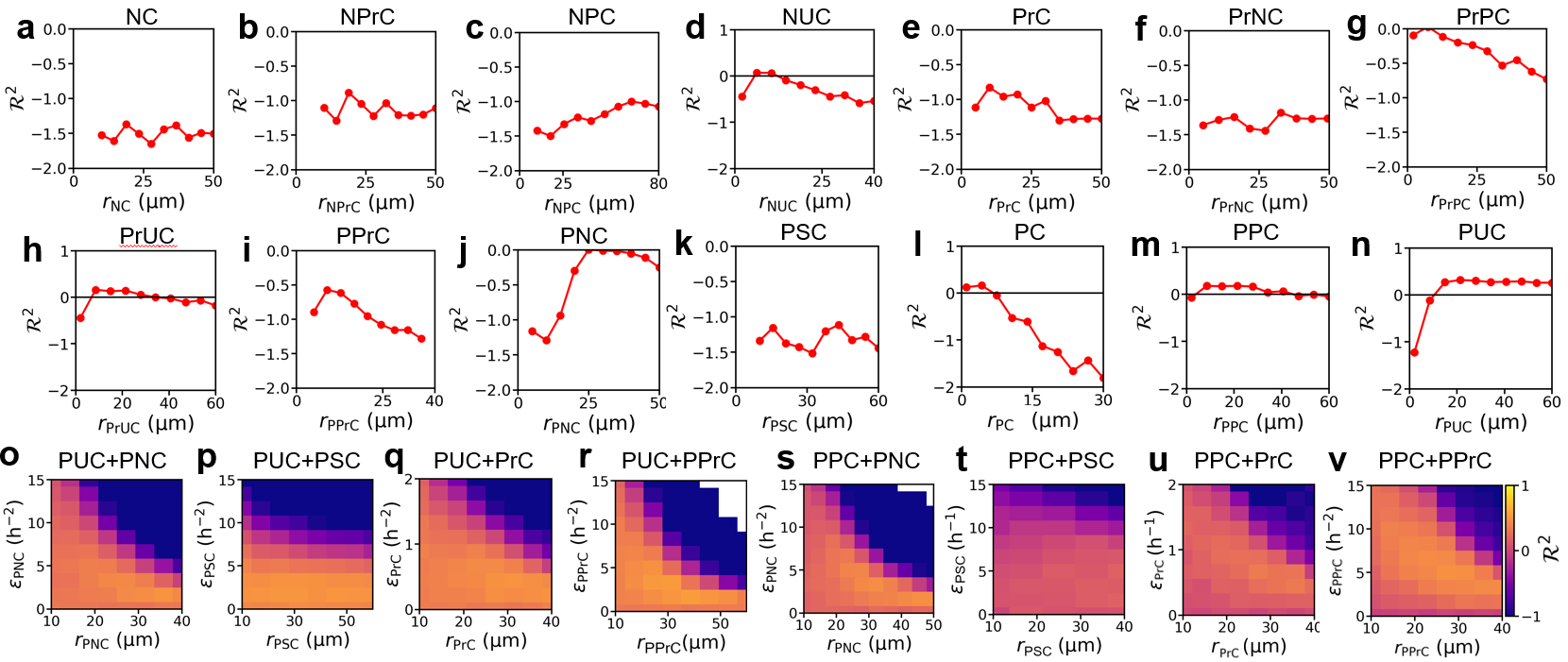}
	\caption{\textbf{Coefficient of determination of the different candidate cell-cell interactions.} \small \textbf{a)}-\textbf{n)} Coefficient of determination $\mathcal{R}^2(r)$ as determined in equation \ref{eq:COD} for the 11 different candidate cell-cell interactions as a function of the interaction range. \textbf{o)}-\textbf{v)} Again the coefficient of determination $\mathcal{R}^2(r)$ as determined in equation \ref{eq:COD} for all the combined candidate cell-cell interactions. As we have three parameters in these combined models which we vary, we now show $\mathcal{R}^2(\epsilon_{2}, r_{2})$ where we vary the strength and range of the additional interaction that we add to polarity alignment interactions. White region in panel r and s correspond to parameter combinations which yield dynamics with no collision events. Thus, our procedure is unable to compute $R^2$ in that region.}
	\label{fig:COD}
\end{figure}

Specifically, Fig. \ref{fig:COD} shows for each candidate cell-cell interaction the resulting $\mathcal{R}^2(r)$ quantifying the global fit performance of our model to the experimental data. Of note is that we are able in most cases to identify a clear single maximum of $\mathcal{R}^2(r)$ in the parameter range we study for our candidate interaction mechanisms. Thus, we are able to mostly fully constrain our candidate interactions using the procedure outlined in section \ref{sec:model_fitting}. We show the maxima of $\mathcal{R}^2(r)$ defined as $\hat{\mathcal{R}}^2$ in Eq. \ref{global_COD} in main text Fig. 6b. Note, that in practice, we are averaging the 3 highest values of $\mathcal{R}^2(r)$ in order to obtain the maximum fit performance $\hat{\mathcal{R}}^2$. 

\subsubsection{Constraining power of inferred underdamped interactions}

For our fitting procedure it is important that inferred interactions and experimental behavior statistics are effectively independent statistics given our overdamped model. To test this, we perform our fitting procedure while considering different total coefficients of determination (COD). Specifically, before, we averaged the different CODs of the various behavior statistics and inferred cell-cell interactions to reach a good fit for all the considered quantities (equation \ref{eq:COD_average}). Now, we fit these quantities separately. This allows us to disentangle the effect of individual quantities on the resulting fit. We consider two cases: 
\begin{enumerate}
    \item Only behavior statistics: $R_k^2 = \frac{1}{2}\left(R^2_{v,k}(\Theta)+R^2_{c,k}(\Theta)\right)$
    \item Only inferred interactions: $R_k^2 = R^2_{d,k}(\Theta)$
\end{enumerate}
If the behavior statistics are really effectively independent of the inferred interactions in our overdamped model, we would expect that both cases should give different final fitting results. 
\begin{figure}[t!]
	\centering
	\includegraphics[width=\textwidth]{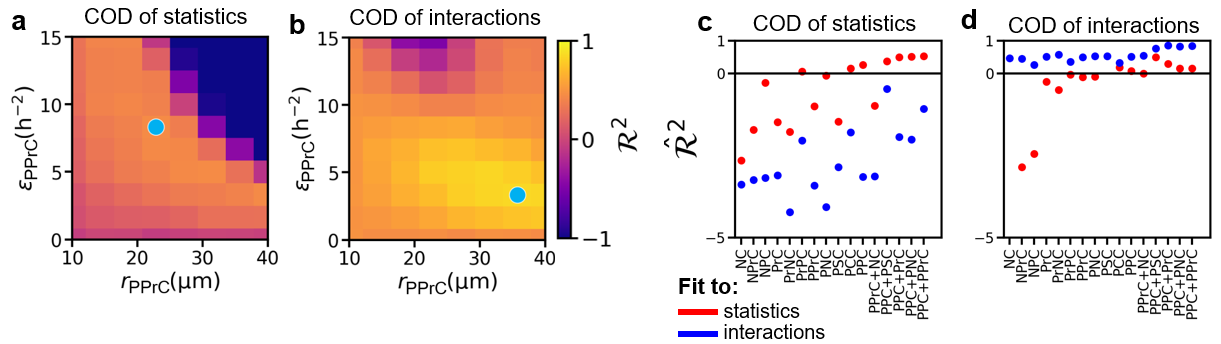}
	\caption{\textbf{Constraining power of inferred underdamped cell-cell interactions.} \small \textbf{a)}-\textbf{b)} Coefficient of determination $\mathcal{R}^2(r)$ as determined in equation \ref{eq:COD} for the candidate cell-cell interaction PPC+PPrC. Panel (a) shows the COD for capturing only the statistics as defined in case 1. Panel (b) shows the COD for capturing only the inferred underdamped cell-cell interactions as in case 2. Blue dots indicate the optimum reached for all of these two cases. \textbf{c)}-\textbf{d)} Maximum COD for all candidate cell-cell interaction as defined in equation \ref{global_COD}. Again we consider the two different cases of only fitting the statistics (red) or only interactions (blue). For each of these cases, we find the maximum COD that we can reach with fitting different quantities.}
	\label{fig:COD_constraints}
\end{figure}
Specifically, we would expect that the optimum $\hat{\mathcal{R}}^2$ is reached at different parameters $\Theta$ and/or the value of $\hat{\mathcal{R}}^2$ is different between the two cases. Both of these cases are generally true for the global fit and all candidate cell-cell interactions. Specifically, we consider for the best fitting candidate cell-cell interaction PPC+PPrC the two cases as defined above. Then, after performing the global fitting procedure, we compute $\mathcal{R}^2(\epsilon_{\mathrm{PPrC}},r_{\mathrm{PPrC}})$ and find that this global COD has optima at different values for $\epsilon_{\mathrm{PPrC}}$ and $r_{\mathrm{PPrC}}$ in the two different definitions for the COD (Fig. \ref{fig:COD_constraints}a,b). This means that only fitting the behavior statistics will yield different final parameters than fitting only the inferred underdamped interactions. Furthermore, we consider for all candidate cell-cell interactions the maximum values of $\mathcal{R}^2(\epsilon_{\mathrm{PPrC}},r_{\mathrm{PPrC}})$, which we defined as $\hat{\mathcal{R}}^2$ in equation \ref{global_COD} and showed in main text Fig. 5b. Again, we consider the two different cases defined above. For all the definitions for $R_k^2$, we perform the global fit and then separately assess how well the statistics and the inferred interactions are captured (Fig. \ref{fig:COD_constraints}c,d). We find that if we fit only the behavior statistics, as expected, the behavior statistics are well captured. However, the inferred interactions are not in this case (red dots in Fig. \ref{fig:COD_constraints}c,d). In contrast, if we fit the inferred interactions, these interactions are well captured while the behavior statistics are not (blue dots in Fig. \ref{fig:COD_constraints}c,d). Thus, taken together, these results reveal that in our overdamped model for all candidate cell-cell interaction, the inferred cell-cell interactions are an effective independent statistics in addition to the behavior statistics. This shows that including the inferred cell-cell interactions does indeed provide additional strong constraints on the overdamped model. 

\subsubsection{Prediction of position correlations}\label{sec:position_correlation_fit}

\begin{figure}[t!]
	\centering
	\includegraphics[width=\textwidth]{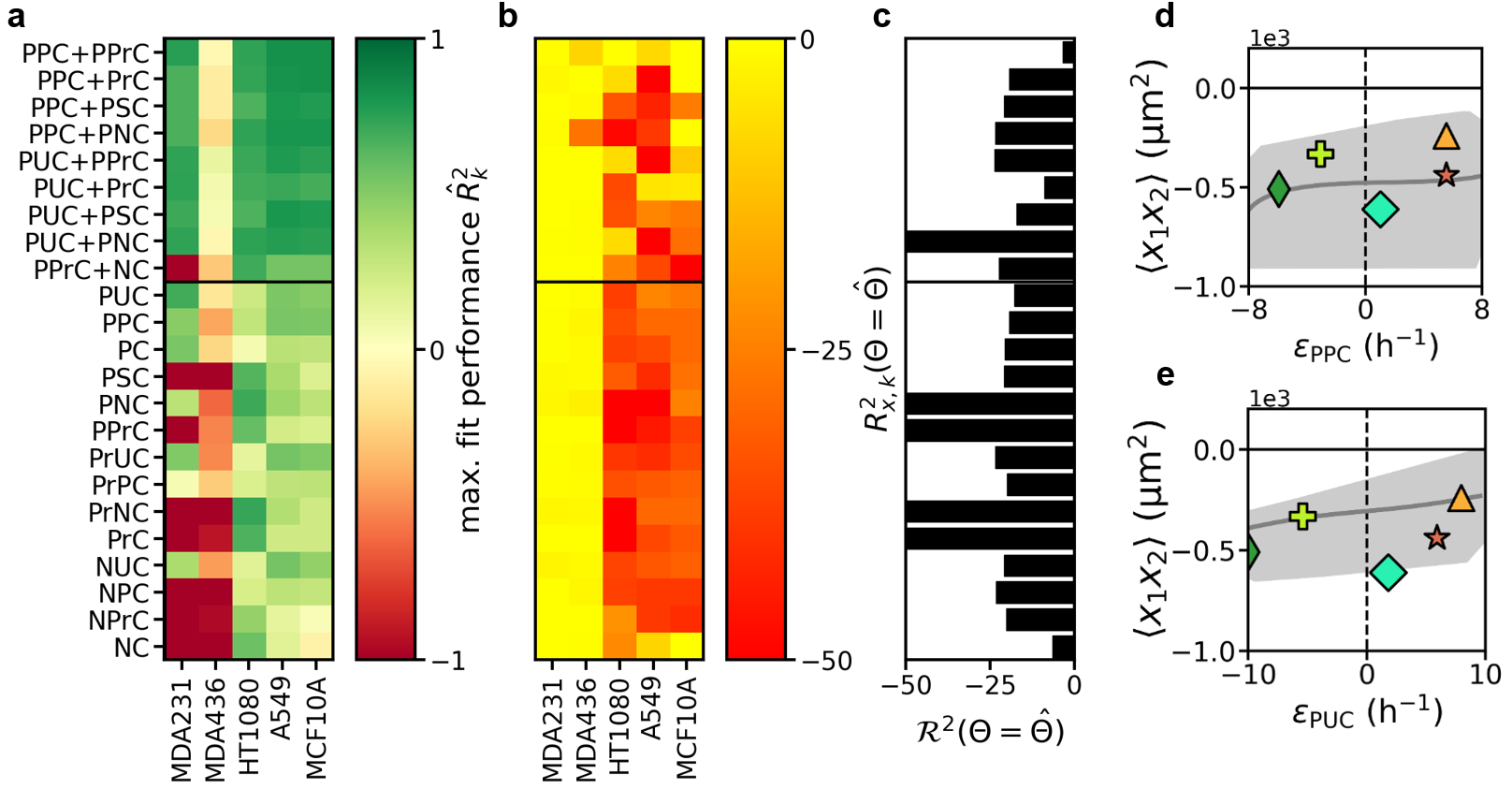}
	\caption{\textbf{Prediction of position correlation function} \small \textbf{a)} Maximum fit performance as quantified by the maximum coefficient of determination $\hat{R}^2$ for each interaction mechanism using an individual fit. \textbf{b)} Coefficient of determination of the position correlation function evaluated at the best fitting parameters $\hat{\Theta}$ obtained from our individual fit $R^2_{x,k}(\Theta = \hat{\Theta})$. \textbf{c)} Coefficient of determination of the position correlation function evaluated at the best fitting parameters $\hat{\Theta}$ obtained from our global fit. \textbf{d)},\textbf{e)} Instantaneous position correlations as predicted by the two models PPC+PPrC (d) and PUC+PPrC (e). Solid lines show model result, symbols show experimental results as also presented in main text Fig. 6(f).}
	\label{fig:position_correlation}
\end{figure}

In this section, we describe to what extend our candidate cell-cell interactions are able to correctly predict the position correlation function $C_X (|t-t'|)$ which has not been used in the fitting procedure to constrain our candidate interactions. To this end, we investigate $R^2_{x,k}(\Theta)$, which quantifies how well our model captures the position correlation function $C_X (|t-t'|)$ for a specific cell type, a specific candidate cell-cell interaction and a specific parameter combination $\Theta$. Thus, of particular interest is the quantity $R^2_{x,k}(\Theta = \hat{\Theta})$, where $\hat{\Theta}$ is the best fitting parameter combination of the individual fit as described in section \ref{sec:fitting}. We find that the best fitting models for each candidate interaction in general do not well capture the position correlation function (Fig. \ref{fig:position_correlation} a,b). In fact, $R^2_{x,k}(\Theta = \hat{\Theta})$ is consistently negative showing that our models struggle to capture these correlations. However, note that combining candidate interactions in general does improve the prediction of the position correlation functions. Specifically, the best working candidate interactions coming out of our inference procedure PPC+PPrC does provide the best prediciton of the position correlation function at its optimal individual and global interaction parameters (Fig. \ref{fig:position_correlation} c). Furthermore, PPC+PPrC outperforms PUC+PPrC (Fig. \ref{fig:position_correlation} d,e). This shows that we can use the predicted position correlation function to further constrain our candidate interactions.  

\subsubsection{Robustness of fitting procedure}\label{sec:robustness_fit}

\begin{figure}[t!]
	\centering
	\includegraphics[width=\textwidth]{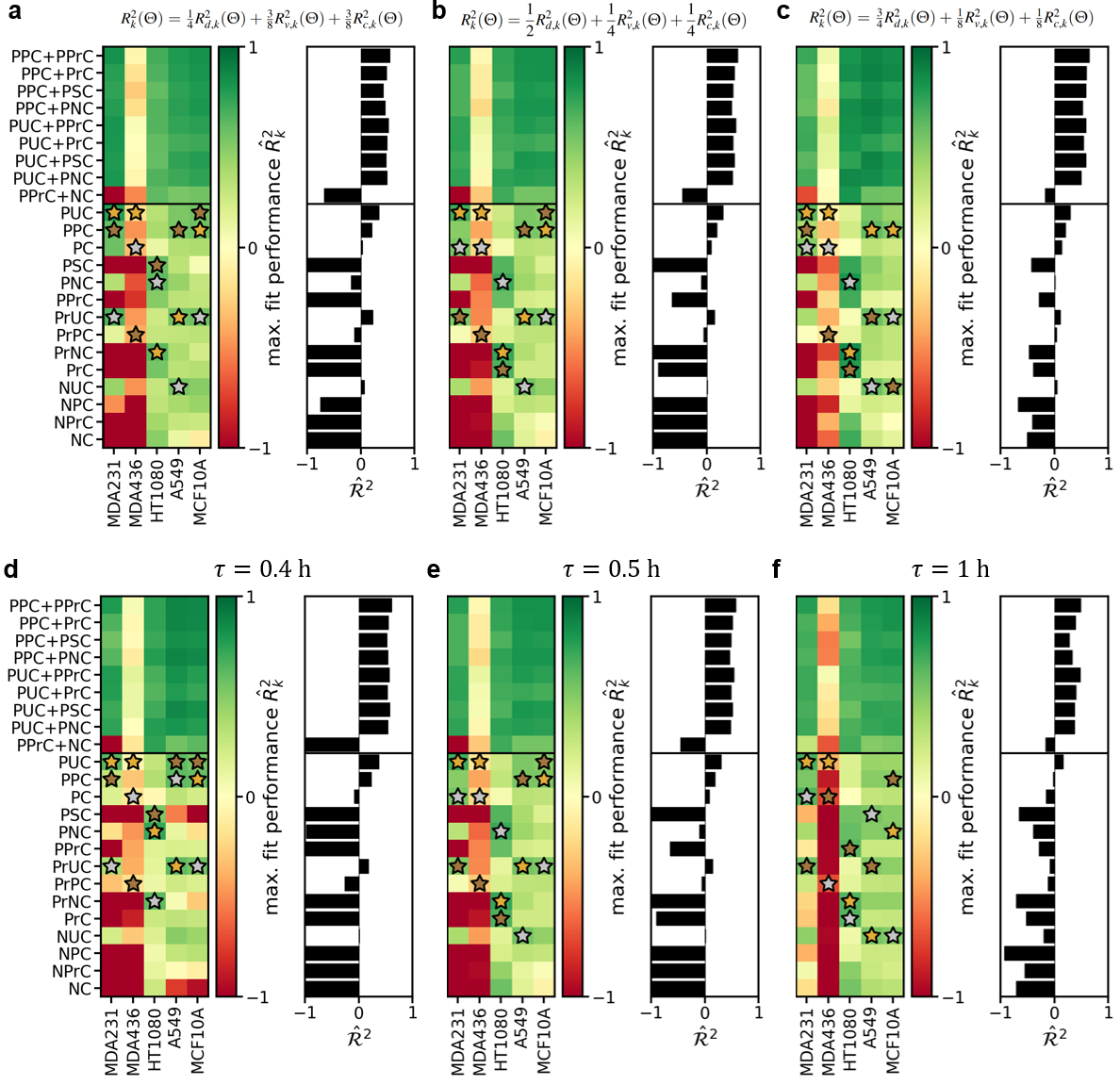}
	\caption{\textbf{Prediction of position correlation function} \small \textbf{a)}-\textbf{c)} Maximum fit performance as quantified by the maximum coefficient of determination $\hat{R}^2$ for each interaction mechanism using an individual (left) and global (right) fit. In panels a)-c), we vary our definition of $R_k^2(\Theta)$, which is the weighted average of the CODs of each individual statistics $R^2_{j,k}(\Theta)$. We vary from more weight on the underdamped interactions (a), to equal weight on the interactions and behavior statistics (b), to more weight on the behavior statistics (c). In panels \textbf{d)}-\textbf{f)} we vary $\tau$, which we use as a threshold in computing the $R^2$ of the velocity correlations.}
	\label{fig:robustness_fit}
\end{figure}

To test the robustness of our statistical inference procedure, we vary our definition of the average COD that we are maximizing throughout this study. Before, we defined $R_k^2(\Theta) = \frac{1}{2} R^2_{d,k}(\Theta) + \frac{1}{4}R^2_{v,k}(\Theta) + \frac{1}{4}R^2_{c,k}(\Theta)$ as the average of all inferred underdamped interactions and behavior statistics, putting equal weight on all the four statistics ($R^2_{d,k}(\Theta)$ already quanitfies the goodness of fit of two statistics). To test how sensitive our final results are with respect to this definition, we consider two alternative definitions: 1) We put more weight on the inferred interactions: $R_k^2(\Theta) = \frac{3}{4} R^2_{d,k}(\Theta) + \frac{1}{8}R^2_{v,k}(\Theta) + \frac{1}{8}R^2_{c,k}(\Theta)$. 2) We put more weight on the behavior statistics: $R_k^2(\Theta) = \frac{1}{4} R^2_{d,k}(\Theta) + \frac{3}{8}R^2_{v,k}(\Theta) + \frac{3}{8}R^2_{c,k}(\Theta)$. The final outcome of the statistical inference of the candidate interaction mechanisms are the maximum CODs $\hat{R}_k^2(\Theta)$ for each cell type and each candidate mechanism as presented in main text Fig. 6a,b. From this we concluded which interaction mechanisms best capture the experimental data. Importantly, our results are not greatly affected by the different definitions of $R_k^2(\Theta)$ (Fig. \ref{fig:robustness_fit}).

\subsubsection{Quantitative fitting of the dynamics with PPC and PUC}\label{sec:model_comparison_fit}

\begin{figure}[t!]
	\centering
	\includegraphics[width=\textwidth]{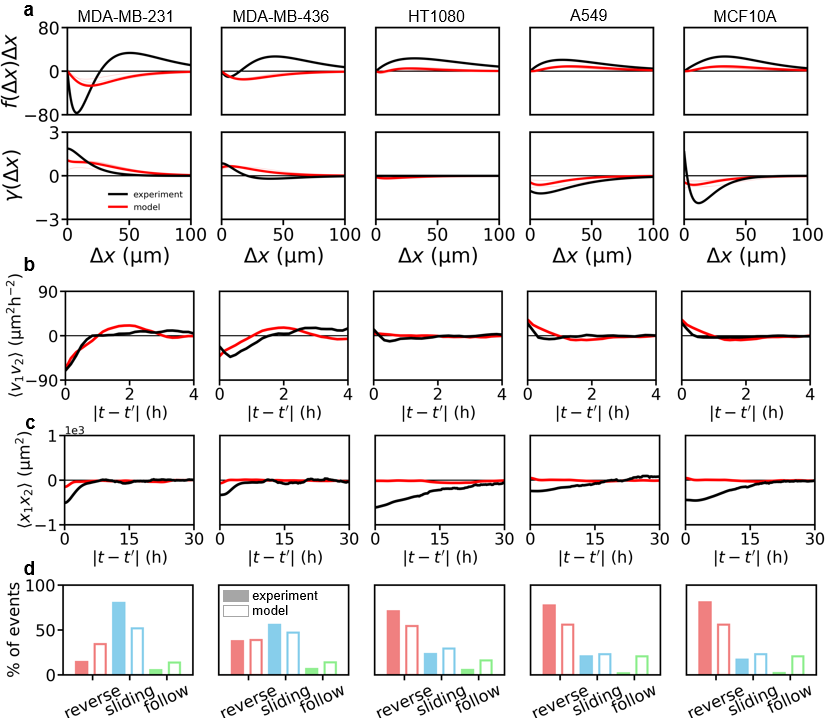}
	\caption{\textbf{Results of the global fit of only PPC to the experimental data.} \small \textbf{a)} Underdamped cell-cell interactions for both experiment (black) and best fitting model (red). Upper row shows the underdamped cohesion interactions, while lower row shows the inferred friction interactions. \textbf{b)}-\textbf{d)} All behavior statistics for both experiment (black, solid bars) and model (red, empty bars). (b) shows the velocity correlation function $\langle v_1(t)v_2(t')\rangle_{\mathrm{same}}$ when cells are on the same island. (c) show the position correlation function $\langle x_1(t)x_2(t')\rangle$. (d) shows the collision statistics for all different cell types.}
	\label{fig:global_fit_PPC}
\end{figure}

\begin{figure}[t!]
	\centering
	\includegraphics[width=\textwidth]{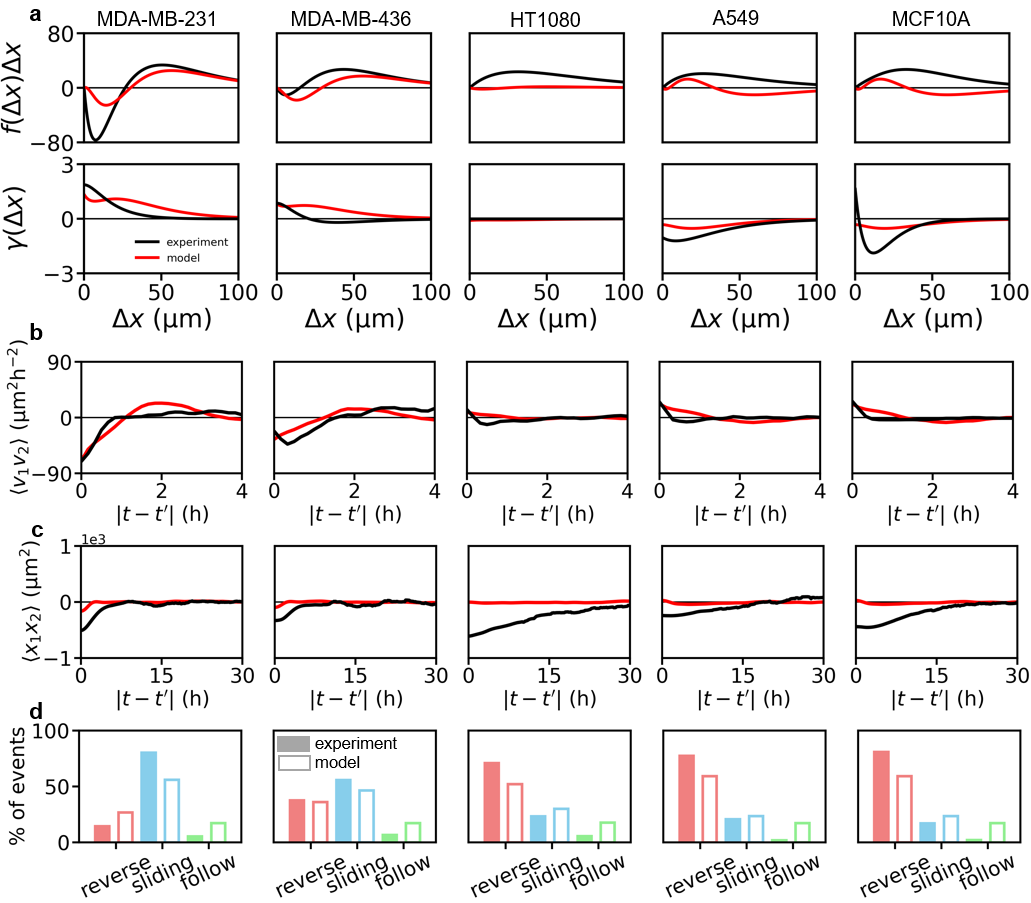}
	\caption{\textbf{Results of the global fit of only PUC to the experimental data.} \small \textbf{a)} Underdamped cell-cell interactions for both experiment (black) and best fitting model (red). Upper row shows the underdamped cohesion interactions, while lower row shows the inferred friction interactions. \textbf{b)}-\textbf{d)} All behavior statistics for both experiment (black, solid bars) and model (red, empty bars). (b) shows the velocity correlation function $\langle v_1(t)v_2(t')\rangle_{\mathrm{same}}$ when cells are on the same island. (c) show the position correlation function $\langle x_1(t)x_2(t')\rangle$. (d) shows the collision statistics for all different cell types.}
	\label{fig:global_fit_PUC}
\end{figure}

Here we describe in detail the fitting result of PPC and PUC to all cell types. First, note that both PPC and PUC yield very similar dynamics. They can capture the experimentally observed correlations of the underdamped interactions (Fig. \ref{fig:global_fit_PPC}a, Fig. \ref{fig:global_fit_PUC}a). Note however, that both interactions underestimate the underdamped repulsion interactions of all cell types. There is an important difference between PPC and PUC: PPC does not capture the long range repulsion of the cancer cells (MDA-MB-231 and MDA-MB-436 in Fig. \ref{fig:global_fit_PPC}a). In contrast, PUC does predict long range attraction of the MCF10A cells (MCF10A and A549 in Fig. \ref{fig:global_fit_PPC}a). Furthermore, PPC and PUC both qualitatively capture the velocity cross-cross correlations and the collision statistics (Fig. \ref{fig:global_fit_PPC}b,d, Fig. \ref{fig:global_fit_PUC}b,d). However, both interactions underestimate for all cell types the dominant collision behavior leading to a quantitatively wrong behavior distribution. Finally, note that both PPC and PUC do not predict significant mutual exclusion behavior (Fig. \ref{fig:global_fit_PPC}c, Fig. \ref{fig:global_fit_PUC}c). 

\newpage
\phantom{blabla}
\newpage

\subsection{Interaction behavior arising from PPC}

In this subsection, we show a detailed analysis of the dynamics arising from PPC and connect the interaction mechanism to the interaction behavior we observe in epithelial and breast cancer cell lines.

\subsubsection{Polarity alignment enables reversal behavior}

\begin{figure}[t!]
	\centering
	\includegraphics[width=\textwidth]{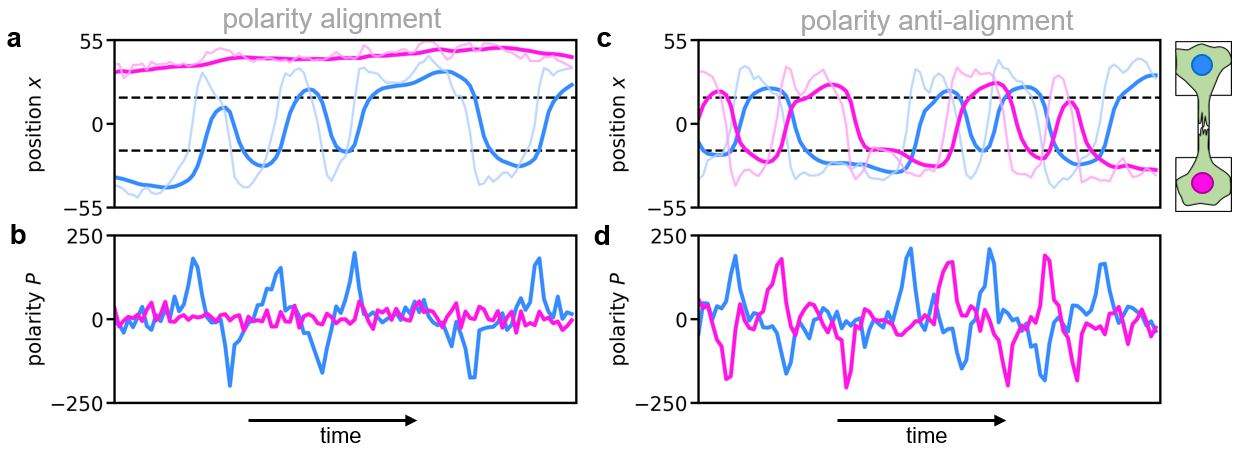}
	\caption{\textbf{Deatiled dynamics of PPC} \small \textbf{a)},\textbf{c)} Trajectories of the two cells in space for both polarity alignment and polarity anti-alignment as allowed by PPC. Thick solid lines show the nucleus trajectories $x_\mathrm{n}(t)$, thin faded lines show the protrusion trajectory $x_\mathrm{p}(t)$. Horizontal dashed lines indicate the boundaries of the two islands of the dumbbell-shaped micropattern. \textbf{b)},\textbf{d)} Trajectories of the polarities $P(t)$ of the two cells.}
	\label{fig:behavior}
\end{figure}

First, we describe how polarity alignment interactions as allowed by PPC can lead to the behavior that we observe in epithelial cells. While it is intuitive that polarity alignment promotes correlated velocities of the two cells as quantified by the velocity cross-correlation function $C_V(|t-t'|)$ ($\epsilon_{\mathrm{PPC}} > 0$ main text Fig. 5c, 6e), the emergence of reversal behavior is more subtle. First, note that we mostly observe reversal events in our model when one cell is on an island, while the other transitions into that island ($\epsilon_{\mathrm{PPC}} > 0$ in main text Fig. 5a, Fig. \ref{fig:behavior}a). This is also typically the case for epithelial cells in the experimental data (main text Fig. 3b iv,v). Importantly, note that the behavior that we observe in our model is a result of both single cell behavior and cell-cell interactions. Specifically, note that the single cell behavior (main text Eq. (3), (B3) in our model yields $\langle P \rangle = 0$ when cells are on the island. Thus, when cells collide on the island, both attain $\langle P \rangle = 0$ and the Gaussian white noise (main text Eq. (3)) together with alignment interactions yield correlated fluctuating polarities (Fig. \ref{fig:behavior}b) ultimately leading to correlated nucleus velocities. However, note that the cell that just transitioned into that island is located closer to the bridge (Fig. \ref{fig:behavior}a). 
Thus, eventually its polarity will grow into the bridge again as promoted by the positive feedback mechanisms in our model (main text Eq. (3)), leading to reversal behavior (Fig. \ref{fig:behavior}a). This illustrates that the presence of the bridge plays an important role in establishing reversal behavior of epithelial cells. Consequently, in the absence of the bridge, our model for epithelial cells shows less reversal behavior but more following behavior (Fig. \ref{fig:rectangular}c) as expected for polarity alignment interactions. Also in the experiment, epithelial cells on rectangular patterns show more following behavior reminiscent of flocking (Fig. \ref{fig:rectangular}b). Taken together, this analysis illustrates that interaction behavior can be influenced by the geometry of the micropattern. However, note that cell-cell interactions themselves are not significantly affected by the geometry of the micropattern as shown in section \ref{sec:rectangular}.\\
\\Finally, note that PPC alone does not quantitatively reproduce the collision statistics of epithelial cells and overestimates following behavior (MCF10A in Fig. \ref{fig:global_fit_PPC}d). These following events arise from alignment interactions promoting flocking of the two cells. This may suggest that the mechanism explained above is not realistic for real cells. This is why we consider combinations of interactions as explained in section \ref{sec:combinations}. Specifically, by combining PPC with mechanisms that also couple the polarity to the position of the cells like PPrC, we do quantitatively capture the collision behavior of epithelial cells (MCF10A in Fig. \ref{fig:individual_fit}d). Specifically, PPrC induces growth of the polarity away from the other cell, robustly promoting reversal behavior. Thus this added CIL-like interaction may be an important component of the interaction mechanisms underlying contact-interactions between epithelial cells.  

\subsubsection{Polarity anti-alignment promotes sliding behavior}

Next, we describe the sliding dominated dynamics arising from anti-alignment. First, it is intuitive that anti-alignment leads to both anti-correlated velocities and sliding behavior ($\epsilon_{\mathrm{PPC}} < 0$ main text Fig. 5a,c). Interestingly, we mostly observe sliding behavior in our model while cells transition between the two islands ($\epsilon_{\mathrm{PPC}} < 0$ in main text Fig. 5a, Fig. \ref{fig:behavior}c). Again, this is a feature we observe for breast cancer cells in our experiment (main text Fig. 3b i,ii). In our model, this happens because as one cell starts to transition, its polarity grows towards the bridge. Consequently, when the other cell is on the opposite island, polarity anti-alignment leads to growth of the polarity of that other cell into the bridge as well. (Fig. \ref{fig:behavior}d). This yields almost simultaneous transitions of both cells including sliding behavior during the transition.

\subsection{Combining candidate cell-cell interactions}\label{sec:combinations}
To potentially overcome the shortcomings of our single candidate cell-cell interactions, we test combinations of the various different candidate interactions. Specifically, as we observe that PPC on its own already provides in general the best fit of the dynamics of the various different cell types, we consider combinations of PPC and one of the other candidate interactions. 

\subsubsection{Underdamped dynamics of combined interactions}

\begin{figure}[t!]
	\centering
	\includegraphics[width=\textwidth]{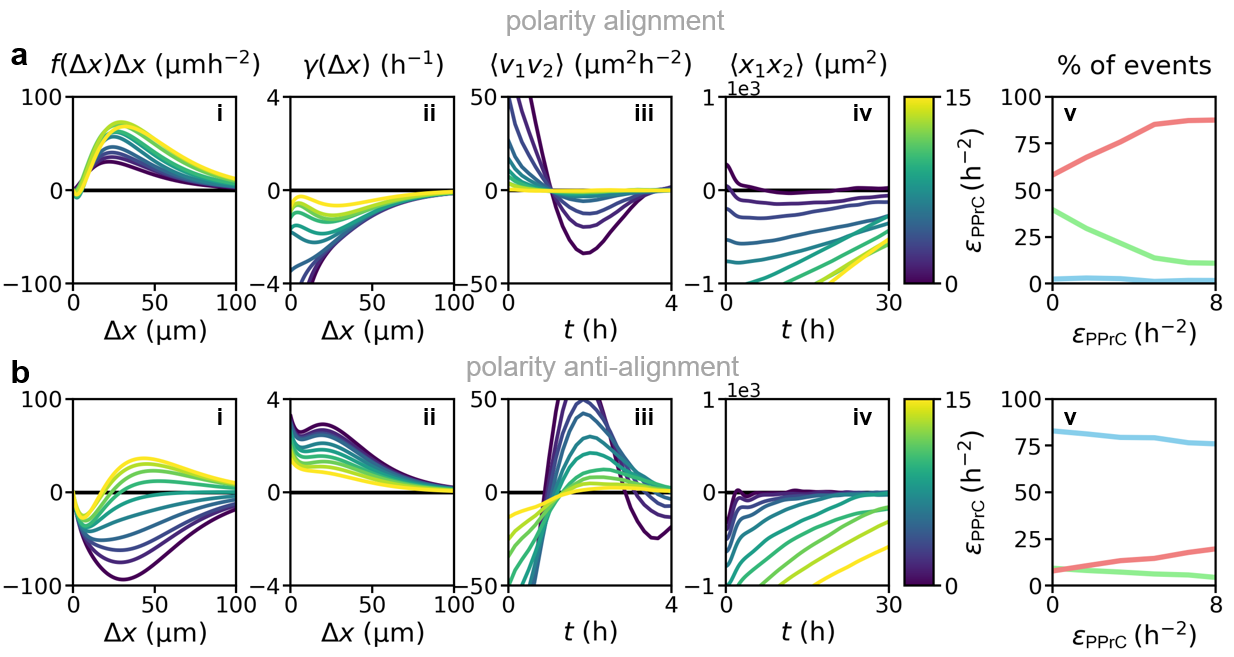}
	\caption{\textbf{Dynamic effects of polarity repulsion on polarity alignment.} \small \textbf{a)} Underdamped dynamics and behavior statistics of the combined candidate cell-cell interactions PPC+PPrC. We show how the dynamics and behavior statistics change when we fix the polarity alignment strength $\epsilon_{\mathrm{PPC}}$ and the two interaction ranges $r_{\mathrm{PPC}}$ and $r_{\mathrm{PPrC}}$ while varying the strength of the additional repulsion interaction $\epsilon_{\mathrm{PPrC}}$. Panel (a) shows polarity alignment ($\epsilon_{\mathrm{PPC}} = 4.3$), panel (b) shows polarity anti-alignment ($\epsilon_{\mathrm{PPC}} = -3.9$)}
	\label{fig:repulsion_addition}
\end{figure}

\begin{figure}[t!]
	\centering
	\includegraphics[width=\textwidth]{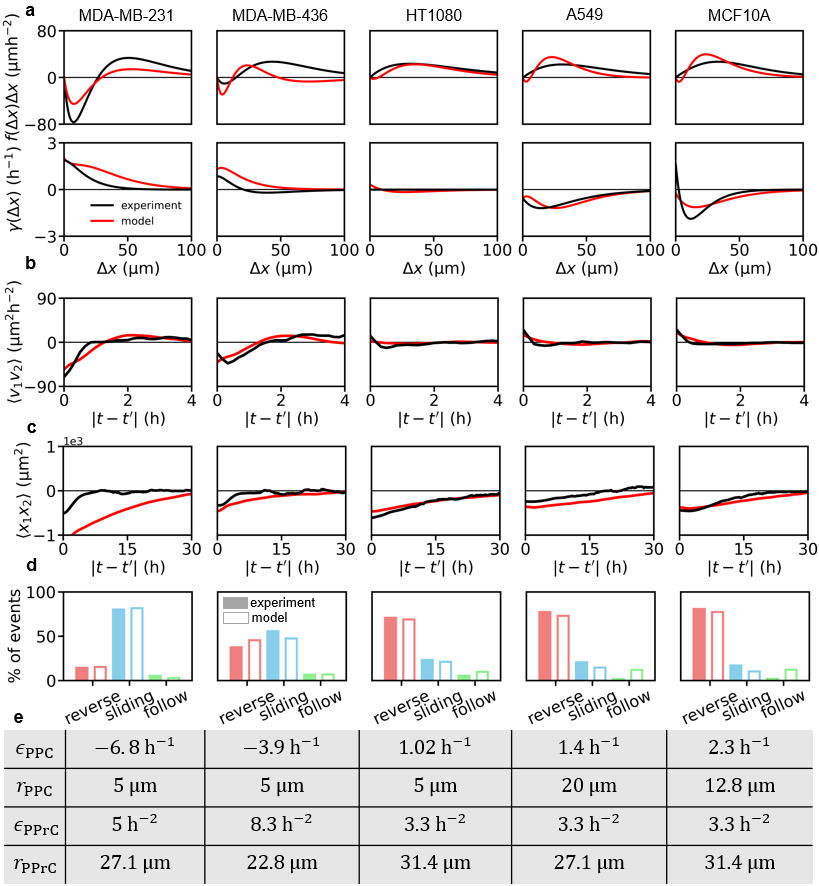}
	\caption{\textbf{Results of individual fit of PPC+PPrC to the experimental data.} \small \textbf{a)} Underdamped cell-cell interactions for both experiment (black) and best fitting model (red). Upper row shows the underdamped cohesion interactions, while lower row shows the inferred friction interactions. \textbf{b)}-\textbf{d)} All behavior statistics for both experiment (black, solid bars) and model (red, empty bars). (b) shows the velocity correlation function $\langle v_1(t)v_2(t')\rangle_{\mathrm{same}}$ when cells are on the same island. (c) show the position correlation function $\langle x_1(t)x_2(t')\rangle$. (d) shows the collision statistics for all different cell types. \textbf{e)} Table shows the final best fitting parameters for the model PPC+PPrC for all the different cell types.}
	\label{fig:individual_fit}
\end{figure}

\begin{figure}[t!]
	\centering
	\includegraphics[width=\textwidth]{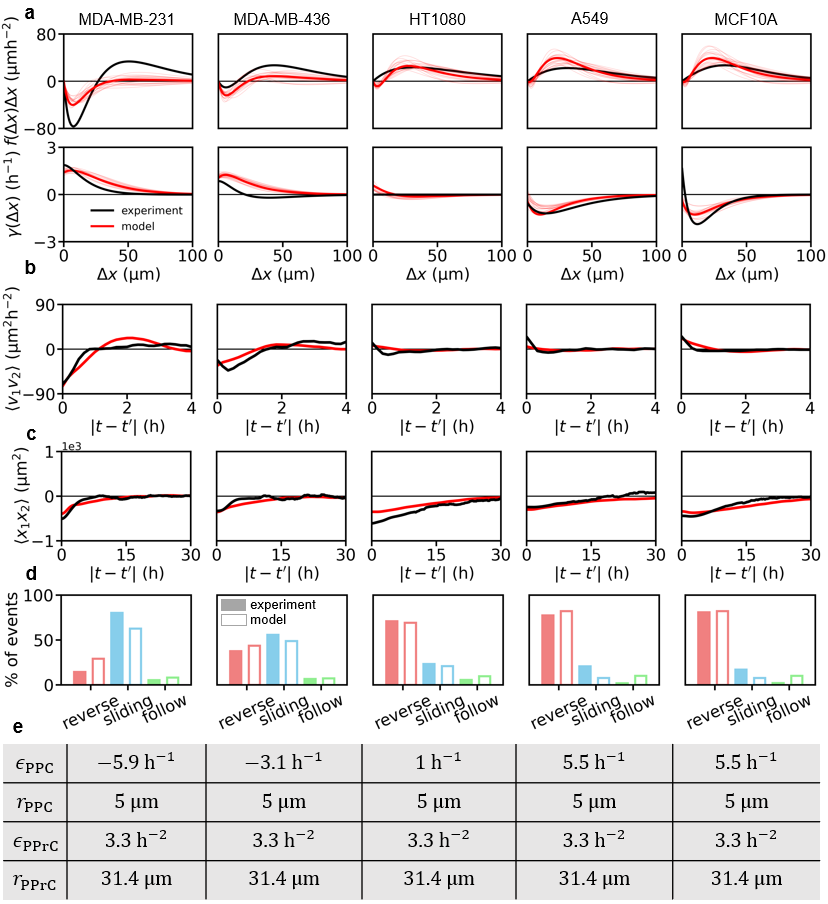}
	\caption{\textbf{Results of the global fit of PPC+PPrC to the experimental data.} \small \textbf{a)} Underdamped cell-cell interactions for both experiment (black) and best fitting model (red). Upper row shows the underdamped cohesion interactions, while lower row shows the inferred friction interactions. \textbf{b)}-\textbf{d)} All behavior statistics for both experiment (black, solid bars) and model (red, empty bars). (b) shows the velocity correlation function $\langle v_1(t)v_2(t')\rangle_{\mathrm{same}}$ when cells are on the same island. (c) show the position correlation function $\langle x_1(t)x_2(t')\rangle$. (d) shows the collision statistics for all different cell types. \textbf{e)} Table shows the final best fitting parameters for the model PPC+PPrC for all the different cell types. In the global fit, we only allow $\epsilon_{\mathrm{PPC}}$ to vary across distinct cell types.}
	\label{fig:global_fit}
\end{figure}

Prime candidates to combine with PPC are interactions that add effective repulsion to the dynamics and generate mutual exclusion behavior. These are features of NC, NPrC, PrC, PPrC, or PNC with $\epsilon > 0$ (Fig. \ref{fig:model_comparison_nucleus}, Fig. \ref{fig:model_comparison_protrusion}). Furthermore, we observed that PPC can not well capture the dynamics of HT1080 cells. Thus, using either PrC, PSC, PPrC, or PNC could allow us to also capture the dynamics of these cells. For instance, combining PPC with PPrC has an interesting effect on the dynamics of polarity alignment interactions: While the underdamped repulsion interactions increase, the underdamped friction interactions get weaker (Fig. \ref{fig:repulsion_addition}ai,ii). Furthermore, the velocity correlations get weaker, while the position correlation functions get increasingly negative (Fig. \ref{fig:repulsion_addition}aiii,iv). These changes indicate that additional repulsion interactions in the form of PPrC weakens the coordination of the two cells but increases the mutual exclusion behavior. Finally, additional PPrC leads to more reversal behavior (Fig. \ref{fig:repulsion_addition}av). In contrast, for polarity anti-alignment, we find that additional polarity repulsion (PPrC) qualitatively changes the dynamics and introduces long range repulsion to the attractive interactions of polarity alignment (Fig. \ref{fig:repulsion_addition}bi). Furthermore, the effect on the inferred underdamped friction interactions, velocity correlations, position correlations, and collision statistics are similar to that of the case with polarity alignment (Fig. \ref{fig:repulsion_addition}bii-iv). Thus, additional PPrC can remedy the shortcomings of PPC discussed above: PPrC adds underdamped repulsion interactions, leads to stronger mutual exclusion behavior and increases the percentage of observed reversal behavior for alignment. Consequently, the combined model can better fit the experimental data of epithelial and breast cancer cells (main text Fig. 6a, Fig. \ref{fig:global_fit},\ref{fig:individual_fit}).

%\subsubsection{Choosing the best combination of interaction mechanisms}

%We note that several combinations including PPC and a second mechanism capture equally well the inferred underdamped interactions and behavior statistics. This is most likely because of the fact that several interactions predict very similar underdamped dynamics. Specifically, we find that NC, NPrC, PrC, PSC, PNC and PPrC alone do all predict no significant underdamped friction interactions but significant underdamped repulsion interactions for $\epsilon > 0$ (main text Fig. 4c).   

\subsubsection{Other combinations without PPC}

Finally, note that the improved performance of combinations of PPC and other interactions is not merely the consequence of including more parameters in the fit. Instead, our results suggest that including PPC in the combination is necessary to reach good fitting performance. We show this by also considering the combination of PPrC with NC. This combination does not provide a significantly improved fit compared to only using PPrC (main text Fig. 6a).

\phantom{blabla}
\newpage

\subsubsection{More than two interaction mechanisms}

\begin{figure}[t!]
	\centering
	\includegraphics[width=\textwidth]{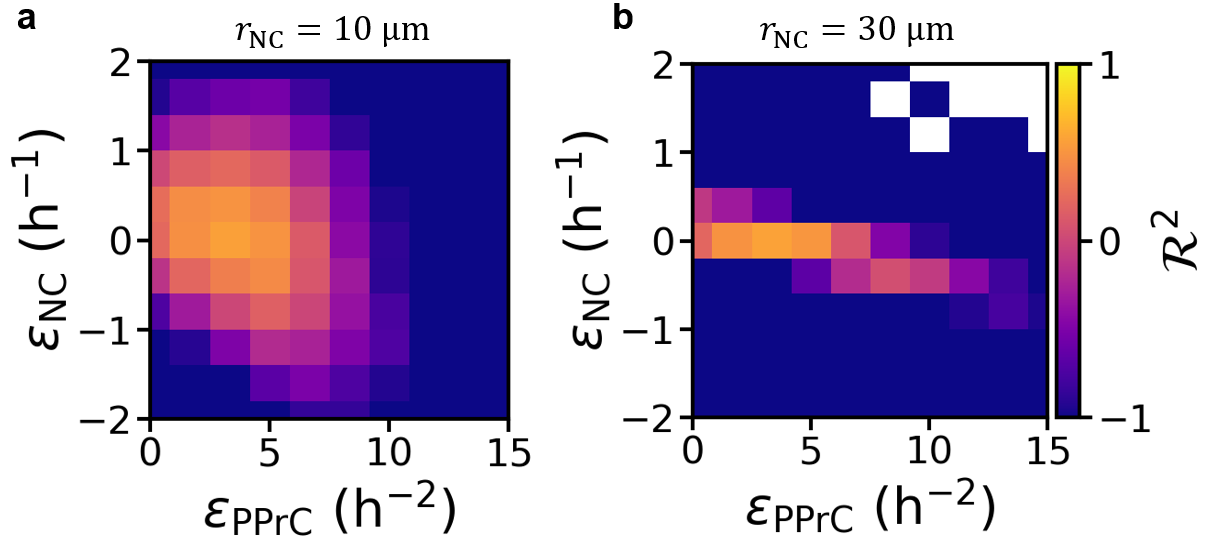}
	\caption{\textbf{Global fit of three interaction mechanisms} \small \textbf{a)},\textbf{b)} Global coefficient of determination $\mathcal{R}(\epsilon_{\mathrm{PPrC}},\epsilon_{\mathrm{NC}})$ for a model with the three interaction mechanisms PPC, PPrC and NC. Panel (a) shows $\mathcal{R}$ at $r_{\mathrm{NC}} = 10\mu m$, panel (b) at $r_{\mathrm{NC}} = 30\mu m$. The maximum is located near $\epsilon_{\mathrm{NC}}=0$ indicating that NC does not improve the fitting performance of the best fitting parameter combination of the model with only PPC and PPrC.}
	\label{fig:three_mechanisms}
\end{figure}

In this subsection, we assess whether adding a third interaction mechanism provides an improved fit of the experimental data compared to our best performing combination of PPC and PPrC. Considering three candidate interactions simultaneously is computationally very expensive. Thus, we consider here an example triplet consisting of PPC, PPrC and NC. Here, NC stand for nucleus coupling allowing for nucleus repulsion or attraction. Such an interaction is potentially critical as the nucleus of cells may contribute crucial excluded volume interactions. To assess whether this third interaction improves the fitting performance, we aim to find the global coefficient of determination $\mathcal{R}( \epsilon_{\mathrm{PPrC}}, \epsilon_{\mathrm{NC}}, r_{\mathrm{PPC}} ,r_{\mathrm{PPrC}}, r_{\mathrm{NC}})$ as defined in equation \ref{eq:COD}. However, as this is numerically costly, we vary only simultaneously the interaction strengths of all three interaction mechanisms $\epsilon_{\mathrm{PPC}}$, $\epsilon_{\mathrm{PPrC}}$, and $\epsilon_{\mathrm{NC}}$. In this procedure, we fix the interaction ranges $r_{\mathrm{PPC}}$ and $r_{\mathrm{PPrC}}$ to the optimal values found through the global fit of PPC and PPrC as described in section \ref{sec:fitting}. We set the interaction range to $r_{\mathrm{NC}} = 10\mu m$ and $r_{\mathrm{NC}} = 30\mu m$. We thus compute the global coefficient of determination $\mathcal{R}(\epsilon_{\mathrm{PPrC}},\epsilon_{\mathrm{NC}})$, which we depict in Fig. \ref{fig:three_mechanisms}a,b. We find that $\mathcal{R}(\epsilon_{\mathrm{PPrC}},\epsilon_{\mathrm{NC}})$ is maximized when $\epsilon_{\mathrm{NC}} \simeq 0$ independent of $r_{\mathrm{NC}}$. This procedure cannot exclude new maxima in the full coefficient of determination $\mathcal{R}( \epsilon_{\mathrm{PPrC}}, \epsilon_{\mathrm{NC}}, r_{\mathrm{PPC}} ,r_{\mathrm{PPrC}}, r_{\mathrm{NC}})$. However, our analysis shows that the maximum we found in $\mathcal{R}(\epsilon_{\mathrm{PPrC}}, r_{\mathrm{PPC}} ,r_{\mathrm{PPrC}})$ in the global fit of PPC and PPrC is at least a local maximum in the more complex $\mathcal{R}(\epsilon_{\mathrm{PPrC}}, \epsilon_{\mathrm{NC}}, r_{\mathrm{PPC}} ,r_{\mathrm{PPrC}}, r_{\mathrm{NC}})$. This indicates that the best fitting parameter combination of PPC and PPrC can not be gradually improved by adding additional nucleus coupling.      

\subsection{The role of single cell behavior}\label{sec:model_single_cells}

In this section, we study the role of the single cell terms in our overdamped model. Specifically, in the absence of cell-cell interactions, our model is consistent with the mechanistic model for MDA-MB-231 cells used in \cite{Bruckner2022}. Given that we here study multiple different cell lines, it is a key question whether the single cell dynamics of our model has a significant impact on the results about cell-cell interactions presented in the main text.

\begin{figure}[t!]
	\centering
	\includegraphics[width=\textwidth]{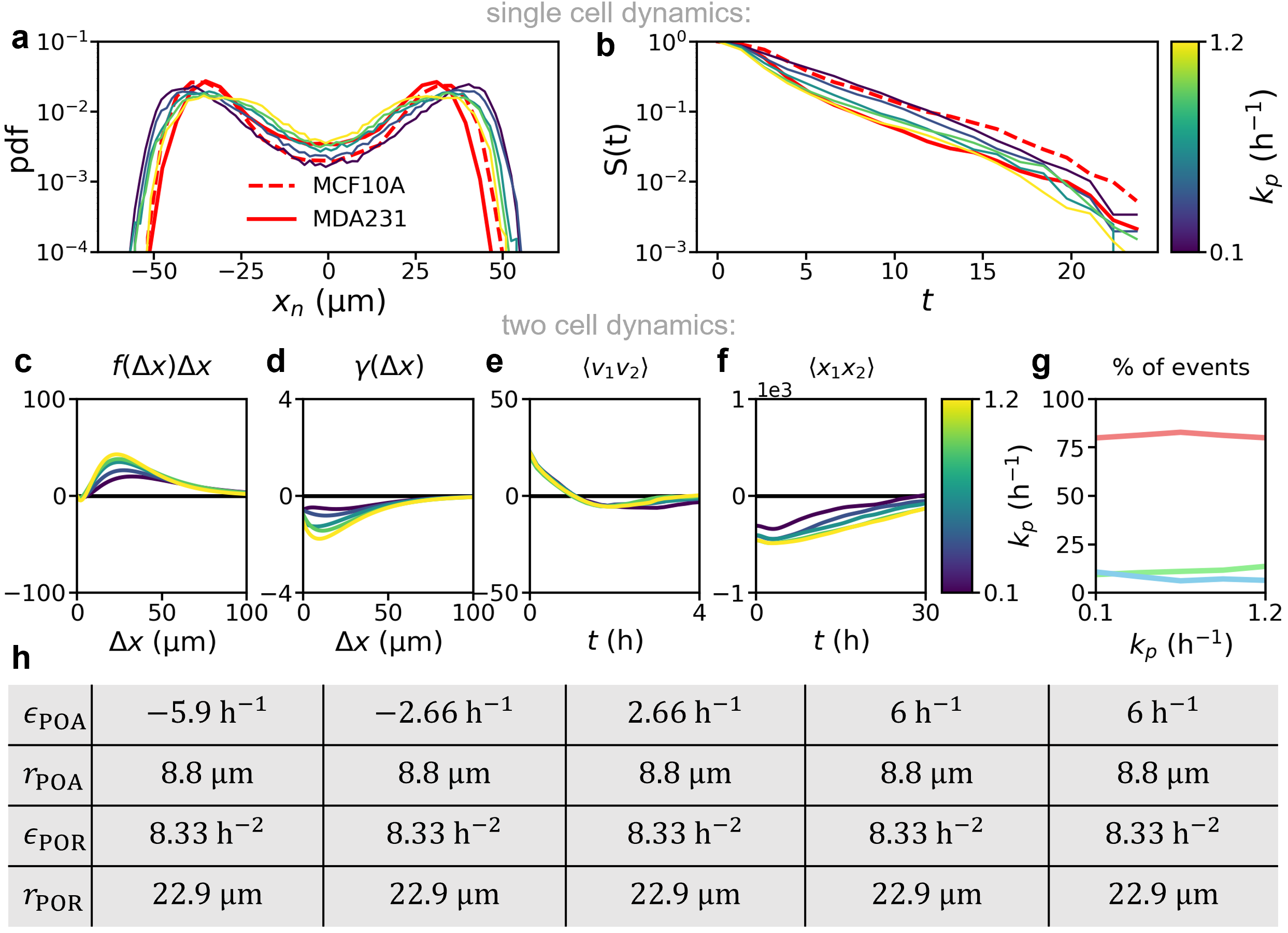}
	\caption{\textbf{Influence of single cell dynamics on two cell dynamics.} \small \textbf{a)} Distribution of nucleus positions for both model and experiment (red) for model and experiments of a single cell on the dumbbell-shaped micropattern. Red solid line shows the experimental result of MDA-MB-231 cells and red dashed line indicates MCF10A cells. Colored lines indicate model results and colorbar for varying $k_\mathrm{p}$ is located next to panel b. \textbf{b)} Survival probability of both model (colored) and experiment (red) for single cells on the pattern. Legend of panel a applies also here. \textbf{c)}-\textbf{g)} Model predictions for the two-cell dynamics dependent on the single cell parameter $k_\mathrm{p}$. Panel (c) shows the underdamped repulsion interactions, panel (d) shows the underdamped friction interactions, panel (e) shows the velocity cross-correlation of two cells on the same island. Panel (f) shows the position correlations of the two cells. Colorbar in panel f applies to panels c-f. Panel (g) shows the percentage of each type of collision event dependent on $k_\mathrm{p}$. \textbf{h)} Resulting parameters in our overdamped model after globally fitting the dynamics of cell-cell interactions.}
	\label{fig:model_single_cell}
\end{figure}

To answer this question, we first determine the range of single cell parameters that is relevant to describe single cell behavior observed in our experiments. Therefore, we find the best single cell parameters for a second cell line, which is significantly different from MDA-MB-231 cells. A prime candidate are MCF10A cells as they show a slightly lower probability to be on the bridge and a significantly higher survival probability (Fig. \ref{fig:single_cell_invasiveness}d i,v). To capture the different single cell features of MCF10A cells, we vary key single cell parameters in our overdamped model. Specifically, the reduced spring constant $k_\mathrm{p} = k/\zeta_\mathrm{p}$ (main text Eq. (2), (B2)) determines how the protrusion of a cell responds to the nucleus position. We find that $k_\mathrm{p}$ influences the probability distribution of the nucleus positions and sets the survival probability of single cells (Fig. \ref{fig:model_single_cell}a,b). Decreasing $k_\mathrm{p}$ from $1.2\ \mathrm{h}^{-1}$ for MDA-MB-231 cells to a value of around $0.1\ \mathrm{h}^{-1}$ allows us to capture the reduced probability of cells being on the bridge and enables us to capture the higher survival probabilities of MCF10A cells. Importantly, this different single cell behavior does not greatly influence the inferred underdamped cell-cell interactions and behavior statistics predicted by our overdamped model (Fig. \ref{fig:model_single_cell}c-g). Nevertheless, we repeat our quantitative fitting procedure with the different single cell parameters for MCF10A cells instead of the single cell parameters of the MDA-MB-231 cells. We find that the best fitting interaction parameters in our overdamped model are only mildly affected by the different single cell behavior (Compare Fig. \ref{fig:model_single_cell}h to Fig. \ref{fig:global_fit}e). Thus, for the sake of simplicity, we opt to keep the single cell behavior constant throughout this study, knowing that the single cell behavior does not qualitatively influence the dynamics emerging from cell-cell interactions. This is supported by the observation that almost all our cells obey qualitatively very similar underdamped single cell dynamics (Fig. \ref{fig:single_cell_invasiveness}e).  

\subsection{Polarity-polarity coupling on rectangular patterns}\label{sec:rectangular}

\begin{figure}[t!]
	\centering
	\includegraphics[width=\textwidth]{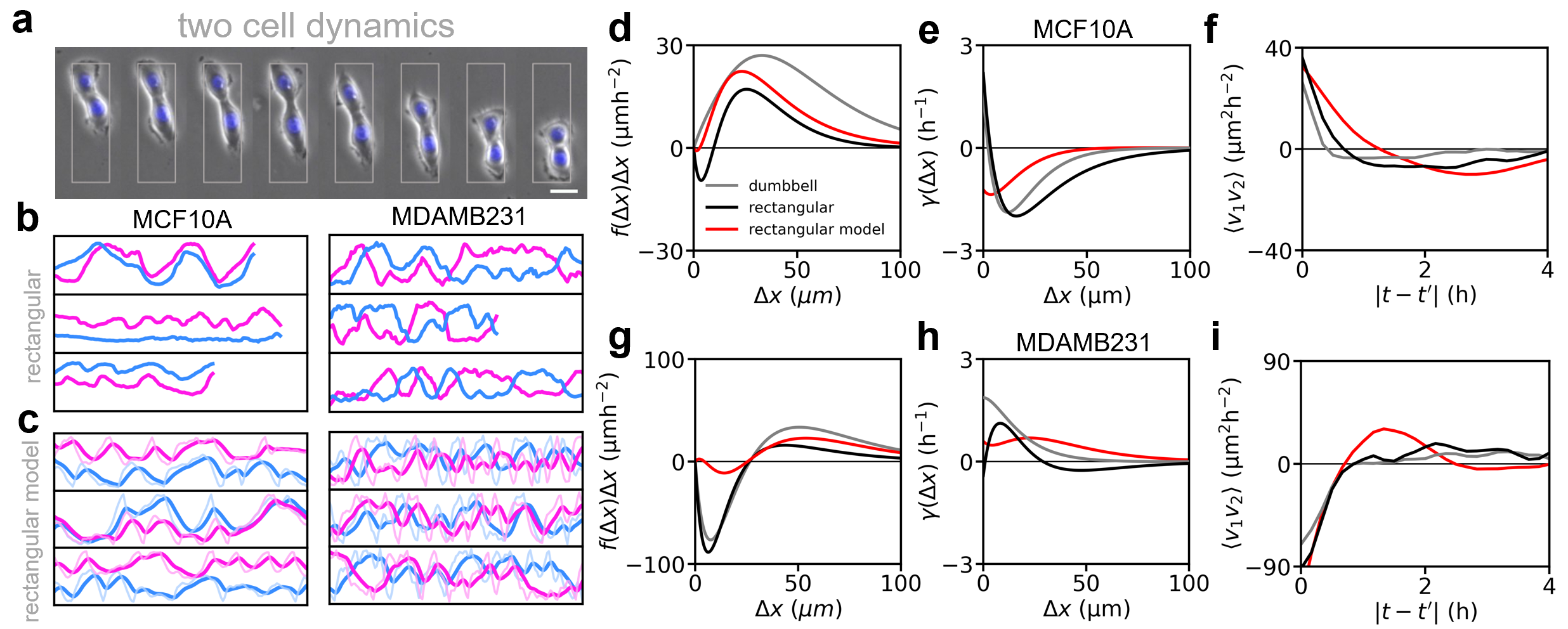}
	\caption{\textbf{Alignment interactions on rectangular patterns.} \textbf{a)} Time series of brightfield images of two MCF10A cells interacting on a rectangular micropattern. Scale bar: 25 µm. \textbf{b)} Sample of cell trajectories obtained from our experiments on a rectangular pattern for both MCF10A and MDA-MB-231 cells. \textbf{c)} Sample of trajectories predicted by our overdamped model with PPC+PPrC as cell-cell interaction. \textbf{d)}-\textbf{e)} Inferred underdamped cohesion interactions (d), inferred underdamped friction interactions (e), and velocity cross-correlation function of two MCF10A cells for the dumbbell pattern (grey), the rectangular pattern (e), and the overdamped model fitted to the rectangular dynamics (red). \textbf{g)}-\textbf{i)} Inferred underdamped cohesion interactions (d), inferred underdamped friction interactions (e), and velocity cross-correlation function of two MDA-MB-231 cells for the dumbbell pattern (grey), the rectangular pattern (e), and the overdamped model fitted to the rectangular dynamics (red).}
	\label{fig:rectangular}
\end{figure}

To test whether insights of our overdamped model sensitively depend on the chosen dumbbell-shaped geometry of our micropattern, we additionally investigate cell-cell interactions on rectangular micropatterns (Fig. \ref{fig:rectangular}a). Specifically, we consider the epithelial cell line MCF10A and the breast cancer cell line MDA-MB-231 and track cell nuclei (Fig. \ref{fig:rectangular}b). We first infer the underdamped cell-cell interactions for these two cell types on the rectangular pattern and compare our results to the inferred underdamped cell-cell interactions on the dumbbell-shaped micropattern. For MCF10A cells, we find that these cells obey qualitatively similar underdamped cell-cell interactions on both geometries (black and grey curves in Fig. \ref{fig:rectangular}d,e). However, the underdamped cohesion interactions are weaker on the rectangular pattern. Furthermore, in agreement with previous work\cite{Bruckner2021}, MDA-MB-231 cells exhibit quantitatively similar underdamped cell-cell interactions (black and grey curves in Fig. \ref{fig:rectangular}g,h). These results show that the geometry of the micropattern has only minor impact on the inferred underdamped cell-cell interactions.

\begin{figure}[t!]
	\centering
	\includegraphics[width=\textwidth]{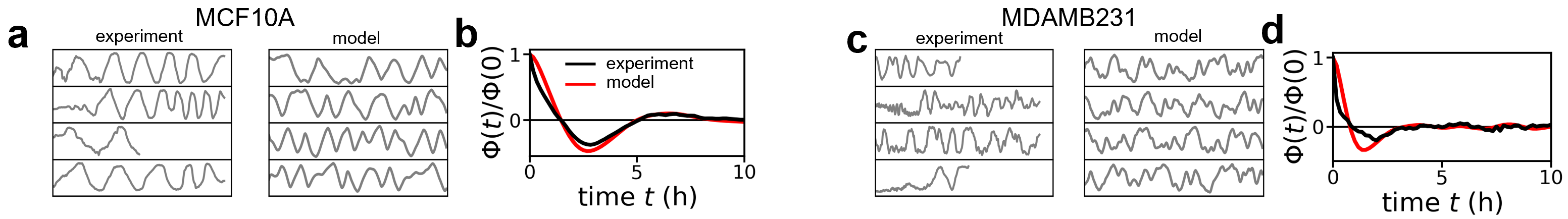}
	\caption{\textbf{Single cell dynamics on rectangular patterns.} \textbf{a)} Sample of cell trajectories of a single MCF10A cell on a rectangular pattern for both experiment and model. \textbf{b)} Velocity auto-correlation function of MCF10A cells on rectangular pattern for both experiment (black) and model (red). \textbf{c)} Sample of cell trajectories of a single MDA-MB-231 cell on a rectangular pattern for both experiment and model. \textbf{d)} Velocity auto-correlation function of MDA-MB-231 cells on rectangular pattern for both experiment (black) and model (red).}
	\label{fig:rectangular_single}
\end{figure}

To test whether our overdamped model can also capture the two-cell dynamics on a rectangular pattern, we perform numerical simulations of our overdamped model on this alternative geometry. Specifically, we consider the best fitting candidate cell-cell interaction PPC+PPrC and use for both MCF10A and MDA-MB-231 cells the best interaction parameters obtained from our global fit on the dumbbell pattern (Fig. \ref{fig:global_fit}e). Importantly, in order to capture the dynamics on a different geometry, we have to adjust the single cell aspects of our overdamped model: We need to remove the geometry dependence of the single cell terms in our overdamped model, which implemented the adaptation of the cell's migratory machinery to the confinement of the bridge in our dumbbell-shaped micropattern (main text Eqs. B4,B5). Specifically, we set $\gamma(x_\mathrm{n}) = 1$ and $\alpha(x_\mathrm{p}) = \alpha_0$. We analyze single cells migrating on rectangular pattern to constrain single cell parameters in our model (Fig. \ref{fig:rectangular_single}). We find that to best capture the dynamics of the cells on the rectangular pattern, we have to set $\alpha_0$ to negative values such that cells remain polarized on the rectangular pattern. We choose $\alpha_0 = -0.3$. To better capture experimental trajectories of single cells on rectangular patterns, we also change the values of $\sigma$ and $k_\mathrm{n}$. Specifically, for MCF10A cells, we choose $\sigma = 40$ and $k_\mathrm{n} = 1.2$. These two changes are necessary to capture the high persistence and frequent turn around events of our MCF10A cells on the rectangular pattern (Fig. \ref{fig:rectangular_single}a) as also quantified by a velocity auto-correlation function $\Phi(t) = \langle v(t')v(t+t')\rangle_{t'}$ (Fig. \ref{fig:rectangular_single}b). Physically, increasing $k_\mathrm{n}$ means that we couple the nucleus closer to the protrusion. This change may be in agreement with the observation of reduced elongation of the cells on rectangular patterns compared to the dumbbell shaped patterns (Fig. \ref{fig:rectangular}a, SI movie). On the rectangular patterns, we observe that when cells hit one of the ends of the rectangular pattern, they polarize away from the end and turn around. We implement this by including a boundary condition where the polarity $P$ of a cell gets flipped to $-P$ as soon as cells hit the boundary of our pattern. For MDA-MB-231 cells, we increase $\sigma$ to $\sigma = 150$ to capture the higher stochasticity of the single cell behavior of these cells on the rectangular pattern (Fig. \ref{fig:rectangular_single}c,d).\\
\\Having constrained the single cell parameters, we can now simulate our overdamped model on a rectangular pattern. We find that using the interaction parameters best describing the dynamics on the dumbbell-shaped geometry, also yields a good description of the dynamics on the rectangular geometry. Specifically, we can qualitatively capture the flocking and frequent reversal behavior of MCF10A cells on the rectangular pattern (Fig. \ref{fig:rectangular}b,c). Furthermore, our overdamped model qualitatively captures the inferred interactions (Fig. \ref{fig:rectangular}d,e) and the positive velocity correlations of MCF10A cells on the rectangular pattern (Fig. \ref{fig:rectangular}f). Moreover, our model captures the frequent sliding events of MDA-MB-231 cells (Fig. \ref{fig:rectangular}b,c) as well as qualitatively the short timescale dynamics of these cells (Fig. \ref{fig:rectangular}g-i). Taken together, these results show that our overdamped model can be generalized to a different geometry. Thus, the identified polarity-polarity coupling interactions do not depend on the specific dumbbell-shaped geometry used in the main text. 

\renewcommand\refname{Supplementary References}
\bibliographystyle{biolett}
\bibliographystyle{unsrt}
\bibliography{library}